%% file: paper.tex
\documentclass{article}
\usepackage{latexsym} 
\usepackage{amstex}  
\usepackage{epsfig}
\usepackage{array}
\usepackage{pstricks,pst-coil,pst-node,pst-plot}

\input defs.tex

\title{Self-consistent Calculation of Real Space\\ 
Renormalization Group Flows\\
and Effective Potentials
\thanks{Supported by Deutsche Forschungsgemeinschaft}
} 
\author{M. Griessl, G.Mack, Y. Xylander \\
   II. Institut f\"ur Theoretische Physik der Universit\"at Hamburg, \\
   D-22761 Hamburg, Luruper Chaussee 149, Germany
\\[6pt]
G. Palma\\
Departmento de F\'{\i}sica, Universidad de Santiago de Chile,\\
Casilla 307, Correo 2, Santiago, Chile
}

\begin{document}
\maketitle
\date{February 9, 1996}

\begin{abstract}
We show how to compute real space renormalization group flows in 
lattice field theory by a self-consistent method which
is designed to preserve the
basic stability properties of a Boltzmann factor.
Particular attention is payed to controlling the errors which come from
truncating the action to a manageable form.
In each step, the integration over the 
fluctuation field (high frequency components of the field) is performed 
by a saddle point method.
The saddle point depends on the block-spin.
Higher powers of derivatives of the field are neglected in the actions, but
no polynomial approximation in the field is made.
The flow preserves a simple parameterization of the action. 
In the first part the method is described and numerical results are
presented. In the second part we  discuss an improvement 
 of the method where
the saddle point approximation
is preceded by self-consistent normal ordering,
i.e. solution of a gap equation. 
In the third part we describe a general procedure 
to obtain higher order corrections with the help of
Schwinger Dyson equations.

In this paper we 
treat scalar field theories as an example. The basic limitations of the 
method are also discussed. They come from a possible breakdown of stability 
which may occur  when
a composite block-spin or block variables for domain walls would be needed. 
\end{abstract}

\input intro.tex
\input saddle.tex
\input numerik.tex

\input normalorder.tex
\input DS.tex
\input rest.tex
\input discreteModels.tex

\bigskip\medskip

\noindent{\Large\bf Acknowledgement}
\medskip

G. P. was partially supported by Fondecyt \#1930067 and Dicyt \#049331PA and
would like to thank II. Institut f\"ur Theoretische Physik der Universit\"at
Hamburg for the kind hospitality.

\input appendix.tex



\end{document}

%% file: defs.tex
\newcommand{\sticky}[1]{%
}

\def\nn{\nonumber}
\def\half{\frac 12}

\def\H{{\cal H}}
\def\tr{\text{tr}}

\newcommand{\scalar}[2]{#1 #2}

\def\Einsch#1{\Bigl\arrowvert\mbox{\lower5pt\hbox{$#1$}}{}}

\newcommand{\Adj}[1]{{#1}{}^{\ast}}

\newcommand{\E}[1]{{#1{}^{i}}}

\newcommand{\HCE}{\H^i_{\AvOp}}

\newcommand{\Z}[1]{{#1{}^{i+1}}}

\newcommand{\D}[1]{{#1{}^{i+2}}}

\newcommand{\Lat}{\Lambda }

\newcommand{\LatNull}{\Lat^0}

\newcommand{\LatN}{\Lat^N}

\newcommand{\LatE}{\E{\Lat}}

\newcommand{\LatZ}{\Z{\Lat}}

\newcommand{\LatSpace}{a}

\newcommand{\LatLength}{L}

\newcommand{\Dim}{D}

\newcommand{\LatSpaceE}{\LatSpace_{i}}
\newcommand{\LatSpaceZ}{\LatSpace_{i+1}}

\newcommand{\LatLengthE}{\LatLength_{i}}

\newcommand{\DualLat}{\tilde{\Lambda} }

\newcommand{\DualLatE}{\E{\DualLat}}

\newcommand{\DualLatZ}{\Z{\DualLat}}
\newcommand{\DualLatEZ}{\DualLat^{i,i+1}}

\newcommand{\DualLatSpace}{\tilde{a}}
\newcommand{\DualLatLength}{\tilde{L}}

\newcommand{\BlockFactor}{s}

\newcommand{\DualLatSpaceE}{\DualLatSpace_{i}}
\newcommand{\DualLatSpaceEZ}{\DualLatSpace_{i,i+1}}
\newcommand{\DualLatSpaceZ}{\DualLatSpace_{i+1}}

\newcommand{\DualLatLengthE}{\DualLatLength_{i}}
\newcommand{\DualLatLengthEZ}{\DualLatLength_{i,i+1}}

\newcommand{\LatExt}{N}
\newcommand{\LatExtE}{\LatExt_{i}}

\newcommand{\xE}{z}
\newcommand{\yE}{w}
\newcommand{\xZ}{x}
\newcommand{\yZ}{y}

\newcommand{\qE}{p}
\newcommand{\qZ}{q}
\newcommand{\lZ}{l}

\newcommand{\p}{p}
\newcommand{\q}{q}
\newcommand{\lM}{l}

\newcommand{\ZE}{\E Z}

\newcommand{\Action}{S}

\newcommand{\ActionNull}{\Action^0}

\newcommand{\ActionE}{\E{\Action}}
\newcommand{\ActionZ}{\Z{\Action}}
\newcommand{\ActionD}{\D{\Action}}
\newcommand{\ActionPE}{{\E{\Action}}{}'}
\newcommand{\ActionPZ}{{\Z{\Action}}{}'}

\newcommand{\ActionPPE}{{\E{\Action}}{}''}
\newcommand{\ActionPPZ}{{\Z{\Action}}{}''}

\newcommand{\ActionPPPE}{{\E{\Action}}{}''' }

\newcommand{\ActionN}{\Action^N}

\newcommand{\PsiderivativeE}[1]{\frac {\delta}{\delta \MFAFieldE (#1)}}

\newcommand{\NAction}{T}
\newcommand{\NActionE}{\E{\NAction}}
\newcommand{\NActionZ}{\Z{\NAction}}

\newcommand{\NActionpE}{{\NActionE_;}}

\newcommand{\NActionppE}{{\NActionE_{;;}}}

\newcommand{\NActionpPE}{{\NActionE{}_{\mathrm{;\,,}}}}

\newcommand{\KAction}{K}
\newcommand{\KActionE}{\E{\KAction}}

\newcommand{\KActionpE}{\E{\KAction}_;}

\newcommand{\KActionppE}{\E{\KAction}_{;;}}

\newcommand{\GAction}{G}
\newcommand{\GActionE}{\E{\GAction}}

\newcommand{\QAction}{Q}
\newcommand{\QActionpE}{\E{\QAction}_;{}}

\newcommand{\QActionppE}{\E{\QAction}_{;;}{}}

\newcommand{\QActionpPE}{\E{\QAction}_{;}{}_{\,,}{}}

\newcommand{\Field}{\phi}

\newcommand{\FieldNull}{\Field^0}

\newcommand{\FieldE}{\phi}
\newcommand{\FieldZ}{\Phi}
\newcommand{\FieldD}{\varphi}
\newcommand{\FieldlocE}{\overline{\FieldE}}
\newcommand{\FieldlocZ}{\overline{\FieldZ}}

\newcommand{\FieldN}{\Phi}

\newcommand{\dEx}[1]{\frac{\delta }{ \delta \FieldE (#1)} }
\newcommand{\dZx}[1]{\frac{\delta }{ \delta \FieldZ (#1)} }

\newcommand{\dZxy}[2]{\frac{\delta^2}{\delta \FieldZ (#1)\delta \FieldZ (#2) }}

\newcommand{\VPot}{V}
\newcommand{\VPotNull}{\VPot^0}

\newcommand{\VPotE}{\E{\VPot}}
\newcommand{\VPotZ}{\Z{\VPot}}

\newcommand{\VPotPE}{{\E{\VPot}}{}'}

\newcommand{\VPotloc}{\overline{V}{}}

\newcommand{\VPotlocPNull}{\VPotloc^{0\prime}}

\newcommand{\VPotlocE}{\E{\VPotloc}}

\newcommand{\VPotlocPE}{{\E{\VPotloc}}{}'}
\newcommand{\VPotlocPZ}{{{\Z{\VPotloc}}{}'}}

\newcommand{\WPot}{W}
\newcommand{\WPotNull}{\WPot^0}
\newcommand{\WPotE}{\E{\WPot}}
\newcommand{\WPotZ}{\Z{\WPot}}

\newcommand{\WPotloc}{\overline{W}{}}
\newcommand{\WPotlocNull}{\WPotloc^0}
\newcommand{\WPotlocPNull}{\WPotloc^{0\prime}}
\newcommand{\WPotlocPPNull}{\WPotloc^{0\prime \prime}}
\newcommand{\WPotlocE}{\E{\WPotloc}}
\newcommand{\WPotlocZ}{\Z{\WPotloc}}

\newcommand{\WPotlocPE}{\E{\WPotloc}{}'}

\newcommand{\WPotlocPPE}{\E{\WPotloc}{}''}

\newcommand{\InOp}{{\cal{A}}}
\newcommand{\InOpE}{\E{\InOp}}

\newcommand{\FlucProp}{\Gamma}

\newcommand{\FlucPropNull}{\FlucProp^0}
\newcommand{\FlucPropE}{\E{\FlucProp}}
\newcommand{\FlucPropZ}{\Z{\FlucProp}}

\newcommand{\FlucPropPE}{\E{\FlucProp}{}'}

\newcommand{\FlucPropVar}{\Gamma_B}
\newcommand{\FlucPropVarE}{\E{\FlucPropVar}}

\newcommand{\Propagator}{v}
\newcommand{\PropagatorE}{\E{\Propagator}}
\newcommand{\PropagatorZ}{\Z{\Propagator}}

\newcommand{\FlucField}{\zeta}
\newcommand{\FlucFieldE}{\E{\FlucField}}

\newcommand{\SourceField}{\xi}
\newcommand{\SourceFieldE}{\E{\SourceField}}

\newcommand{\AvOp}{{\cal{C}}}
\newcommand{\AvOpE}{\E{\AvOp}}

\newcommand{\AvOpTildeE}{\E{\widetilde{\AvOp}}}

\newcommand{\av}[1]{\underset{#1}{\text{av}}}

\newcommand{\Laplace}{\Delta}

\newcommand{\mean}[1]{\langle{#1}\rangle}
\newcommand{\meanE}[1]{\E{\mean{#1}}}

\newcommand{\MFAField}{\Psi }

\newcommand{\MFAFieldE}{\E{\Psi}}
\newcommand{\MFAFieldZ}{\Z{\Psi}}

\newcommand{\MFAFieldPE}{{\E{\Psi}}{}'}

\def\idE{1} 
\def\idZ{1} 

\newcommand{\expdiff}{
e^{\left( \half \scalar{\frac{\delta}{\delta \xi}}{\FlucPropE[\MFAField] 
\frac{\delta}{\delta \xi}\right)}}}

\input graphen.tex


%% file: graphen.tex
\newlength{\scale} 
\psset{linewidth=1.7pt,unit=\scale}

\def\SnodeE#1#2{\cnode[fillstyle=solid,fillcolor=lightgray,linewidth=0.8pt](#1){0.5}{#2}}
\def\SnodeZ#1#2{\cnode[fillstyle=solid,fillcolor=lightgray,linewidth=0.8pt](#1){0.5}{#2}}
\def\SnodeD#1#2{\cnode[fillstyle=solid,fillcolor=lightgray,linewidth=0.8pt](#1){0.8}{#2}}

\def\NnodeE#1#2{\cnode(#1){0.5}{#2}\rput(#1){N}}
\def\NnodeZ#1#2{\cnode(#1){0.5}{#2}\rput(#1){N}}
\def\KnodeE#1#2{\cnode[fillstyle=crosshatch,hatchsep=2pt](#1){0.5}{#2}}
\def\KnodeZ#1#2{\cnode[fillstyle=crosshatch,hatchsep=2pt](#1){0.5}{#2}}
\def\psivertex#1{\cnode[fillstyle=solid,fillcolor=black](#1){0.15}{Blob}}
\def\vertex#1#2{\cnode[fillstyle=solid,fillcolor=black](#1){0.2}{#2}}
\def\psiline{\psCoil[coilaspect=0,coilheight=2,coilwidth=0.15]{-90}{1840}}
\def\psiliner{\psCoil[coilaspect=0,coilheight=2,coilwidth=0.15]{-40}{1890}}

\newcommand{\BildAa}[1]{
\begin{figure}
\begin{eqnarray*}
-\frac{\delta^n}{\delta\FieldE(\xE_1)\cdots\delta\FieldE(\xE_n)}
\ActionE[\FieldE]
&=&
\pspicture[0.45](0,0)(2,2)
\SnodeZ{1,1}{Vertex1}
\rput(0.3,0.3){\rnode{z1}{$\xE_1$}}
\rput(1.7,0.3){\rnode{z2}{$\xE_n$}}
\rput(1.0,0.3){\rnode{zz}{$\cdots$}}
\ncline[nodesepB=1pt]{Vertex1}{z1}
\ncline[nodesepB=1pt]{Vertex1}{z2}
\endpspicture
\\
 - \frac{\delta^n}{\delta\FieldZ(\xZ_1)\cdots\delta\FieldZ(\xZ_n)}
\ActionZ[\FieldZ]
&=&
\pspicture[0.45](0,0)(2,2)
\SnodeD{1,1}{Vertex1}
\rput(0.0,0.0){\rnode{z1}{$\xZ_1$}}
\rput(2.0,0.0){\rnode{z2}{$\xZ_n$}}
\rput(1.0,0.0){\rnode{zz}{$\cdots$}}
\ncline[nodesepB=1pt]{Vertex1}{z1}
\ncline[nodesepB=1pt]{Vertex1}{z2}
\endpspicture
\\
 \frac{\delta}{\delta\FieldZ(\xZ)}\MFAFieldE[\FieldZ](z)
&=&
\ 
\pspicture[0.29](0,-1)(5,2)
\put(0.3,0){\psiline}\psivertex{1.9,0}
\rput(0.0,0.0){\rnode{z1}{$\xZ$}}
\rput(2.3,0.0){\rnode{z1}{$\xE$}}
\endpspicture
\end{eqnarray*}
   \caption{#1}
 \label{BildAa}
\end{figure}
}
\newcommand{\BildAb}[1]{
\begin{figure}
\begin{eqnarray*}
 - \frac{\delta^n}{\delta\SourceFieldE(\xE_1)\cdots\delta\SourceFieldE(\xE_n)}
\NActionE[\MFAFieldE,\SourceFieldE]\Einsch{\SourceFieldE=0}
&=&
\pspicture[0.45](0,0)(2,2)
\NnodeE{1,1}{Vertex1}
\rput(0.3,0.3){\rnode{z1}{$\xE_1$}}
\rput(1.7,0.3){\rnode{z2}{$\xE_n$}}
\rput(1.0,0.3){\rnode{zz}{$\cdots$}}
\ncline[nodesepB=1pt]{Vertex1}{z1}
\ncline[nodesepB=1pt]{Vertex1}{z2}
\endpspicture
\end{eqnarray*}
   \caption{#1}
 \label{BildAb}
\end{figure}
}

\newcommand{\BildB}[1]
{
\begin{figure}
\begin{equation*}
\pspicture[0.45](0,0)(2,2)
\NnodeE{1,1}{Vertex1}
\rput(0.3,0.3){\rnode{z1}{$\xE_1$}}
\rput(1.7,0.3){\rnode{z2}{$\xE_n$}}
\rput(1.0,0.3){\rnode{zz}{$\cdots$}}
\ncline[nodesepB=1pt]{Vertex1}{z1}
\ncline[nodesepB=1pt]{Vertex1}{z2}
\endpspicture
=
\pspicture[0.45](0,0)(2,2)
\SnodeZ{1,1}{Vertex1}
\rput(0.3,0.3){\rnode{z1}{$\xE_1$}}
\rput(1.7,0.3){\rnode{z2}{$\xE_n$}}
\rput(1.0,0.3){\rnode{zz}{$\cdots$}}
\ncline[nodesepB=1pt]{Vertex1}{z1}
\ncline[nodesepB=1pt]{Vertex1}{z2}
\endpspicture
+
\frac12
\pspicture[0.45](0,0)(2,2)
\SnodeZ{1,1}{Vertex1}
\rput(0.3,0.3){\rnode{z1}{$\xE_1$}}
\rput(1.7,0.3){\rnode{z2}{$\xE_n$}}
\rput(1.0,0.3){\rnode{zz}{$\cdots$}}
\nccircle{Vertex1}{0.5\scale}
\ncline[nodesepB=1pt]{Vertex1}{z1}
\ncline[nodesepB=1pt]{Vertex1}{z2}
\endpspicture
+\frac18
\pspicture[0.45](0,0)(2,2)
\SnodeZ{1,1}{Vertex1}
\rput(0.3,0.3){\rnode{z1}{$\xE_1$}}
\rput(1.7,0.3){\rnode{z2}{$\xE_n$}}
\rput(1.0,0.3){\rnode{zz}{$\cdots$}}
\ncline[nodesepB=1pt]{Vertex1}{z1}
\ncline[nodesepB=1pt]{Vertex1}{z2}
\nccircle{Vertex1}{0.5\scale}
\nccircle{Vertex1}{0.7\scale}
\endpspicture
+
\cdots
\end{equation*}
\caption{#1}
 \label{BildB}
\end{figure}
}

\newcommand{\BildC}[1]
{
\begin{figure}
\begin{equation*}
\begin{split}
\pspicture[0.465](0,-1)(2.2,1)
\psline(0.0,0)(0.2,0)
\psline(2.0,0)(1.8,0)
\SnodeD{1,0}{Vertex1}
\endpspicture
= & 
\pspicture[0.465](0,-1)(5,1)
\put(0.3,0){\psCoil[coilaspect=0,coilheight=2,coilwidth=0.15]{-90}{1840}}
\cnode[fillstyle=solid,fillcolor=black](1.9,0){0.15}{Blob}
\cnode[fillstyle=solid,fillcolor=black](3.1,0){0.15}{Blob1}
\put(3.2,0){\psCoil[coilaspect=0,coilheight=2,coilwidth=0.15]{-40}{1890}}
\cnode[fillstyle=solid,fillcolor=lightgray,linewidth=0.8pt](2.5,0){0.5}{Vertex1}
\endpspicture
+ \frac12
\pspicture[0.465](0,-1)(5,1)
\put(0.3,0){\psCoil[coilaspect=0,coilheight=2,coilwidth=0.15]{-90}{1840}}
\cnode[fillstyle=solid,fillcolor=black](1.9,0){0.15}{Blob}
\cnode[fillstyle=solid,fillcolor=black](3.1,0){0.15}{Blob1}
\put(3.2,0){\psCoil[coilaspect=0,coilheight=2,coilwidth=0.15]{-40}{1890}}
\cnode[fillstyle=solid,fillcolor=lightgray,linewidth=0.8pt](2.5,0){0.5}{Vertex1}
\nccircle{Vertex1}{0.7\scale}
\endpspicture \\
& + \frac12
\pspicture[0.465](0,-1)(7.6,1)
\put(0.3,0){\psCoil[coilaspect=0,coilheight=2,coilwidth=0.15]{-90}{1840}}
\cnode[fillstyle=solid,fillcolor=black](1.9,0){0.15}{Blob}
\cnode[fillstyle=solid,fillcolor=black](5.6,0){0.15}{Blob1}
\put(5.7,0){\psCoil[coilaspect=0,coilheight=2,coilwidth=0.15]{-40}{1890}}
\cnode[fillstyle=solid,fillcolor=lightgray,linewidth=0.8pt](2.5,0){0.5}{Vertex1}
\cnode[fillstyle=solid,fillcolor=lightgray,linewidth=0.8pt](5,0){0.5}{Vertex2}
\ncarc[arcangleA=45,arcangleB=45]{Vertex1}{Vertex2}
\ncarc[arcangleA=-45,arcangleB=-45]{Vertex1}{Vertex2}
\endpspicture \\
& + \frac12
\pspicture[0.465](0.3,-1.5)(6.7,1.5)
\rput[r]{-30}(2.5,0){\put(-2.2,0){\psCoil[coilaspect=0,coilheight=2,coilwidth=0.15]{-90}{1840}}%
          \cnode[fillstyle=solid,fillcolor=black](-0.6,0){0.15}{Blob}%
         }
\rput[r]{30}(2.5,0){\put(-2.2,0){\psCoil[coilaspect=0,coilheight=2,coilwidth=0.15]{-90}{1840}}%
          \cnode[fillstyle=solid,fillcolor=black](-0.6,0){0.15}{Blob1}%
         }
\cnode[fillstyle=solid,fillcolor=lightgray,linewidth=0.8pt](2.5,0){0.5}{Vertex1}
\cnode[fillstyle=solid,fillcolor=lightgray,linewidth=0.8pt](5,0){0.5}{Vertex2}
\ncline{Vertex1}{Vertex2}
\nccircle[angleA=-90]{Vertex2}{0.7\scale}
\endpspicture \\
\end{split}
\end{equation*}
   \caption{#1}
 \label{BildC}
\end{figure}
}

\newcommand{\BildD}[1]
{
\begin{figure}
\begin{equation*}
\pspicture[0.465](0,-1)(2.0,1)
\psline(0.0,0)(0.5,0)
\SnodeD{1,0}{Vertex1}
\endpspicture
= 
\pspicture[0.465](0,-1)(3.2,1)
\put(0.3,0){\psCoil[coilaspect=0,coilheight=2,coilwidth=0.15]{-90}{1840}}
\cnode[fillstyle=solid,fillcolor=black](1.9,0){0.15}{Blob}
\SnodeZ{2.5,0}{Vertex1}
\endpspicture
+ \frac12
\pspicture[0.465](0,-1)(4.1,1)
\put(0.3,0){\psCoil[coilaspect=0,coilheight=2,coilwidth=0.15]{-90}{1840}}
\cnode[fillstyle=solid,fillcolor=black](1.9,0){0.15}{Blob}
\SnodeZ{2.5,0}{Vertex1}
\nccircle[angle=-90]{Vertex1}{0.7\scale}
\endpspicture
\end{equation*}
  \caption{#1}
\label{BildD}
\end{figure}
}

\newcommand{\BildE}[1]
{
\begin{figure}
\begin{equation*}
\pspicture[0.45](0,0)(2,2)
\KnodeE{1.5,1}{Vertex1}
\pnode(0.0,1){z1}
\ncline{Vertex1}{z1}
\endpspicture
= 
0
\end{equation*}
\begin{equation*}
\pspicture[0.45](0,0)(3,2)
\KnodeE{1.5,1}{Vertex1}
\pnode(0.0,1){z1}
\pnode(3,1){z2}
\ncline{Vertex1}{z1}
\ncline{Vertex1}{z2}
\endpspicture
= 
0
\end{equation*}
\caption{#1}
\label{BildE}
\end{figure}
}

\newcommand{\BildF}[1]
{
\newcounter{saveeqn}
\setcounter{saveeqn}{\value{equation}}
\setcounter{equation}{0}
\renewcommand{\theequation}{\alph{equation}}
\begin{figure}
  \begin{equation}
\pspicture[0.45](0,0)(2.5,2)
\NnodeE{1.5,1}{Vertex1}
\pnode(0.0,1){z1}
\ncline{Vertex1}{z1}
\endpspicture
+
\frac16
\pspicture[0.45](0,0)(4.5,2)
\NnodeE{1.5,1}{Vertex1}
\KnodeE{3.5,1}{Vertex2}
\pnode(0.0,1){z1}
\ncline{Vertex1}{z1}
\ncline{Vertex1}{Vertex2}
\ncarc[arcangle=25]{Vertex2}{Vertex1}
\ncarc[arcangle=25]{Vertex1}{Vertex2}
\endpspicture
+
\cdots
= 0 \label{BildFa}
\end{equation}
\begin{eqnarray}
\pspicture[0.45](0,0)(1,2)
\pnode(0.0,1){z3}
\pnode(1.0,1){z4}
\ncline{z3}{z4}
\endpspicture
&=&
\pspicture[0.45](0,0)(3,2)
\NnodeE{1.5,1}{Vertex1}
\pnode(0.0,1){z1}
\pnode(3,1){z2}
\ncline{Vertex1}{z1}
\ncline{Vertex1}{z2}
\endpspicture
+
\frac12
\pspicture[0.45](0,0)(5,2)
\NnodeE{1.5,1}{Vertex1}
\KnodeE{3.5,1}{Vertex2}
\pnode(0.0,1){z1}
\pnode(5.0,1){z2}
\ncline{Vertex1}{z1}
\ncline{Vertex2}{z2}
\ncarc[arcangle=25]{Vertex2}{Vertex1}
\ncarc[arcangle=25]{Vertex1}{Vertex2}
\endpspicture
\nonumber \\
&+&
\frac16
\pspicture[0.45](0,0)(5,2)
\NnodeE{1.5,1}{Vertex1}
\KnodeE{3.5,1}{Vertex2}
\pnode(0.0,1){z1}
\pnode(5.0,1){z2}
\ncline{Vertex1}{z1}
\ncline{Vertex2}{z2}
\ncline{Vertex1}{Vertex2}
\ncarc[arcangle=25]{Vertex2}{Vertex1}
\ncarc[arcangle=25]{Vertex1}{Vertex2}
\endpspicture
+
\frac16
\pspicture[0.25](0,0)(3,3.5)
\NnodeE{1.5,1}{Vertex1}
\pnode(0.0,1){z1}
\pnode(3,1){z2}
\ncline{Vertex1}{z1}
\ncline{Vertex1}{z2}
\KnodeE{1.5,3}{Vertex2}
\pnode(0.0,1){z1}
\pnode(5.0,1){z2}
\ncline{Vertex1}{Vertex2}
\ncarc[arcangle=25]{Vertex2}{Vertex1}
\ncarc[arcangle=25]{Vertex1}{Vertex2}
\endpspicture
+ \cdots
\label{BildFb}
\end{eqnarray}

\begin{multline}
-\ 
\pspicture[0.45](0,0)(4.5,2)
\NnodeE{1.5,1}{Vertex1}
\KnodeE{3.5,1}{Vertex2}
\pnode(0.0,1){z1}
\ncline{Vertex1}{z1}
\ncline{Vertex2}{Vertex1}
\pnode(4.5,0.0){z1}
\pnode(4.5,2){z2}
\ncline[nodesepB=1pt]{Vertex2}{z1}
\ncline[nodesepB=1pt]{Vertex2}{z2}
\endpspicture
\quad=\qquad
\pspicture[0.45](0,0)(2.5,2)
\NnodeZ{1.5,1}{Vertex1}
\pnode(0.0,1){z1}
\ncline{Vertex1}{z1}
\pnode(2.5,0){z3}
\pnode(2.5,2){z2}
\ncline[nodesepB=1pt]{Vertex1}{z3}
\ncline[nodesepB=1pt]{Vertex1}{z2}
\endpspicture  \\
\begin{split}
&+\ 
\frac12
\pspicture[0.45](0,0)(4.5,2)
\NnodeE{1.5,1}{Vertex1}
\KnodeE{3.5,1}{Vertex2}
\pnode(0.0,1){z1}
\ncline{Vertex1}{z1}
\ncarc[arcangle=25]{Vertex2}{Vertex1}
\ncarc[arcangle=25]{Vertex1}{Vertex2}
\pnode(4.5,0.0){z1}
\pnode(4.5,2){z2}
\ncline[nodesepB=1pt]{Vertex2}{z1}
\ncline[nodesepB=1pt]{Vertex2}{z2}
\endpspicture
+ 
\frac16
\pspicture[0.45](0,0)(4.5,2)
\NnodeE{1.5,1}{Vertex1}
\KnodeE{3.5,1}{Vertex2}
\pnode(0.0,1){z1}
\ncline{Vertex1}{z1}
\ncarc[arcangle=25]{Vertex2}{Vertex1}
\ncarc[arcangle=25]{Vertex1}{Vertex2}
\ncline{Vertex1}{Vertex2}
\pnode(4.5,0.0){z1}
\pnode(4.5,2){z2}
\ncline[nodesepB=1pt]{Vertex2}{z1}
\ncline[nodesepB=1pt]{Vertex2}{z2}
\endpspicture
\nonumber \nonumber \\
&+\ 
\frac12
\pspicture*[0.45](0,0)(4.8,4)
\NnodeE{1.5,2}{Vertex1}
\KnodeE{3.5,1}{Vertex2}
\pnode(0.0,2){z1}
\ncline{Vertex1}{z1}
\ncarc[arcangle=25]{Vertex2}{Vertex1}
\ncarc[arcangle=25]{Vertex1}{Vertex2}
\pnode(5.5,0){z1}
\pnode(2.5,3){z2}
\ncline[nodesepB=1pt]{Vertex1}{z2}
\ncline[nodesepB=1pt]{Vertex2}{z1}
\endpspicture
+
\frac12
\pspicture*[0.45](0,0)(4.8,4)
\NnodeE{1.5,2}{Vertex1}
\KnodeE{3.5,3}{Vertex2}
\pnode(0.0,2){z1}
\ncline{Vertex1}{z1}
\ncarc[arcangle=25]{Vertex2}{Vertex1}
\ncarc[arcangle=25]{Vertex1}{Vertex2}
\pnode(5.5,4){z1}
\pnode(2.5,1){z2}
\ncline[nodesepB=1pt]{Vertex1}{z2}
\ncline[nodesepB=1pt]{Vertex2}{z1}
\endpspicture
\nonumber \\
&+\ 
\frac16
\pspicture*[0.45](0,0)(4.8,4)
\NnodeE{1.5,2}{Vertex1}
\KnodeE{3.5,1}{Vertex2}
\pnode(0.0,2){z1}
\ncline{Vertex1}{z1}
\ncarc[arcangle=25]{Vertex2}{Vertex1}
\ncarc[arcangle=25]{Vertex1}{Vertex2}
\ncline{Vertex1}{Vertex2}
\pnode(5.5,0){z1}
\pnode(2.5,3){z2}
\ncline[nodesepB=1pt]{Vertex1}{z2}
\ncline[nodesepB=1pt]{Vertex2}{z1}
\endpspicture
+
\frac16
\pspicture*[0.45](0,0)(4.8,4)
\NnodeE{1.5,2}{Vertex1}
\KnodeE{3.5,3}{Vertex2}
\pnode(0.0,2){z1}
\ncline{Vertex1}{z1}
\ncarc[arcangle=25]{Vertex2}{Vertex1}
\ncarc[arcangle=25]{Vertex1}{Vertex2}
\ncline{Vertex1}{Vertex2}
\pnode(5.5,4){z1}
\pnode(2.5,1){z2}
\ncline[nodesepB=1pt]{Vertex1}{z2}
\ncline[nodesepB=1pt]{Vertex2}{z1}
\endpspicture
\nonumber \\
&+\ 
\frac{1}{24}
\pspicture*[0.45](0,0)(4.8,4)
\NnodeE{1.5,2}{Vertex1}
\KnodeE{3.5,3}{Vertex2}
\KnodeE{3.5,1}{Vertex3}
\pnode(0.0,2){z1}
\ncline{Vertex1}{z1}
\ncarc[arcangle=25]{Vertex2}{Vertex1}
\ncarc[arcangle=25]{Vertex1}{Vertex2}
\ncarc[arcangle=25]{Vertex3}{Vertex1}
\ncarc[arcangle=25]{Vertex1}{Vertex3}
\pnode(5.5,0){z1}
\pnode(5.5,4){z2}
\ncline[nodesepB=1pt]{Vertex3}{z1}
\ncline[nodesepB=1pt]{Vertex2}{z2}
\endpspicture
+\cdots \\
\end{split}
\label{BildFc}
\end{multline}
\caption{#1}
\label{BildF}
\end{figure}
\setcounter{equation}{\value{saveeqn}}
\renewcommand{\theequation}{\arabic{equation}}
}

\newcommand{\BildG}[1]
{
\begin{figure}
\begin{equation*}
\pspicture[0.45](0,0)(1,2)
\pnode(0.0,1){z3}
\pnode(1.0,1){z4}
\ncline{z3}{z4}
\endpspicture
=
\pspicture[0.45](0,0)(3,2)
\NnodeE{1.5,1}{Vertex1}
\pnode(0.0,1){z1}
\pnode(3,1){z2}
\ncline{Vertex1}{z1}
\ncline{Vertex1}{z2}
\endpspicture
+
\frac12
\pspicture[0.45](0,0)(5,2)
\NnodeE{1.5,1}{Vertex1}
\NnodeE{3.5,1}{Vertex2}
\pnode(0.0,1){z1}
\pnode(5.0,1){z2}
\ncline{Vertex1}{z1}
\ncline{Vertex2}{z2}
\ncarc[arcangle=25]{Vertex2}{Vertex1}
\ncarc[arcangle=25]{Vertex1}{Vertex2}
\endpspicture
\end{equation*}
\caption{#1}
\label{BildG}
\end{figure}
}

\newcommand{\BildH}[1]
{
\begin{figure}
\begin{eqnarray*}
\pspicture[0.45](0,0)(2,2)
\pnode(0.0,1){z1}
\pnode(2.0,1){z2}
\SnodeD{1,1}{Vertex1}
\ncline{Vertex1}{z1}
\ncline{Vertex1}{z2}
\endpspicture
&=&  
\pspicture[0.29](0,-1)(5,2)
\put(0.3,0){\psiline}\psivertex{1.9,0}
\psivertex{3.1,0}\put(3.2,0){\psiliner}
\NnodeE{2.5,0}{Vertex1}
\put(3.2,0){\Huge{\bf'}}
\endpspicture
\\
&=&
\pspicture[0.29](0,-1)(5,2)
\put(0.3,0){\psiline}\psivertex{1.9,0}
\psivertex{3.1,0}\put(3.2,0){\psiliner}
\NnodeE{2.5,0}{Vertex1}
\endpspicture
+
\frac12
\pspicture[0.45](0,0)(5.5,2)
\put(0.3,1){\psiline}\psivertex{1.9,1}
\psivertex{5.1,1}\put(5.2,1){\psiliner}
\put(5.1,1.0){\Huge{\bf'}}
\NnodeE{2.5,1}{Vertex1}
\NnodeE{4.5,1}{Vertex2}
\pnode(0.0,1){z1}
\pnode(6.0,1){z2}
\ncarc[arcangle=25]{Vertex2}{Vertex1}
\ncarc[arcangle=25]{Vertex1}{Vertex2}
\endpspicture
\end{eqnarray*}
\caption{#1}
\label{BildH}
\end{figure}
}

\newcommand{\BildI}[1]
{
\begin{figure}
\begin{eqnarray*}
\pspicture[0.45](0,0)(1,2)
\pnode(0.0,1){z3}
\pnode(1.0,1){z4}
\ncline{z3}{z4}
\endpspicture
&=&
\pspicture[0.45](0,0)(3,2)
\NnodeE{1.5,1}{Vertex1}
\pnode(0.0,1){z1}
\pnode(3,1){z2}
\ncline{Vertex1}{z1}
\ncline{Vertex1}{z2}
\endpspicture
\\
&=&
\pspicture[0.45](0,0)(3,2)
\SnodeE{1.5,1}{Vertex1}
\pnode(0.0,1){z1}
\pnode(3,1){z2}
\ncline{Vertex1}{z1}
\ncline{Vertex1}{z2}
\endpspicture
+
\frac12
\pspicture[0.45](0,0)(3,2)
\SnodeE{1.5,1}{Vertex1}
\pnode(0.0,1){z1}
\pnode(3,1){z2}
\ncline{Vertex1}{z1}
\ncline{Vertex1}{z2}
\nccircle{Vertex1}{0.5\scale}
\endpspicture
+
\frac18
\pspicture[0.45](0,0)(3,2)
\SnodeE{1.5,1}{Vertex1}
\pnode(0.0,1){z1}
\pnode(3,1){z2}
\ncline{Vertex1}{z1}
\ncline{Vertex1}{z2}
\nccircle{Vertex1}{0.5\scale}
\nccircle{Vertex1}{0.7\scale}
\endpspicture
+\cdots
\end{eqnarray*}
\caption{#1}
\label{BildI}
\end{figure}
}

\newcommand{\BildJ}[1]
{
\begin{figure}
  \begin{eqnarray*}
\pspicture[0.465](0,-1)(2.0,1)
\psline(0.0,0)(0.5,0)
\SnodeD{1,0}{Vertex1}
\endpspicture
&=&
\pspicture[0.29](0,-1)(3,2)
\put(0.3,0){\psiline}\psivertex{1.9,0}
\NnodeE{2.5,0}{Vertex1}
\endpspicture
+
\frac16
\pspicture[0.45](0,0)(5,2)
\put(0.3,1){\psiline}\psivertex{1.9,1}
\NnodeE{2.5,1}{Vertex1}
\KnodeE{4.5,1}{Vertex2}
\pnode(0.0,1){z1}
\ncline{Vertex1}{Vertex2}
\ncarc[arcangle=25]{Vertex2}{Vertex1}
\ncarc[arcangle=25]{Vertex1}{Vertex2}
\endpspicture
+\cdots
\end{eqnarray*}
\caption{#1}
\label{BildJ}
\end{figure}
}

\newcommand{\BildK}[1]
{
\begin{figure}
  \begin{eqnarray*}
\pspicture[0.45](0,0)(2,2)
\pnode(0.0,1){z1}
\pnode(2.0,1){z2}
\SnodeD{1,1}{Vertex1}
\ncline{Vertex1}{z1}
\ncline{Vertex1}{z2}
\endpspicture
&=&  
\pspicture[0.29](0,-1)(5,2)
\put(0.3,0){\psiline}\psivertex{1.9,0}
\put(3.1,0){\Huge{\bf'}}
\psivertex{3.1,0}\put(3.2,0){\psiliner}
\NnodeE{2.5,0}{Vertex1}
\endpspicture
+
\frac16
\pspicture[0.45](0,0)(7,2)
\put(0.3,1){\psiline}\psivertex{1.9,1}
\psivertex{5.1,1}\put(5.2,1){\psiliner}
\NnodeE{2.5,1}{Vertex1}
\put(5.1,1.0){\Huge{\bf'}}
\KnodeE{4.5,1}{Vertex2}
\ncline{Vertex1}{Vertex2}
\ncarc[arcangle=25]{Vertex2}{Vertex1}
\ncarc[arcangle=25]{Vertex1}{Vertex2}
\endpspicture
\\
&+&
\frac16
\pspicture[0.25](0,0)(5,3.5)
\NnodeE{2.5,1}{Vertex1}
\KnodeE{2.5,3}{Vertex2}
\put(3.1,1.0){\Huge{\bf'}}
\put(0.3,1){\psiline}\psivertex{1.9,1}
\psivertex{3.1,1}\put(3.2,1){\psiliner}
\ncline{Vertex1}{Vertex2}
\ncarc[arcangle=25]{Vertex2}{Vertex1}
\ncarc[arcangle=25]{Vertex1}{Vertex2}
\endpspicture
+ \cdots
\end{eqnarray*}
\caption{#1}
\label{BildK}
\end{figure}
}

\newcommand{\BildN}[1]
{
\begin{figure}
\begin{equation*}
\pspicture[0.45](0,0)(2,2)
\pnode(0.0,1){z1}
\SnodeD{1,1}{Vertex1}
\ncline{Vertex1}{z1}
\endpspicture
=  
\pspicture[0.29](0,-1)(5,2)
\put(0.3,0){\psiline}\psivertex{1.9,0}
\NnodeE{2.5,0}{Vertex1}
\endpspicture
\end{equation*}
\caption{#1}
\label{BildN}
\end{figure}
}

\newcommand{\BildO}[1]
{
\setcounter{saveeqn}{\value{equation}}
\setcounter{equation}{0}
\renewcommand{\theequation}{\alph{equation}}
\begin{figure}
  \begin{equation}
\pspicture[0.45](0,0)(2.5,2)
\NnodeE{1.5,1}{Vertex1}
\pnode(0.0,1){z1}
\ncline{Vertex1}{z1}
\endpspicture
= 0 \label{BildOa}
\end{equation}
\begin{eqnarray}
\pspicture[0.45](0,0)(1,2)
\pnode(0.0,1){z3}
\pnode(1.0,1){z4}
\ncline{z3}{z4}
\endpspicture
&=&
\pspicture[0.45](0,0)(3,2)
\NnodeE{1.5,1}{Vertex1}
\pnode(0.0,1){z1}
\pnode(3,1){z2}
\ncline{Vertex1}{z1}
\ncline{Vertex1}{z2}
\endpspicture
\label{BildOb}
\end{eqnarray}

\begin{eqnarray}
\pspicture[0.45](0,0)(2.5,2)
\KnodeZ{1.5,1}{Vertex1}
\pnode(0.0,1){z1}
\ncline{Vertex1}{z1}
\pnode(2.5,0){z3}
\pnode(2.5,2){z2}
\ncline[nodesepB=1pt]{Vertex1}{z3}
\ncline[nodesepB=1pt]{Vertex1}{z2}
\endpspicture
&=&
-\ 
\pspicture[0.45](0,0)(2.5,2)
\NnodeZ{1.5,1}{Vertex1}
\pnode(0.0,1){z1}
\ncline{Vertex1}{z1}
\pnode(2.5,0){z3}
\pnode(2.5,2){z2}
\ncline[nodesepB=1pt]{Vertex1}{z3}
\ncline[nodesepB=1pt]{Vertex1}{z2}
\endpspicture
\label{BildOc}
\end{eqnarray}
\caption{#1}
\label{BildO}
\end{figure}
\setcounter{equation}{\value{saveeqn}}
\renewcommand{\theequation}{\arabic{equation}}
}

\newcommand{\MexicanHat}[1]
{
\begin{figure}
\begin{center}
\psset{unit=0.4cm}
\psset{xunit=1.1cm}
\pspicture(-5,-17)(5,8)
\psplot[linecolor=black,linewidth=1.5pt,plotstyle=curve,plotpoints=50]%
     {-5}{5}{x dup mul -1.333 mul x dup mul dup mul 0.074 mul add 0.5 mul}
\psline(-5,0)(5,0)
\psline(0,-4)(0,6.5)
\put(-1.2,6){$V$}
\psplot[linecolor=black,linewidth=1.5pt,plotstyle=curve,plotpoints=50]%
{-4}{4.5}{x -2.666 mul x dup dup mul mul 0.074 mul 4 mul add 0.5 mul -13 add}
\psline(-5,-13)(5,-13)
\psline(0,-17)(0,-8)
\put(-1.2,-8.5){$V'$}
\psline[linestyle=dashed](-1.7,-1.6)(-1.7,-11.5)
\psline[linestyle=dashed](1.7,-1.6)(1.7,-14.5)
\psline[linestyle=dashed](3,-3)(3,-13)
\psline{<->}(-1.7,-6)(1.7,-6)
\psline{<->}(1.7,-6)(3,-6)
\put(-3,-5.5){instable region}
\put(5.2,-4.7){meta-}
\put(5.2,-5.5){stable}
\put(9,-4){stable region}
\endpspicture
\end{center}
\caption{#1}
\label{mexican}
\end{figure}
}

\newcommand{\GraphPert}[1]{
\begin{figure}
\psset{unit=0.5cm}
\begin{equation*}
\begin{split}
  &\VPotZ =
    \pspicture[0.45](0,-2)(3,2)
    \vertex{1.5,0}{Blob}
    \psline(0.2,0)(2.8,0)
    \endpspicture
    +
    \pspicture[0.45](0,-2)(3,2)
    \vertex{1.5,0}{Blob}
    \psline(0.2,0)(2.8,0)
    \pscurve[linestyle=dashed,linewidth=1pt](1.5,0)(2,0.7)(1.5,1.6)(1,0.7)
            (1.5,0)
    \endpspicture
    +
    \pspicture[0.45](0,-2)(3,2)
    \vertex{1.5,0}{Blob}
    \vertex{1.5,1.5}{Blob1}
    \psline(0.2,0)(2.8,0)
    \ncarc[linestyle=dashed,linewidth=1pt,arcangle=50]{Blob}{Blob1}
    \ncarc[linestyle=dashed,linewidth=1pt,arcangle=50]{Blob1}{Blob}
    \endpspicture
    +
    \pspicture[0.45](0,-2)(3,2)
    \vertex{1.5,0}{Blob}
    \vertex{0.75,1.5}{Blob1}
    \vertex{2.25,1.5}{Blob2}
    \psline(0.2,0)(2.8,0)
    \ncline[linestyle=dashed,linewidth=1pt]{Blob}{Blob1}
    \ncline[linestyle=dashed,linewidth=1pt]{Blob1}{Blob2}
    \ncline[linestyle=dashed,linewidth=1pt]{Blob2}{Blob}
    \endpspicture
\\ &+
    \pspicture[0.45](0,-2)(3,2)
    \vertex{1.5,0}{Blob}
    \psline(0.4,1.1)(2.6,-1.1)
    \psline(0.4,-1.1)(2.6,1.1)
    \endpspicture
    +
    \pspicture[0.45](0,-2)(5,2)
    \vertex{1.5,0}{Blob}
    \vertex{3.5,0}{Blob1}
    \psline(0.4,1.1)(1.5,0)
    \psline(3.5,0)(4.6,-1.1)
    \psline(0.4,-1.1)(1.5,0)
    \psline(3.5,0)(4.6,1.1)
    \ncarc[linestyle=dashed,linewidth=1pt,arcangle=50]{Blob}{Blob1}
    \ncarc[linestyle=dashed,linewidth=1pt,arcangle=50]{Blob1}{Blob}
    \endpspicture
    +
    \pspicture[0.45](0,-2)(5,2)
    \vertex{1.5,0}{Blob}
    \vertex{3.5,0}{Blob1}
    \vertex{2.5,1.5}{Blob2}
    \psline(0.4,1.1)(1.5,0)
    \psline(3.5,0)(4.6,-1.1)
    \psline(0.4,-1.1)(1.5,0)
    \psline(3.5,0)(4.6,1.1)
    \ncline[linestyle=dashed,linewidth=1pt]{Blob}{Blob1}
    \ncline[linestyle=dashed,linewidth=1pt]{Blob1}{Blob2}
    \ncline[linestyle=dashed,linewidth=1pt]{Blob2}{Blob}
    \endpspicture
    +
    \pspicture[0.45](0,-2)(3,2)
    \vertex{1.5,0}{Blob}
    \psline(0.3,0.9)(2.7,-0.9)
    \psline(0.3,-0.9)(2.7,0.9)
    \pscurve[linestyle=dashed,linewidth=1pt](1.5,0)(2,0.7)(1.5,1.6)(1,0.7)
            (1.5,0)
    \endpspicture
    +
    \pspicture[0.45](0,-2)(3,2)
    \vertex{1.5,0}{Blob}
    \vertex{1.5,1.5}{Blob1}
    \psline(0.3,0.9)(2.7,-0.9)
    \psline(0.3,-0.9)(2.7,0.9)
    \ncarc[linestyle=dashed,linewidth=1pt,arcangle=50]{Blob}{Blob1}
    \ncarc[linestyle=dashed,linewidth=1pt,arcangle=50]{Blob1}{Blob}
    \endpspicture
\\  & +
    \pspicture[0.45](0,-2)(3,2)
    \vertex{1.5,0}{Blob}
    \psline(0.3,0.9)(2.7,-0.9)
    \psline(0.3,-0.9)(2.7,0.9)
    \psline(1.5,-1.3)(1.5,1.3)
    \endpspicture
    +
    \pspicture[0.37](0,-2)(5,3)
    \vertex{1.5,0}{Blob}
    \vertex{3.5,0}{Blob1}
    \vertex{2.5,1.5}{Blob2}
    \psline(0.4,1.1)(1.5,0)
    \psline(3.5,0)(4.6,-1.1)
    \psline(0.4,-1.1)(1.5,0)
    \psline(3.5,0)(4.6,1.1)
    \psline(2.5,1.5)(1.4,2.6)
    \psline(2.5,1.5)(3.6,2.6)
    \ncline[linestyle=dashed,linewidth=1pt]{Blob}{Blob1}
    \ncline[linestyle=dashed,linewidth=1pt]{Blob1}{Blob2}
    \ncline[linestyle=dashed,linewidth=1pt]{Blob2}{Blob}
    \endpspicture
    +
    \pspicture[0.37](0,-2)(3,3)
    \vertex{1.5,0}{Blob}
    \vertex{1.5,1.5}{Blob1}
    \psline(0.3,0.9)(2.7,-0.9)
    \psline(0.3,-0.9)(2.7,0.9)
    \psline(0.4,2.6)(1.5,1.5)
    \psline(2.6,2.6)(1.5,1.5)
    \ncarc[linestyle=dashed,linewidth=1pt,arcangle=50]{Blob}{Blob1}
    \ncarc[linestyle=dashed,linewidth=1pt,arcangle=50]{Blob1}{Blob}
    \endpspicture
    + \ \ldots
\end{split}
\end{equation*}
\caption{#1}
\label{graphPert}
\end{figure}
}
\newcommand{\GammaDecay}[1]
{
\begin{figure}
\begin{center}
\psset{unit=3\scale}
\pspicture(0,-1.5)(4,2.7)
\psline[linewidth=1.4pt](0.5,2.56647)(1,0.444131)(1.5,0.124268)
  (2,0.0384542)(2.5,0.0103965)(3,-0.00363696)(3.5,-0.0168331)
\psline[linestyle=dashed,linewidth=1.4pt](3.5,1.332585)(3,1.407656)
  (2.5,1.467079)(2,1.490601)(1.5,1.470914)(1,1.489894)(0.5,1.541754)
\put(3.7,0){$\Gamma(z,w)$}
\put(3.7,1.3){$\MFAField$}
\put(0.5,-0.4){$z$}
\put(3.5,-0.4){$w$}
\psline[linestyle=dotted,linewidth=0.7pt](0.5,2.56647)(0.5,-0.2)
\psline[linestyle=dotted,linewidth=0.7pt](3.5,1.33)(3.5,-0.2)
\qdisk(0.5,-0.6){0.08}
\qdisk(1,-0.6){0.08}
\qdisk(1.5,-0.6){0.08}
\qdisk(2,-0.6){0.08}
\qdisk(2.5,-0.6){0.08}
\qdisk(3,-0.6){0.08}
\qdisk(3.5,-0.6){0.08}
\qdisk(0,-0.6){0.08}
\qdisk(0.75,-1.2){0.08}
\qdisk(2.75,-1.2){0.08}
\put(0.75,-1.5){$x$}
\put(2.75,-1.5){$y$}
\psline[linewidth=1pt,linearc=0.15](0,-0.75)(0,-0.9)(0.75,-0.9)(0.75,-1.05)
\psline[linewidth=1pt,linearc=0.15](1.5,-0.75)(1.5,-0.9)(0.75,-0.9)(0.75,-1.05)
\psline[linewidth=1pt,linearc=0.15](2,-0.75)(2,-0.9)(2.75,-0.9)(2.75,-1.05)
\psline[linewidth=1pt,linearc=0.15](3.5,-0.75)(3.5,-0.9)(2.75,-0.9)(2.75,-1.05)
\endpspicture
\end{center}
\caption{#1}
\label{gammaDecay}
\end{figure}
}


%% file: intro.tex
%
%
\section{Renormalization group flow on the lattice}
\label{introSection}
Suppose we start from the Euclidean action $\ActionNull[\FieldNull]$ of a 
field theory
which lives on some lattice $\LatNull $ or on the continuum. It depends on a 
field $\FieldNull $ on $\LatNull$. We wish to compute a sequence of effective 
actions $\ActionE[\E{\Field}]$ which live on lattices
$\LatE$ of increasing lattice spacing $\LatSpaceE $, i=0,1,2,... .

In principle the sequence of actions  is defined once we fix a block-spin 
definition. The block-spin definition determines a field
$\FieldZ=\Z{\Field}$
on the coarser lattice $\LatZ$ as some kind of average
\begin{equation*}
    \FieldZ = \AvOpE \FieldE
\end{equation*}
of the field $\FieldE=\E{\Field}$ on the lattice $\LatE$.  The actions are
defined recursively by
\begin{equation}
   \ZE[\FieldZ] = e^{-\ActionZ[\FieldZ]}
                = \int D\FieldE \,\delta (\AvOpE \FieldE - \FieldZ)
                  e^{-\ActionE[\FieldE]}. \label{defEffective}
\end{equation}
For given $\FieldZ$, $\ZE[\FieldZ]$ is the partition function of an
{\em auxiliary theory} in which only those variables are
 integrated out which are to be interpreted later as 
 high frequency modes of the field. 

We will also need certain expectation
 values of this auxiliary theory. 
If $A=A [\FieldE ]$  is an observable, we set
\begin{equation}
        \meanE{A} = \ZE[\FieldZ ]^{-1}
           \int  D\FieldE  A [\FieldE ] \delta (\AvOpE \FieldE - \FieldZ)
                  e^{-\ActionE[\FieldE]}. \label{defExpVal} 
\end{equation}
These expectation values depend on $\FieldZ$, but we neglect to indicate this
dependence explicitly. 
In the calculations, two auxiliary $\FieldZ$-dependent quantities 
will play
an important role, the {\em background field} $\MFAFieldE [\FieldZ ]$ and
the {\em fluctuation propagator} $\FlucPropE [\MFAFieldE ]$. 
The fluctuation propagator is 
considered to depend on the block-spin $\FieldZ$ through the background field
$\MFAFieldE = \MFAFieldE [\FieldZ ]$. 
Both auxiliary quantities are defined as expectation values of the
 auxiliary theory.
\begin{eqnarray}
    \MFAFieldE(\xE) &\equiv & \MFAFieldE [\FieldZ ](\xE ) =
          \meanE{\FieldE (\xE )},  \label{MFAFieldDef} \\
    \FlucPropE [\MFAFieldE ](\xE , \yE ) &=& 
          \meanE{ \FieldE (\xE )\FieldE (\yE )}
    - \meanE{\FieldE (\xE )}\meanE{\FieldE (\yE )}.  \label{FlucPropDefIntro}
\end{eqnarray}
These definitions differ from those of Gawedzki and Kupiainen, cp. 
\cite{GawKup,MackCargese}. In their work, expectation values of a free theory
were used. The present approach is more in the spirit of Balaban's work on 
gauge theories, where expansions of the full action around its minimum 
were used for reasons of gauge invariance \cite{balabanGauge}. 

The validity of the method requires that the auxiliary theory has an 
infrared cutoff of order $\LatSpace_{i+1}^{-1}$. Therefore the fluctuation
 propagator must decay exponentially with decay length no larger than about
$\LatSpace_{i+1}$. When this condition is violated, it signals a bad choice
of the block-spin definition.  

We are not able to calculate the functional integrals
(\ref{defEffective},\ref{defExpVal}) 
exactly. Therefore approximations are necessary. 
The calculation is done in a  self-consistent way. Both 
the background field and the fluctuation propagator are used in the
 calculation. Conditions will be imposed on them which ensure that
 equations (\ref{MFAFieldDef}) and (\ref{FlucPropDefIntro}) are
fulfilled within the accuracy of the approximation. 
We discuss in subsection \ref{localApproximation} how a truncation of 
the effective action
to a manageable form can be justified. No expansion in powers of fields
is involved.

It follows from the definition
(\ref{FlucPropDefIntro}) that the fluctuation propagator satisfies the
constraints
\begin{equation}
    \AvOpE \FlucPropE [\MFAFieldE] = 0 = \FlucPropE [\MFAFieldE ] \Adj{\AvOpE}.
    \label{GammaConstraint0}
\end{equation}

The simplest choice of block-spin for a theory of scalar fields is  as follows.
One identifies the sites $\xZ$ of the coarser lattice $\LatZ$ with disjoint 
hypercubes in the lattice $\LatE$, and one chooses the block-spin as
the average of the original field on these hypercubes
\begin{equation*}
  \FieldZ(\xZ )  = \int_{\xE }\AvOpE (\xZ ,\xE )\FieldE(\xE) = 
  \av{\xE \in \xZ} \FieldE (\xE ) .
\end{equation*}
One can try to improve on the locality properties of the 
effective actions by modifying the block-spin procedure \cite{Hasenfratz}.

We may imagine starting from a lattice of finite volume.  
After a finite number of steps we arrive at a lattice $\LatN$ which 
consists of only a single site. The field $\FieldN $ 
on this lattice is some
average of an average ... of an average of the original field $\FieldNull$.
Let us interpret it as magnetization. The action
$\ActionN[\FieldN]$ will give us the constraint effective potential 
-- i.e. the free energy --
as a function of the magnetization. 

We will make the following assumptions on the choice of block-spin in this 
paper.
We  assume that $\AvOpE $ is a linear map, and  that $\AvOpE \Adj{\AvOpE}  $
is invertible. (For gauge theories, linearity will have to be given up.)
We assume also that $\Adj{\AvOpE}  $ interpolates constant fields to constant 
fields, and we impose the following normalization condition
\begin{equation*}
\int_{\xZ\in \LatZ} \AvOpE {\FieldE} (\xZ ) = \int_{\xE\in \LatE}
\FieldE (\xE ) 
\end{equation*} 
for constant fields. For the above mentioned choice of block-spin, 
the equation is an identity for arbitrary fields $\FieldE $.  
 
Let $\H^i$ be the Hilbert space of  square summable functions on $\LatE$, 
and let $\H^i_{\AvOp } $
be the subspace of functions $\FlucFieldE $
with vanishing block average $\AvOpE \FlucFieldE $ = 0.
Invertibility of  $\AvOpE \Adj{\AvOpE}  $ ensures the existence of the
orthogonal decomposition
$$
    \E\H = \Adj {\AvOpTildeE} \Z{\H} \oplus \HCE,
$$
where $\AvOpTildeE=(\AvOpE\Adj{\AvOpE})^{-1}\AvOpE$.
 This is true because every $\FieldE \in \E{\H }$ can be decomposed as
$$
    \FieldE = \Adj{\AvOpTildeE}\FieldZ+\FlucFieldE
$$
with $\FieldZ=\AvOpE\FieldE\in\Z\H$ and
$\FlucFieldE=\FieldE-\Adj{\AvOpTildeE}\FieldZ$.
The first summand is obviously in $\Adj {\AvOpTildeE} \Z{\H}$ and 
the second is in $\HCE$ because it vanishes when we apply $\AvOpE$:
$
    \AvOpE \FlucFieldE = \AvOpE \FieldE - \AvOpE \Adj {\AvOpTildeE} \FieldZ
    = \FieldZ - \FieldZ = 0.
$
 Also the orthogonal decomposition is true
because the scalar product vanishes:
$
    (\FlucFieldE,\Adj {\AvOpTildeE} \FieldZ) =
    ((\AvOpE\Adj{\AvOpE})^{-1}\AvOpE\FlucFieldE,\FieldZ)
    = 0.
$

{\em Notation:} We will use letters $\xE , \yE ...$ for sites in 
$\LatE $ and $\xZ , \yZ ... $ for sites in $\LatZ$.  


%% file: saddle.tex
\section{The saddle point approximation}
\label{GaussSection}

\subsection{Parameterization of the action}
\label{introductionSaddle}

We wish to compute the action $\ActionZ[\FieldZ]$
which depends on a  field $\FieldZ $  on the lattice
$\LatZ$ from the action $\ActionE[\FieldE] $ on the finer lattice $\LatE$.
We will make some
approximations to perform the calculation.  We present a parameterization
of the actions which will be preserved by the approximate renormalization
group flow.

The action $\ActionZ[\FieldZ]$ will depend on the field $\FieldZ$ through
a functional $\MFAFieldE[\FieldZ]$.
The field $\MFAFieldE $
lives on the lattice $\LatE $. It is called the background field.

The recursion formula will have the form
\begin{eqnarray}
\ActionZ[\FieldZ] &=&
 \ActionE[\MFAFieldE] -\half \tr \ln \FlucPropE [\MFAFieldE ] , \label{action}\\
\MFAFieldE &=& \MFAFieldE [\FieldZ ] .
\end{eqnarray}

Approximations will be made such that
\begin{enumerate}
\item
    The parameterization of the action preserves its form
\item
    The fluctuation propagator, the effective action and its first two
    field derivatives, and 
    the background field $\MFAFieldE[\FieldZ]$ and its derivative
    $\MFAFieldPE[\FieldZ]$
    with respect to the field $\FieldZ $ will enter into the recursion
    relations, but they
    need only be calculated for  constant fields. This
    computational  problem is fit for a PC.
\end{enumerate}
We will use a dual notation for derivatives with respect to fields
\begin{equation*}
    \ActionPE[\FieldE](\xE) \equiv
    \ActionE_{,\xE }[\FieldE] = \dEx {\xE} \ActionE [\FieldE] \ ,
\end{equation*}
and similarly for second derivatives $\ActionE_{, \xE \yE} $.

The background field $\MFAFieldE $
 will be determined as a functional of $\FieldZ $ by
the saddle point condition which involves the previous action
\begin{equation*}
   \ActionE [\MFAFieldE ] = min \ \ \mbox{subject\ to}\ \
   \AvOpE \MFAFieldE = \FieldZ  \ .
\end{equation*}

 The fluctuation field propagator $\FlucPropE $ lives on lattice
$\LatE $. It is a selfadjoint integral operator with kernel
$$ \FlucPropE [\FieldE] (\xE , \yE) \ , \ \ \xE , \yE \in \LatE  $$
and is defined as pseudoinverse of $\ActionPPE[\FieldE]$ for all $\FieldE$ 
in the sense that
\begin{equation}
    \FlucPropE [\FieldE ]\ActionPPE [\FieldE ]
    \FlucPropE [\FieldE ]=  \FlucPropE [\FieldE ]. \label{FlucPropQinv}
\end{equation}
In the case where the field $\FieldE$ is equal to the background field 
$\MFAFieldE[\FieldZ]$, 
the fluctuation field propagator  is positive definite on $\HCE$. It 
satisfies the constraints (\ref{GammaConstraint0}), viz. 
\begin{equation}
    \AvOpE \FlucPropE [\MFAFieldE] = 0 = \FlucPropE [\MFAFieldE ] \Adj{\AvOpE}.
    \label{GammaConstraint}
\end{equation}

Therefore the space $ \E\H $ is mapped into $ \HCE$ by $\FlucPropE $,
and  $\FlucPropE $ vanishes on $\Adj {\AvOpTildeE}\Z\H $.
The trace $\tr $ which appears in the action (\ref{action}) is
to be understood as a trace over $\HCE$.

We will explain later on how the background field and the fluctuation
propagator can be evaluated by a self
consistent approximation which involves neglect of higher order terms in
gradients of fields.

It is the crux of any real space renormalization group method to find
truncated forms of the effective actions which can be parameterized in a
 manageable form.

Often this is done in an ad hoc fashion which throws away some pieces which
are irrelevant in a {\em perturbative} sense - higher powers of fields, 
for instance. This is not really justified because 
existing irrelevant terms get suppressed in the
next RG-step. But new irrelevant terms are created by the marginal and
relevant ones at the same time.
As a consequence, there is a kind of  equilibrium, so that
(along the renormalized trajectory) the
irrelevant terms are determined by the relevant and marginal ones. They are
not necessarily very small and they influence the flow of the marginal and
relevant coupling constants. Throwing them away after each renormalization
group step introduces therefore a systematic error which accumulates.

In principle there is a better way. One could determine the irrelevant pieces
as a function of the marginal and relevant ones by solving a fixed point
equation. But until now this is practical only in simplified (hierarchical)
models \cite{Pordt}.

The motivation for our method of truncation will be described below. We will
argue that our approximation becomes more and more accurate the larger
the scaling factor (ratio of lattice spacings) $s=\LatSpaceZ/\LatSpaceE $.
We will present
numerical evidence which confirms this.

On the other hand, the saddle point approximation becomes exact in the limit
$s\mapsto 1 $ (when the nature of the cutoff permits such a limit). It becomes
less accurate with increasing scaling factor because the phase space for
high frequency modes increases.

The truncation consists of a local approximation to the second derivative
$\ActionPPE=-W $ of the action. For more precise notation, set
\begin{eqnarray}
    -\ActionE_{,\xE \yE}[\FieldE ] &=& \WPotE [\FieldE](\xE , \yE) \\
    \left( \WPotE [\FieldE ] f \right)(\xE ) &=& \int_{\yE }
    \WPot [\FieldE] ( \xE , \yE) f (\yE )
\end{eqnarray}
We will approximate $\WPotE$ by a function $\WPotlocE(\xE , \yE |\xi ) $
which
depends only on the value $\xi $ of the field at one site of the lattice
$\LatE $. If $f\in \E\H $ is a function
on $\LatE $ then we approximate
\begin{equation}
    \left( \WPotE [\Field ]f\right) (\xE ) = \half
    \int_{\yE} \left[ \WPotlocE\left( \xE ,\yE |\Field (\xE )\right) +
    \WPotlocE\left( \xE ,\yE |\Field (\yE )\right) \right] f (\yE ) \ .
    \label{approximationIntro}
\end{equation}
$\WPotlocE $ can be determined from the knowledge of $\WPotE $
for {\em constant} fields $\Field $. Symmetrization is performed to maintain
hermiticity of $\WPot$.

To understand the meaning of the approximation, suppose that
$\ActionE $ is the action of a $\phi^4$-theory. Then
\begin{equation}
  \left( \ActionPPE f \right) (\xE ) = \left[ -\Delta + m^2 
    + \half \lambda \phi (\xE)^2
  \right] f(\xE). \label{startingAction}
\end{equation}
We see that our approximation is exact in this case because the field
 dependent part of $\ActionPPE  $ only involves the field
 $\Field $ at a single point $\xE$.

In particular, the approximation is exact for free field theories. Since
the saddle point approximation is also exact for free fields,
our method is exact for free field theory.

A general action can be decomposed into a potential and a term 
whose first derivative with respect to the field 
vanishes for constant fields. We call it a generalized kinetic term.
\begin{equation}
\ActionE [\Field ]= \mbox{generalized kinetic term} + \VPotE[\Field] \ .
\end{equation}
A field independent contribution to $\VPotE $ is of no interest. Therefore it
suffices to know the derivative $\VPotPE $. It is uniquely determined
if we know the derivative $\ActionPE $ of the action for {\em constant} fields.

Our method consists in deriving recursion relations for
$\VPotE $ and $\WPotE$ i.e. for the first and second derivatives
$\ActionPE $ and $\WPotE = -\ActionPPE $
of the action evaluated at {\em constant} fields.

Moreover we wish to obtain the effective actions as a function of the
block-spin and not only as a function of the background field. Therefore
we will need information on the block-spin dependence of the
background field.
It turns out that the use of the recursion formulae
requires knowledge of the derivative of the background field with respect to the
block-spin.
There is a formula for this:
\begin{equation}
    \MFAFieldE_{,\xZ }[\FieldZ](\xE ) \equiv 
    \dZx{\xZ} \MFAFieldE[\FieldZ] = 
        (1 - \FlucPropE \ActionPPE )\Adj{ \tilde \AvOpE}(\xE,\xZ).
\end{equation}

For a block-spin $\FieldlocZ$ which lives on a lattice made of a single site,
translation invariance implies that
there exists a saddle point $\MFAFieldE = \MFAFieldE [\FieldlocZ ]$
of the action which is also constant. Because of the
constraint $\AvOpE \MFAFieldE = \FieldZ $ and our normalization conventions
it follows that
\begin{equation}
    \MFAFieldE [\FieldlocZ ] = \Adj{\AvOpTildeE} \FieldlocZ \ .
    \label{Fieldloc}
\end{equation}
So we know the background field in this case, assuming the saddle point is the
minimum.
 If it is not, this signals a breakdown of stability.
 We will come back to such possibilities. They are familiar from
the Maxwell construction in thermodynamics.

The argument can be extended to show that to any block-spin
$\FieldlocZ$ which is constant on $\LatZ$ there exists a saddle point
 (\ref{Fieldloc}) which is constant, assuming that translational invariance 
is not spontaneously broken. If
the invariance under translations by block lattice vectors is not
spontaneously broken in
the auxiliary theory with expectation values (\ref{defExpVal}), then the
appropriate saddle point $\MFAField $ will be invariant under block lattice
translations. Therefore it is equal to the constraint minimum of the
action on a single block with periodic boundary conditions. Now we can appeal
to translation invariance of the action under translations on the fine lattice
$\LatE$ again to conclude that there exists a constant saddle point 
unless translational symmetry is spontaneously broken.
It may happen that it is spontaneously broken or the
saddle point is not a minimum. Again,   this signals a breakdown of stability.

In view of its interpretation as a propagator for the high frequency modes
of the field which are integrated out in a renormalization group step,
 the kernel $\FlucPropE[\FieldE] (\xE,\yE)$ of the fluctuation field
propagator
must decay exponentially with distance $|\xE -\yE |$ with decay length
at most one block lattice spacing $\LatSpaceZ $ (see \cite{balaban}).
This is a condition which limits
the range of applicability of the present method. It could be monitored during
the computation of the renormalization group flow. When it is violated,
this will typically entail a violation of the locality properties 
of the next action $\ActionZ$ as well. Locality requires that 
$\ActionE_{, \xE \yE} $ decays exponentially with the distance between 
$\xE$ and $\yE $ with decay length no larger than one lattice spacing
 $\LatSpaceE $. 

When this is violated 
it is typically a sign that the choice of the block-spin definition
is not appropriate. This can happen for several reasons, see section
\ref{locSection}.

\subsection{The detailed procedure of saddle point approximation}

Let $\ActionE$ be a generic action.
We now want to evaluate the effective action $\ActionZ$
\begin{equation}
    e^{-\ActionZ[\FieldZ]}
        = \int D\FieldE \delta (\AvOpE \FieldE - \FieldZ)
          e^{-\ActionE[\FieldE]} 
    \label{defRG}
\end{equation}
in a saddle point approximation.
For this purpose we split the field
\begin{equation*} 
   \FieldE (\xE )= \MFAFieldE[\FieldZ](\xE) + \FlucFieldE (\xE) \ \ \
   (\xE \in \LatE).
\end{equation*}
into the background field $\MFAFieldE[\FieldZ]$ and a
fluctuation field $\FlucFieldE$. The background field has also the meaning
of a mean field, because in the Gaussian approximation that we will use
for the action $\ActionE$ eq.(\ref{MFAFieldDef}) holds

The mean field $\MFAFieldE$ is determined as a functional of the block-spin
field $\FieldZ$ by
a saddle point condition which involves the previous action
\begin{equation*} 
   \ActionE [\MFAFieldE ] = min \ \ \mbox{subject\ to}\ \
   \AvOpE \MFAFieldE = \FieldZ  \ .
\end{equation*}
The result is a nonlinear equation  for $\MFAFieldE[\FieldZ]$
\begin{eqnarray}
\ActionPE [\MFAFieldE] &=& \Adj {\AvOpE }\E\lambda , \label{MFAFieldEq1} \\
\AvOpE \MFAFieldE &=& \FieldZ \ . \label{MFAFieldEq2}
\end{eqnarray}
$\E\lambda [\FieldZ](\xZ) $ are  Lagrange multipliers, $\xZ\in \LatZ $.

Expanding the action around the mean field to second order in $\FlucFieldE$ yields
\begin{equation}
  \ActionE[\MFAFieldE+\FlucFieldE] = \ActionE [\MFAFieldE] +
  \int_{\xE } \ActionE_{,\xE}[\MFAFieldE]\FlucFieldE(\xE)
  + \half   
  \int_{\xE,\yE} \ActionE_{,\xE \yE}[\MFAFieldE]\FlucFieldE(\xE)
  \FlucFieldE (\yE ). 
\label{expand2Order}
\end{equation}
The linear term in the expansion vanishes
$$
    (\ActionPE,\FlucFieldE)=(\Adj{\AvOpE}\E\lambda,\FlucFieldE)
    =(\E\lambda,\AvOpE\FlucFieldE)=0,
$$
since due to (\ref{MFAFieldEq2}) $\AvOpE\FlucFieldE=0$.
Therefore  we get
\begin{align}
  e^{-\ActionZ[\FieldZ]} &\approx \int_{\E\H} 
        D\FlucFieldE \delta (\AvOp\FlucFieldE)
          e^{-\ActionE[\MFAFieldE[\FieldZ]]
             -\half\scalar{\FlucFieldE}
                          {\ActionPPE[\MFAFieldE[\FieldZ]]\FlucFieldE}
            }
 \label{MFAintegral}\\
 &=e^{-\ActionE[\MFAFieldE[\FieldZ]]} \int_{\HCE} D\FlucFieldE 
      e^{-\half\scalar{\FlucFieldE}
                      {\FlucPropE[\MFAFieldE[\FieldZ]]^{-1}\FlucFieldE}
        }
\label{GammaInvInt}
\end{align}
Note that $\ActionPPE$ is a linear operator on $\E\H$. 
In appendix \ref{GammaFormAp}
we show that $\FlucPropE$ can be written as
\begin{equation}
    \FlucPropE = {\ActionPPE}^{-1} -
      {\ActionPPE}^{-1}\Adj{\AvOpE}(\AvOpE{\ActionPPE}^{-1}\Adj{\AvOpE})^{-1}
           \AvOpE{\ActionPPE}^{-1}\,.
  \label{gammaform}  
\end{equation}
Because of $\AvOpE\FlucPropE=0=\FlucPropE\Adj{\AvOpE}$ the fluctuation 
propagator vanishes on $\Adj {\AvOpE}\Z\H$ and is a linear operator
$\E\H\rightarrow\HCE$. Therefore we can use the quantity
$(\FlucPropE)^{-1}=\ActionPPE\Einsch{\HCE}$ in (\ref{GammaInvInt}).
The extension of $\FlucPropE$ to all of $\E\H$ is a pseudoinverse of 
$\ActionPPE $ as described by (\ref{FlucPropQinv}).

The remaining fluctuation integral
is then Gaussian with covariance $\FlucPropE[\MFAFieldE[\FieldZ]]$ and may 
easily be performed.  The result is 
\begin{equation} 
\ActionZ [\FieldZ ] = \ActionE [\MFAFieldE [\FieldZ ]]
 -\half \tr \ln \FlucPropE [\MFAFieldE [\FieldZ ] ] , \label{actionPrel}
\end{equation} 
The trace $\tr $ is to be understood as a trace over  $\HCE$. 

If one uses in (\ref{MFAintegral})
$$
    \delta(\FieldE) = \lim\limits_{\kappa\rightarrow\infty}
    e^{-\frac\kappa2 \int_\xE\FieldE(\xE)^2}\,,
$$
one obtains the following alternative expression for the
 fluctuation propagator. 
\begin{equation}
  \FlucPropE[\MFAFieldE [\FieldZ ] ] = \lim_{\kappa \rightarrow \infty }\left(
  \ActionPPE  [\MFAFieldE [\FieldZ ]] + \kappa \Adj{\AvOpE}\AvOpE   
  \right)^{-1}\,.
  \label{GammaWithKappa}
\end{equation}

\subsection{Recursion relations}

By iterating the saddle point approximation we obtain
 $\ActionD$ from $\ActionZ$:
$$
   \ActionD [\FieldD ] = \ActionZ [\MFAFieldZ [\FieldD ]]
   -\half \tr \ln \FlucPropZ [\MFAFieldZ [\FieldD ] ].
$$
Now we need $\FlucPropZ$. It is defined in terms of $\ActionPPZ$
by an equation similar to (\ref{gammaform}). Therefore we try to
get a recursion relation for
\begin{equation*}
    \WPotE[\FieldE] = -\ActionPPE[\FieldE]\,.
\end{equation*}
In appendix \ref{recursionRelation}  the following
recursion relation  will be deduced by differentiating eq.(\ref{action}).
(We neglect to write functional 
$\MFAFieldE[\FieldZ]$-dependencies on the right hand side.) 

\begin{multline}
  \WPotZ [\FieldZ](\xZ,\yZ) =  \\
\begin{split}
& \int_{\xE,\yE\in\LatE} 
    \WPotE(\xE,\yE)  
    \MFAFieldE_{,{\yZ}}(\xE)
    \MFAFieldE_{,{\xZ}}(\yE)  \\
& + \ \half\int\limits_{\xE,\yE,\xE_i\in\LatE }
    \FlucPropE (\xE_1,\xE_2)
    \WPotE_{,\xE} (\xE_2,\xE_3)
    \FlucPropE (\xE_3,\xE_4)
    \WPotE_{,\yE} (\xE_4,\xE_1)
    \MFAFieldE_{,\xZ} (\xE)  \MFAFieldE_{,\yZ} (\yE)
    \\[8pt]
& + \ \half\int\limits_{\xE,\yE,\xE_1,\xE_2\in\LatE }
    \FlucPropE (\xE_1,\xE_2)
    \WPotE_{,\xE\yE}(\xE_2,\xE_1)
    \MFAFieldE_{,\xZ }(\xE)
    \MFAFieldE_{,\yZ }(\yE) \\
& + \ \half\int\limits_{\xE,\xE_1,\xE_2\in\LatE}
    \FlucPropE (\xE_1,\xE_2)
    \WPotE_{,\xE} (\xE_2,\xE_1)
    \FlucPropE(\xE,\xE_3) \WPotE_{,\yE}(\xE_3,\xE_4)
    \MFAFieldE_{,\yZ}(\yE) \MFAFieldE_{,\xZ}(\xE_4)
    . \\
\end{split}
\label{recursionW}
\end{multline}

This  functional relation is an exact consequence of eq.(\ref{action}),
 no localization approximation has been made yet. 

We see that we need an equation for the derivative   $\MFAFieldPE $
of the background field. In appendix \ref{proofOfLemma} it will be shown that
\begin{equation}
    \MFAFieldPE = (1-\FlucPropE \ActionPPE) \Adj{\widetilde{\AvOpE}}.
    \label{psiDerivative}
\end{equation}

The recursion relation is shown in graphical form in figure
\ref{BildC}. The graphical notation is described in figure \ref{BildAa}.

We see that the recursion relation for $\ActionPPE$  contains a 1-particle 
reducible term. This is an artifact of the simple saddle point 
approximation which will disappear  in the improved version of section
\ref{FBapproximation}.

\BildAa{Definition of graphical notation for the derivatives of the
  action and of the background field}

\BildC{Recursion relation for $\WPotE=-\ActionPPE$ in saddle point
       approximation}

As described in section \ref{introductionSaddle}  the action
$\ActionE$  can be split into a potential term
$\VPotE[\FieldE]$ and 
a generalized kinetic term whose first field derivative vanishes for 
constant fields; the potential term is determined by the derivative
$\ActionPE$ evaluated at constant field. 

 We give
 the recursion relation for the first derivative $\ActionPE$ in a general form.
Using (see appendix \ref{FlucPropDer})
\begin{equation}
   \FlucPropE_{,\xE} = \FlucPropE\WPotE_{,\xE}\FlucPropE\,.
\end{equation}
one gets by differentiation of (\ref{actionPrel})
\begin{equation}
    \ActionZ_{,\xZ}[\FieldZ] = \int\limits_{\xE} \MFAFieldE_{,\xZ}[\FieldZ](\xE)
    \Bigl\{ \ActionE_{,\xE}[\MFAFieldE[\FieldZ]]
        - \half\tr \FlucPropE [\MFAFieldE[\FieldZ]]
             \WPotE_{,\xE}[\MFAFieldE[\FieldZ]] \Bigr\}\,.
    \label{SPrecursion}
\end{equation}
The graphical illustration is shown in figure \ref{BildD}.

\BildD{Recursion relation for $-\ActionPZ$ in saddle point approximation}

For constant block-spin field the derivative of the 
generalized kinetic term vanishes.
Therefore the recursion relation for the potential term has
the same form as (\ref{SPrecursion}):
\begin{equation}
    \VPotZ_{,\xZ}[\FieldZ] = \int\limits_{\xE} \MFAFieldE_{,\xZ}[\FieldZ](\xE)
    \Bigl\{ \VPotE_{,\xE}[\MFAFieldE[\FieldZ]]
        - \half\tr \FlucPropE [\MFAFieldE[\FieldZ]]
             \WPotE_{,\xE}[\MFAFieldE[\FieldZ]] \Bigr\}\,.
    \label{VPrecursion}
\end{equation}


\subsection{Justification of the localization approximation;
Application to constant fields}
\label{localApproximation}

Let us summarize what we have achieved so far. Given the functionals 
$\VPotE$ and $\WPotE$ we can compute the fluctuation propagator 
$\FlucPropE$ via equation (\ref{gammaform}) and the derivative of the 
background field $\MFAFieldPE$ via equation (\ref{psiDerivative}). 
This enables us to calculate 
$\WPotZ$ and $\VPotZ$ by means of recursion relations (\ref{recursionW}) and
(\ref{VPrecursion}). At least this can be done in principle.
Because of the functional nature of these equations the calculation 
is  too complicated for numerical purposes.  We will arrive at a  
significant simplification by arguing that we may focus attention on constant 
fields.
\sticky
{
For constant block-spin $\FieldlocZ$ the mean field $\MFAFieldE[\FieldlocZ]$
is constant too and has the same value. To see this let us make an ansatz.
We split the field as
\begin{equation}
    \Field = \Adj{\AvOpTildeE} \FieldlocZ + \FlucFieldE\,.
\end{equation}
It fulfills the constraint (\ref{MFAFieldEq2}) and the first term on the
right-hand side is constant for our choice of the average operator.
Inserting this into (\ref{MFAFielDef}) leads to
\begin{equation}
    \MFAFieldE [\FieldlocZ ] = \Adj{\AvOpTildeE} \FieldlocZ \ .
    \label{Fieldloc}
\end{equation}
The expectation value of the fluctuation field is zero because the
auxiliary theory is even in the fluctuation field, therefore it
vanishes:
\begin{equation}
    \mean{\FlucFieldE} = \frac
               { \int_{\HCE} D\FlucFieldE \,\FlucFieldE
                 e^{-\half\scalar{\FlucFieldE}
                 {\FlucPropE[\MFAFieldE[\FieldZ]]^{-1}\FlucFieldE} } }
               { \int_{\HCE} D\FlucFieldE e^{-\half\scalar{\FlucFieldE}
                 {\FlucPropE[\MFAFieldE[\FieldZ]]^{-1}\FlucFieldE} } }
    = 0
\end{equation}
}

Fields $\FieldZ $ with very large derivatives have very small probability
because of the kinetic term in the action. We may therefore assume that
$\FieldZ $ does not have such violent behavior. 

$\MFAFieldE [\FieldZ ](\xE )$ is an interpolation of the field $\FieldZ$ from
the coarser lattice $\LatZ$ to the lattice $\LatE$. The interpolation is
determined by a minimality condition on the action.
Because of the kinetic term
in the action, we expect that the result of the interpolation is 
a smooth function on $\LatE $. 
This means 
that we may consider $\MFAFieldE[\FieldZ ]$ as nearly  constant over 
distances $\LatSpaceE$, assuming  that the scaling factor 
$s=\LatSpaceZ/\LatSpaceE$  is big enough

By locality $\ActionE_{, \xE \yE }[\MFAFieldE[\FieldZ ] ] $
is very nearly zero if
$\xE $ is not within about one lattice spacing 
of $\yE $, and it depends only
on the field within a neighbourhood of about one lattice spacing 
$\LatSpaceE $ of $\xE $ and $\yE $. Therefore we may approximate
$\WPotE [\MFAFieldE]$ by a hermitian operator $\WPotlocE$
which involves only the value $\xi = \MFAFieldE (\xE )$ of the field
at one point (see also (\ref{approximationIntro})),
\begin{equation}
   \WPotE[\MFAFieldE](\xE,\yE) \approx 
        \half\Bigl(
          \WPotlocE(\xE,\yE|\MFAFieldE(\xE))+\WPotlocE(\xE,\yE|\MFAFieldE(\yE))
         \Bigr),
    \label{Wloc}
\end{equation}
This justifies the localization approximation for $\WPotZ$. 
$\WPotlocE$  has locality properties similar to a Laplacian.

To determine $\WPotZ[\FieldZ]$ we need $\FlucPropE[\FieldE]$ for
$\FieldE=\MFAFieldE [\FieldZ]$.
$\FlucPropE (\xZ , \yZ) $ depends on the field $\FieldE$ on a domain of
diameter about $\LatSpaceZ$. Unless $\FieldZ $ is constant, the field 
$\FieldE $ need not be nearly constant over such a big domain. But  
 $\FlucPropE[\FieldE]$ is determined by
$\ActionPPE[\FieldE]=-\WPotE $ and we can use the localization approximation
for this latter quantity in order to determine $\FlucPropE$ for general field
$\FieldZ$ if we need it.    

Looking now at the recursion relation for $\WPotZ(\xZ,\yZ)$ we see that three
different quantities are involved:
\begin{itemize} 
  \item The fluctuation propagator has 
    an exponential decay of one block lattice spacing 
    $\LatSpaceZ$ (see \cite{balaban} and figure \ref{gammaDecay}).
  \item The operator $\WPotE(\xE,\yE)$ and its derivatives are very
    local by assumption and have therefore a range of one lattice spacing 
    $\LatSpaceE$. 
  \item The derivative of the background field $\MFAFieldPE$
    is also determined by these quantities and has exponential decay 
    with decay length $\LatSpaceZ$.
\end{itemize}
\GammaDecay{The decay of the fluctuation propagator.}
Combining these arguments we see 
 that  $\WPotZ[\FieldZ]$ has 
similar locality properties on the scale of the new lattice spacing 
$\LatSpaceZ$ as $\WPotE$ had on the scale $\LatSpaceE$. 

Therefore we will be entitled to 
 make similar approximations on the next level as before.

Now we can perform functional derivatives of $\WPotE$ for almost constant fields:
\begin{align}
    \WPotE_{,\xE}[\FieldE](\xE_1,\xE_2) = \half\Bigl(&
      \WPotlocPE(\xE_1,\xE_2|\FieldE(\xE_1))\delta(\xE-\xE_1) \nonumber\\
     & + \WPotlocPE(\xE_1,\xE_2|\FieldE(\xE_2))\delta(\xE-\xE_2)\Bigr)
     \label{firstDerivWloc} \\
    \WPotE_{,\xE\yE}[\FieldE](\xE_1,\xE_2) = \half\Bigl(&
      \WPotlocPPE(\xE_1,\xE_2|\FieldE(\xE_1))\delta(\xE-\xE_1)\delta(\yE-\xE_1)
      \nonumber \\
     & +\WPotlocPPE(\xE_1,\xE_2|\FieldE(\xE_2))\delta(\xE-\xE_2)\delta(\yE-\xE_2)
     \Bigr)\,,
     \label{secondDerivWloc}
\end{align}
where $'$ denotes the derivative of $\WPotlocE(\xE_1,\xE_2|\xi)$ with respect to
the single variable $\xi$. These formulas are valid when smeared with test functions in variables $\xE $ and $\yE$ which are nearly constant over 
distances $\LatSpaceE$.  

Setting the block-spin field constant the fluctuation propagator becomes a
function of the field value.
By virtue of (\ref{psiDerivative}) the derivative of the 
background field can be written
as a simple derivative:
\begin{equation}
    \MFAFieldE_{,\xZ}[\,\FieldlocZ\,](\xE) \equiv \MFAFieldPE(\xE,\xZ|\FieldlocZ)
    = \int\limits_{\xE_1}\Bigl(\delta(\xE,\xE_1)+\int\limits_{\yE}
      \FlucPropE(\xE,\yE|\FieldlocZ) \WPotlocE(\yE,\xE_1|\FieldlocZ)\Bigr) 
      \Adj{\widetilde{\AvOpE}}(\xE_1,\xZ)\,.
  \label{PsiDerivativeCoord}
\end{equation}
For constant block-spin fields,  inserting (\ref{firstDerivWloc}) and
(\ref{secondDerivWloc}) into the recursion relation (\ref{recursionW}) for 
$\WPot$ gives therefore a recursion relation for $\WPotlocE$:

\begin{multline}
  \WPotlocZ (\xZ,\yZ|\FieldlocZ) =
  \int_{\xE,\yE\in\LatE}
    \WPotlocE(\xE,\yE|\FieldlocZ)
    \MFAFieldPE (\xE,\yZ|\FieldlocZ)
    \MFAFieldPE (\yE,\xZ|\FieldlocZ)  \\
\begin{split}
& + \ \frac18\int\limits_{\xE_i\in\LatE }
    \FlucPropE (\xE_1,\xE_2|\FieldlocZ)
    \WPotlocPE (\xE_2,\xE_3|\FieldlocZ)
    \FlucPropE (\xE_3,\xE_4|\FieldlocZ)
    \WPotlocPE (\xE_4,\xE_1|\FieldlocZ) \\[-8pt]
& \qquad\qquad\quad
     \Bigl\{  \MFAFieldPE(\xE_2,\xZ|\FieldlocZ) \MFAFieldPE(\xE_4,\yZ|\FieldlocZ)
            + \MFAFieldPE(\xE_2,\xZ|\FieldlocZ) \MFAFieldPE(\xE_1,\yZ|\FieldlocZ) \\
& \qquad\qquad\qquad
            + \MFAFieldPE(\xE_3,\xZ|\FieldlocZ) \MFAFieldPE(\xE_4,\yZ|\FieldlocZ)
            + \MFAFieldPE(\xE_3,\xZ|\FieldlocZ) \MFAFieldPE(\xE_1,\yZ|\FieldlocZ)
     \Bigr\} \\
& + \ \frac14\int\limits_{\xE_1,\xE_2\in\LatE }
    \FlucPropE (\xE_1,\xE_2|\FieldlocZ)
    \WPotlocPPE(\xE_2,\xE_1|\FieldlocZ) \\[-8pt]
&\qquad\qquad\qquad
   \Bigl\{  \MFAFieldPE (\xE_1,\xZ|\FieldlocZ) \MFAFieldPE (\xE_1,\yZ|\FieldlocZ)
          + \MFAFieldPE (\xE_2,\xZ|\FieldlocZ) \MFAFieldPE (\xE_2,\yZ|\FieldlocZ)
   \Bigr\} \\
& + \ \frac18\int\limits_{\xE_1,\xE_2\in\LatE}
    \FlucPropE (\xE_1,\xE_2|\FieldlocZ)
    \WPotlocPE(\xE_2,\xE_1|\FieldlocZ)
    \WPotlocPE(\xE_3,\xE_4|\FieldlocZ) \MFAFieldPE (\xE_4,\xZ|\FieldlocZ) \\[-8pt]
&\qquad\qquad\qquad
    \Bigl\{  \FlucPropE(\xE_2,\xE_3|\FieldlocZ)\MFAFieldPE (\xE_3,\yZ|\FieldlocZ)
           + \FlucPropE(\xE_2,\xE_3|\FieldlocZ)\MFAFieldPE (\xE_4,\yZ|\FieldlocZ) \\
&\qquad\qquad\qquad\quad
           + \FlucPropE(\xE_1,\xE_3|\FieldlocZ)\MFAFieldPE (\xE_3,\yZ|\FieldlocZ)
           + \FlucPropE(\xE_1,\xE_3|\FieldlocZ)\MFAFieldPE (\xE_4,\yZ|\FieldlocZ)
    \Bigr\}    . \\
\end{split}
\label{recursionWloc}
\end{multline}

For constant fields we have translation symmetry. Therefore the derivative of the 
potential must be independent of the coordinate:
\begin{equation}
        \VPotE_{,\xE}[\FieldlocE] = \VPotlocPE(\FieldlocE) = \text{const.}
\end{equation}
To obtain the recursion relation of the effective potential we insert the 
local approximation for $\WPotE$:
\begin{multline}
            \VPotlocPZ(\FieldlocZ) = \int\limits_{\xE} \MFAFieldE_{,\xZ}(\xE|\FieldlocZ)
                                        \VPotlocPE(\FieldlocZ) \\
        - \frac14 \int\limits_{\xE_1,\xE_2} \FlucPropE (\xE_1,\xE_2|\FieldlocZ)
             \WPotlocPE(\xE_2,\xE_1|\FieldlocZ) 
                \Bigl(\MFAFieldE_{,\xZ}(\xE_1|\FieldlocZ) 
                      + \MFAFieldE_{,\xZ}(\xE_2|\FieldlocZ)\Bigr)\,.
    \label{VPrecursionLocalPre}
\end{multline}

We simplify this. By assumption of section \ref{introSection}
$$
        \int_{\xZ\in\LatZ}\AvOpE(\xZ,\xE) = 1 \,.
$$
Differentiating the constraint (\ref{MFAFieldEq2})
\begin{equation}
        \int\limits_{\xE} \AvOpE(\xZ',\xE) \MFAFieldE_{,\xZ}(\xE) = \delta(\xZ'-\xZ)\,,
\end{equation}
and integrating over $\xZ'$
we see that 
$$
        \int_{\xE\in\LatE} \MFAFieldE_{,\xZ}[\FieldZ](\xE) = 1 \,.
$$
This leads to the simpler recursion relation
\begin{multline}
            \VPotlocPZ(\FieldlocZ) = \VPotlocPE(\FieldlocZ) \\
        - \frac14 \int\limits_{\xE_1,\xE_2} \FlucPropE (\xE_1,\xE_2|\FieldlocZ)
             \WPotlocPE(\xE_2,\xE_1|\FieldlocZ) 
                \Bigl(\MFAFieldE_{,\xZ}(\xE_1|\FieldlocZ) 
                      + \MFAFieldE_{,\xZ}(\xE_2|\FieldlocZ)\Bigr)\,.
    \label{VPrecursionLocal}
\end{multline}

One can use this to determine the constraint effective potential, i.e. 
the potential on 
the last lattice $\Lat^N$ which consists of one point only.


%% file: numerik.tex
\section{Numerical results}

\subsection{Getting started}

We want to calculate the constraint effective potential on the lattice
numerically. 
We consider an example. 
For starting action we take the  standard $\varphi^4$-theory in two
dimensions, 
\begin{equation}
    \ActionNull(\FieldNull) = \,-\!\!\!\int\limits_{\xE,\yE\in\LatNull} 
             \half\FieldNull(\xE)\Laplace(\xE,\yE)\FieldNull(\yE)
             \ +\!\! \int\limits_{\xE\in\LatNull}\Bigl( \half m_0^2 \FieldNull(z)^2
               + \frac {\lambda_0} {4!} \FieldNull(z)^4 \Bigr) .
\label{startAction}
\end{equation}
To calculate $\WPotloc^1[\FieldZ]$ and $\VPotloc^1[\FieldZ]$ we need 
$\WPotlocNull[\FieldNull]$, $\WPotlocPNull[\FieldNull]$, $\WPotlocPPNull[\FieldNull]$, 
$\VPotlocPNull[\FieldNull]$ and $\FlucPropNull[\FieldNull]$ for constant fields 
$\FieldNull=\FieldlocE$. 
We write $\WPotNull$ (see also (\ref{startingAction})) in a symmetric form:
\begin{eqnarray}
        \WPotNull(\xE_1,\xE_2|\FieldNull) &=& \Laplace(\xE_1,\xE_2) 
                -(m_0^2 +\frac{\lambda_0}2 \FieldNull(\xE_1)^2)\delta(\xE_1-\xE_2) \\
                &=& \Laplace(\xE_1,\xE_2) 
                -\Bigl(m_0^2 +\frac{\lambda_0}4 (\FieldNull(\xE_1)^2+\FieldNull(\xE_2)^2)\Bigr)
                \delta(\xE_1-\xE_2) 
\end{eqnarray}
Therefore we have for $\WPotlocNull$
\begin{equation}
        \WPotlocNull (\xE_1,\xE_2|\xi) = \Laplace(\xE_1,\xE_2) 
                -\Bigl(m_0^2 +\frac{\lambda_0}2 \xi^2 \Bigr) \delta(\xE_1-\xE_2)
\end{equation}
which is not an approximation because equation (\ref{Wloc}) is {\em exact} in
this case. Now it follows for the constant field $\FieldlocE$
\begin{eqnarray}
        \WPotlocNull (\xE_1,\xE_2|\FieldlocE) &=& \Laplace(\xE_1,\xE_2) 
                -\Bigl(m_0^2 +\frac{\lambda_0}2 \FieldlocE^2 \Bigr) \delta(\xE_1-\xE_2) \\
        \WPotlocPNull(\xE_1,\xE_2|\FieldlocE) &=& -\lambda \FieldlocE\delta(\xE_1-\xE_2) \\
        \WPotlocPPNull(\xE_1,\xE_2|\FieldlocE) &=& -\lambda \delta(\xE_1-\xE_2)
\end{eqnarray}
Similarly, the potential
\begin{equation}
        \VPotlocPNull(\FieldlocE) \, = \, \VPotNull_{,\xE}[\FieldNull]
                                  \Einsch{\FieldNull=\FieldlocE} 
          \, = \, m_0^2\FieldlocE +\frac\lambda{3!} \FieldlocE^3
\end{equation}

\subsection{Relations in momentum space}

We have fixed the starting point. Next we want to calculate the recursion relations for 
$\WPotlocE$ and $\VPotlocE$. To reduce the amount of work we switch to momentum space. 
Because we insert constant fields we have translation symmetry. Therefore the 
Fourier
transformation looks quite simple. For the notation see appendix \ref{GammaLattice}.
\begin{eqnarray}
        \WPotlocE(\xE_1,\xE_2|\FieldlocE) &=& \int\limits_{\p} \WPotlocE(\p|\FieldlocE)
                e^{i\p(\xE_1-\xE_2)} \\
        \MFAFieldE_{,\xZ}(\xE|\FieldlocE) &=& \int\limits_{\p} \MFAFieldE(\p|\FieldlocE)
                e^{i\p(\xE-\xZ)} \\
        \FlucPropE(\xE_1,\xE_2|\FieldlocE) &=& \int\limits_{\lM,\q,\lM'} 
                \FlucPropE(\lM,\q,\lM'|\FieldlocE) e^{i((\q+\lM)\xE_1-(\q+\lM')\xE_2)}
                \label{flucmorep}
\end{eqnarray}
and similar for the derivatives, which are now simple derivatives with respect to $\FieldlocE$.
$\FlucPropE$ has only translation symmetry on the block lattice. Therefore it is not 
diagonal in momentum space (see appendix \ref{GammaLattice}).

Transforming the recursion relations into momentum space leads to
\begin{multline}
  \WPotlocZ(\qZ|\FieldlocE)  = \int\limits_{\lZ}
  \Adj{\MFAFieldPE}(\qZ+\lZ|\FieldlocE)
  \WPotlocE(\qZ+\lZ|\FieldlocE)\MFAFieldPE(\qZ+\lZ|\FieldlocE) \\
  \begin{split}
    & + \frac18\int\limits_{\qZ',\lZ_1,\ldots,\lZ_4}
        \FlucPropE(\lZ_1,\qZ',\lZ_2|\FieldlocE)
        \FlucPropE(\lZ_3,\qZ+\qZ',\lZ_4|\FieldlocE)
        \Adj{\MFAFieldPE}(\qZ-\lZ_2+\lZ_3|\FieldlocE)
        \MFAFieldPE(\qZ-\lZ_1+\lZ_4|\FieldlocE) \\
    & \quad
    \Bigl\{ \WPotlocPE(\qZ+\qZ'+\lZ_3|\FieldlocE)
           +\WPotlocPE(\qZ'+\lZ_2|\FieldlocE) \Bigr\} 
    \Bigl\{ \WPotlocPE(\qZ'+\lZ_1|\FieldlocE)
           +\WPotlocPE(\qZ+\qZ'+\lZ_4|\FieldlocE) \Bigr\} \\
    & + \frac14\int\limits_{\qZ',\lZ_1,\lZ_2,\lZ_3}
        \FlucPropE(\lZ_1,\qZ',\lZ_2|\FieldlocE)
        \Adj{\MFAFieldPE}(\qZ-\lZ_2+\lZ_3|\FieldlocE)
        \MFAFieldPE(\qZ-\lZ_1+\lZ_3|\FieldlocE) \\[-8pt]
    & \qquad\qquad\qquad\quad
    \Bigl\{ \WPotlocPPE(\qZ'+\lZ_1|\FieldlocE)
           +\WPotlocPPE(\qZ'+\lZ_2|\FieldlocE) \Bigr\}  \\
    & + \frac18\int\limits_{\qZ',\lZ_1,\ldots,\lZ_4}
        \FlucPropE(\lZ_1,\qZ',\lZ_2|\FieldlocE)
        \FlucPropE(-\lZ_1+\lZ_2,0,\lZ_4|\FieldlocE)
        \Adj{\MFAFieldPE}(\qZ+\lZ_3|\FieldlocE)
        \MFAFieldPE(\qZ+\lZ_3+\lZ_4|\FieldlocE) \\
    & \quad\qquad\qquad\quad
    \Bigl\{ \WPotlocPE(\qZ'+\lZ_1|\FieldlocE)
           +\WPotlocPE(\qZ'+\lZ_2|\FieldlocE) \Bigr\} 
    \Bigl\{ \Adj{\WPotlocPE}(\qZ+\lZ_3|\FieldlocE)
           +\WPotlocPE(\lZ_4|\FieldlocE) \Bigr\} \\
  \end{split}
\end{multline}
and
\begin{multline}
  \label{VPrecursionMom}
  \VPotlocPZ(\FieldlocE) = \VPotlocPE(\FieldlocE)
     - \frac14 \int\limits_{\qZ,\lZ,\lZ'}
         \FlucPropE(\lZ,\qZ,\lZ'|\FieldlocE)  \MFAFieldPE(-\lZ+\lZ'|\FieldlocE)  \\[-6pt]
               \Bigl\{\WPotlocPE(\qZ+\lZ|\FieldlocE) + \WPotlocPE(\qZ+\lZ'|\FieldlocE)\Bigr\}\,.
\end{multline}
We need also the formula (\ref{psiDerivative}) in momentum space:
\begin{equation}
        \MFAFieldPE(\qZ+\lZ|\FieldlocE) = \Adj{\AvOpE}(\qZ+\lZ) + \int\limits_{\lZ'}
                \FlucPropE(\lZ,\qZ,\lZ'|\FieldlocE)\WPotlocE(\qZ+\lZ'|\FieldlocE) 
                \Adj{\AvOpE}(\qZ+\lZ')
        \label{psiDerivativeMomentum}
\end{equation}

\subsection{Algorithm}

We are now able to calculate all 
quantities needed.  The following algorithm gives an overview
of the procedure:
\begin{itemize}
        \item[0.] $i=0:$ Given the starting action fix the values of 
                $\WPotlocNull(\p|\FieldlocE_j)$ and 
                $\VPotlocPNull(\FieldlocE_j)$ for a number of constant field 
                values $\FieldlocE_1,\ldots,\FieldlocE_n$ 
                and for  all momenta $\p$.
        \item[1.] Calculate $\WPotlocPE(\p|\FieldlocE_j)$ and $\WPotlocPPE(\p|\FieldlocE_j)$
                by numerical differentiation. 
        \item[2.] Determine $\FlucPropE(\lZ,\qZ,\lZ'|\FieldlocE_j)$
                with the help of the explicit formula (\ref{gammaform}).
        \item[3.] Calculate $\MFAFieldPE(\qZ+\lZ|\FieldlocE_j)$ by (\ref{psiDerivativeMomentum}).
        \item[4.] Insert all into recursion relations for $\WPotlocZ(\qZ|\FieldlocE_j)$ and
                $\VPotlocPZ(\FieldlocE_j)$.
        \item[5.] Increment $i$ and goto 1 until $i=N$.
\end{itemize}

We want to see numerically how this procedure works to produce the correct -- derivative of the --
effective
potential. Therefore we wrote three programs to compare various methods. A Monte-Carlo 
program gives us reference values. A combined heatbath and metropolis 
algorithm is used.
The heatbath is used with the kinetic term of the action to produce a Gaussian distributed
random value for the fluctuation field value. The following metropolis decides  on the basis of the whole action if this value
is accepted.

As an alternative we calculated  the effective potential perturbatively, using
the Gawedzki-Kupianen formalism \cite{GawKup}.
The action $\ActionE[\FieldE]$ is split into a kinetic term
$\half\scalar{\FieldE}{(\PropagatorE)^{-1}\FieldE}$ and a potential $\VPotE[\FieldE]$.
$\PropagatorE$ is the massless 
free propagator.
Again the notion of fluctuation field $\FlucFieldE$ with $\AvOpE\FlucFieldE=0$
is used 
and we have a background field $\MFAFieldE$:
\begin{equation}
    \FieldE=\MFAFieldE+\FlucFieldE\,.
    \label{fieldSplit}
\end{equation}
 The background field is an
interpolation of the block-spin field:
$$
    \MFAFieldE=\InOpE\FieldZ\,.
$$
The interpolation operator $\InOpE$ is defined as
\begin{equation}
    \InOpE = \PropagatorE\Adj{\AvOpE}(\PropagatorZ)^{-1}
    \label{InOperator}
\end{equation}
here $\PropagatorZ$ is the free propagator for the block-spin field:
\begin{equation}
    \PropagatorZ = \AvOpE\PropagatorE\Adj{\AvOpE}
    \label{Propa}
\end{equation}
Inserting the split of the field $\FieldE$ (\ref{fieldSplit}) into (\ref{defRG})
we get for the fluctuation integral
\begin{equation}
    e^{-\ActionZ[\FieldZ]}
     = e^{-\half\scalar{\FieldZ}{(\PropagatorZ)^{-1}\FieldZ} -\VPotZ[\FieldZ]}
     =  \int_{\HCE} D\FlucFieldE
        e^{-\half\scalar{\FlucFieldE}{\FlucPropE^{-1}\FlucFieldE}-\VPotE[\MFAFieldE+\FlucFieldE]}
\end{equation}
where the fluctuation propagator is defined as\footnote{Inserting (\ref{InOperator})
    and (\ref{Propa}) one sees the same formula for $\FlucPropE$
    as in (\ref{gammaform}). For free fields $\PropagatorE$ and ${\ActionPPE}^{-1}$ are identical.}
$$
    \FlucPropE = \PropagatorE - \InOpE\PropagatorZ\Adj{\InOpE}\,.
$$
The fluctuation integral can be rewritten as
\begin{equation}
   e^{-\VPotZ[\FieldZ]} = e^{\half\scalar{\frac{\delta}{\delta\MFAFieldE}}
                             {\FlucPropE\frac{\delta}{\delta\MFAFieldE}}}
                          e^{-\VPotE[\MFAFieldE]}\,.
\end{equation}
This can be expanded. All graphs in one loop
order and up to three fluctuation propagators are taken into account.
This leads to the graphs in figure \ref{graphPert}.

\GraphPert{Perturbative expansion of the effective potential. External
  lines represent $\InOpE\FieldZ$ and internal lines $\FlucPropE$.
  2-point vertices have the weight $m_i^2$, 4-point vertices have the
  weight $\lambda_i$ and 6-point vertices have the weight $\gamma_i$.}

The scale factor (block length) is the same as for the saddle point
approximation and we used a polynomial parameterization of the potentials:
$$
    \VPotE[\Field] = \int\limits_{\xE\in\LatE}\Bigl(
    \half{m_i}^2\Field(\xE)^2 + \frac{\lambda_i}{4!}\Field(\xE)^4
    + \frac{\gamma_i}{6!}\Field(\xE)^6\Bigr) \,.
$$

\subsection{Result}

As a result we get the following picture:
\begin{itemize}
\item For small field values the perturbative calculation is comparable
in accuracy  to   our method.
\item For larger field values the perturbative calculation goes rapidly wrong,
while the saddle point approximation remains accurate and has
the correct   asymptotics (see figure \ref{overview}).
\item  For the values of coupling constants which we consider,
larger blocks are better than small blocks (see figure \ref{largeBlock}).
This confirms our expectation from the discussion of the localization
approximation. 
\end{itemize}

\newlength{\pswidth}
\setlength{\pswidth}{0.7\textwidth}
\newcounter{Angle}
\setcounter{Angle}{270}

\begin{figure}
    \begin{center}
        \epsfig{file=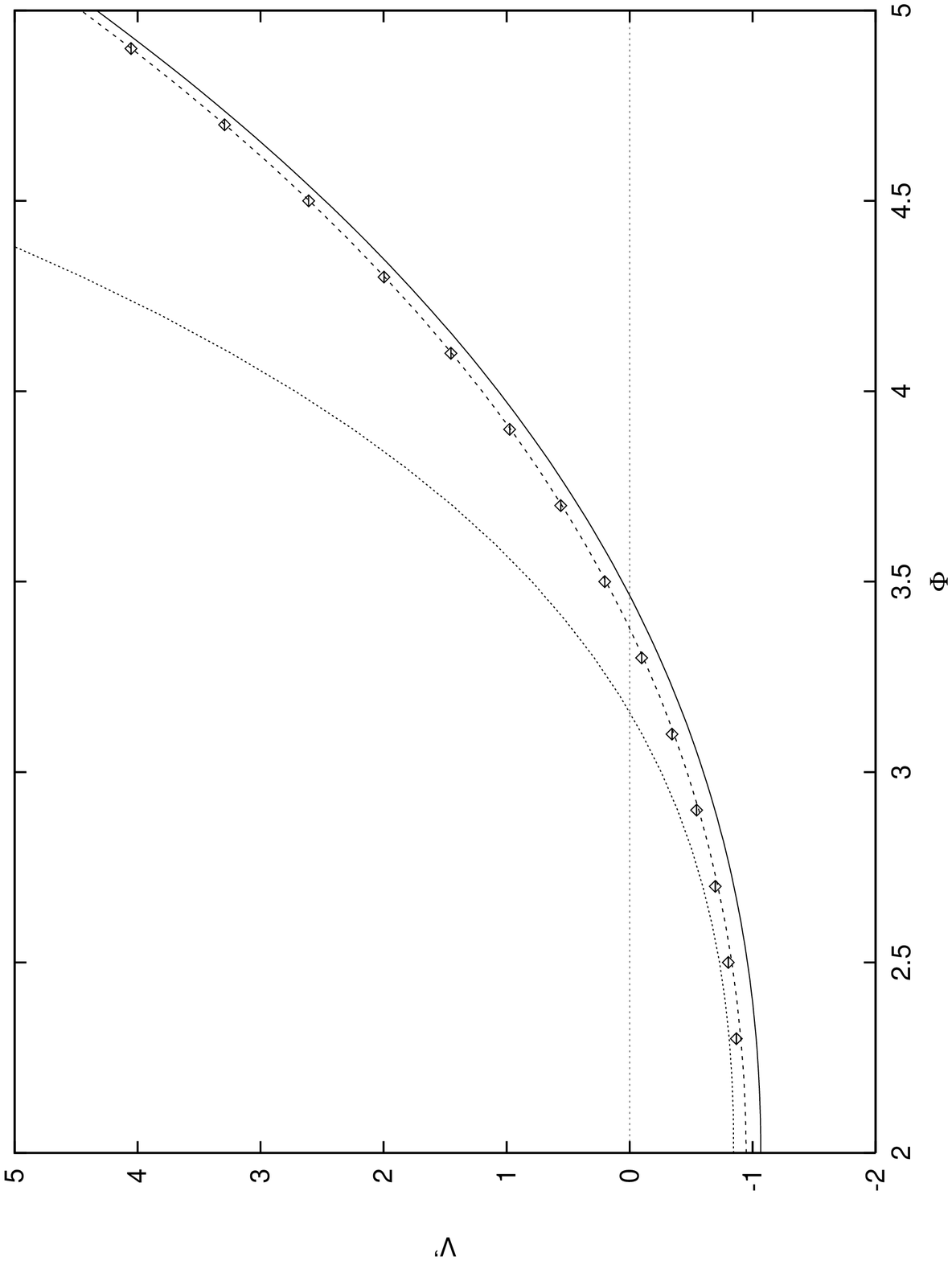,width=\pswidth,angle=\theAngle}
    \caption{The derivative of the constraint effective potential 
        is plotted against the block-spin $\FieldZ$. 
        The bottom line is the original potential. The next upper line
        is the saddle point approximation and the top line 
        the perturbative calculation.
        The open squares represent the Monte Carlo
        results. This was calculated on a $16\times16$
        lattice with blocking factor 2, $m_0^2=-0.8$
        and $\lambda_0=0.4$. 
Note that the minimum of the potential is where $V'=0$ (see figure
\ref{mexican}.)
}
    \label{overview}
    \end{center}
\end{figure}

\begin{figure}
    \begin{center}
        \epsfig{file=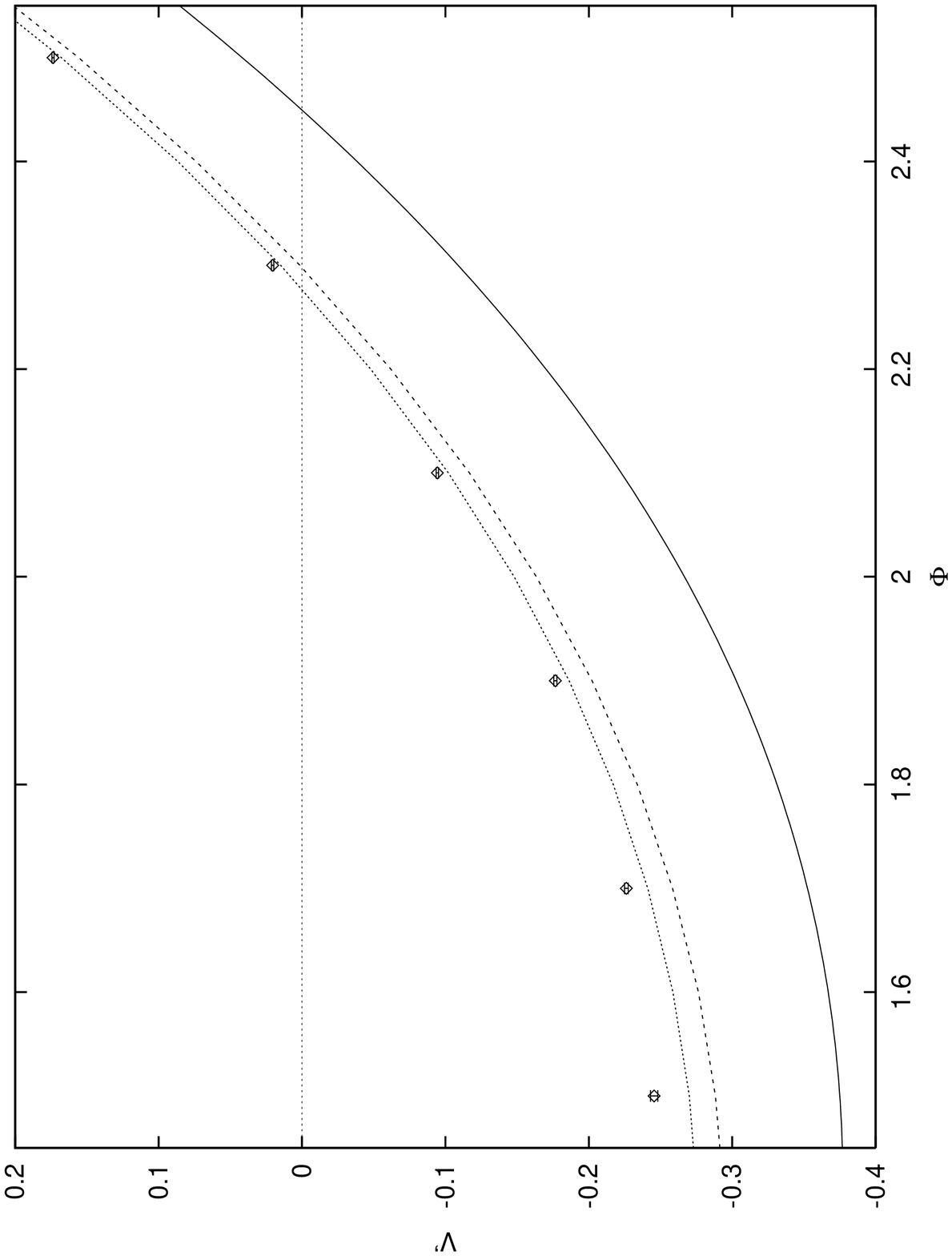,width=\pswidth,angle=\theAngle}
    \caption{The derivative of the constraint effective potential 
        is plotted against the block-spin $\FieldZ$.  
        The bottom line is the original potential. \sticky{The next upper line 
        consists actually of two lines. The saddle point approximation
        and the perturbative calculation (both with blocking factor 2) coincide
        for small block-spin field values.}
        The next upper line is the saddle point approximation with blocking
        factor 2.
        The top line is the saddle point approximation with blocking factor 4.
        It is better than with blocking factor 2.
        The open squares represent the Monte Carlo
        results. This was calculated on a $16\times16$ for $m_0^2=-0.4$
        and $\lambda_0=0.4$.}
    \label{largeBlock}
    \end{center}
\end{figure}

\begin{figure}
    \begin{center}
        \epsfig{file=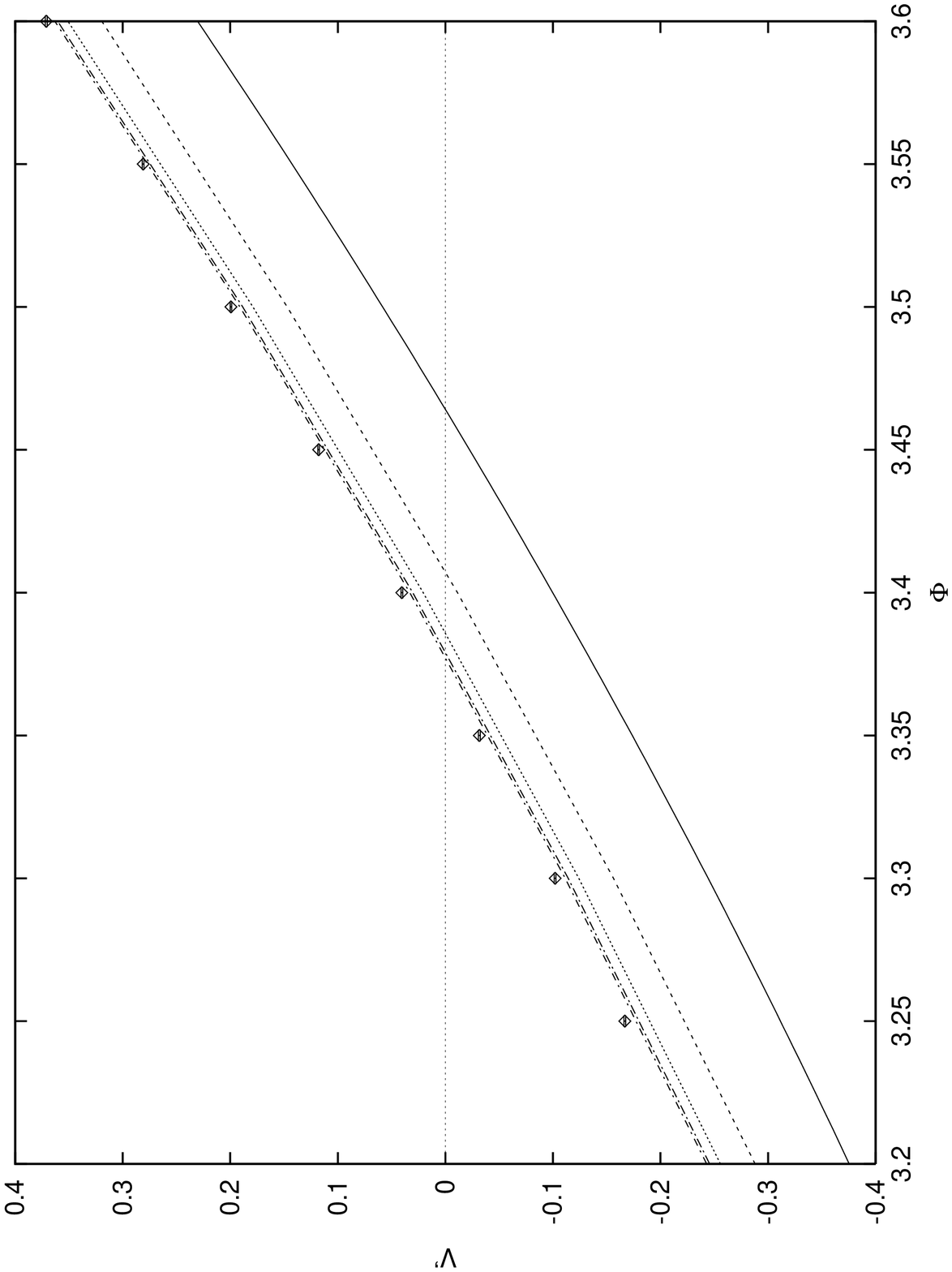,width=\pswidth,angle=\theAngle}
    \caption{The flow of the derivative of the constraint effective potential 
        is plotted against the block-spin $\FieldZ$ 
        in the vicinity of the minimum of the potential.   
        The lines represent $\VPotlocPE$ for $i=0,1,2,3,4$ from the bottom
        and the open squares the Monte Carlo
        results. This was calculated on a $16\times16$
        lattice with blocking factor 2, $m_0^2=-0.4$
        and $\lambda_0=0.4$. The extension of the lattice is getting smaller 
        after every blocking step by factor 2. On the other side the
        lattice spacing grows by factor 2, so that the lattice
        volume stays constant. No rescaling to the unit lattice is performed.}
    \label{flowOfPotential}
    \end{center}
\end{figure}

\noindent In figure \ref{flowOfPotential} the flow of the potential is shown.

%% file: normalorder.tex
\section{Improved saddle point method with normal ordering}
\label{FBapproximation}

\subsection{Advanced method with normal ordering}

The localization approximation becomes more and more accurate the larger
the scaling factor (ratio of lattice spacings) $s=\LatSpaceZ/\LatSpaceE $.
On the other hand, the saddle point approximation becomes exact in the limit
$s\mapsto 1 $ (when the nature of the cutoff permits such a limit). It becomes
less accurate with increasing scaling factor because the phase space for
high frequency modes increases. This is one reason why we choose to consider
a fairly sophisticated improvement of the saddle point method.

The self consistent improvement of the saddle point
integration consists of two steps.
In the first step, the old action $\ActionE [\FieldE ]=
  \ActionE [\MFAFieldE + \FlucFieldE ]$ is self-consistently normal ordered in
the fluctuation field $\FlucFieldE$; at the same time the background field
$\MFAFieldE=\MFAFieldE[\FieldZ]$ is determined. Symbolically
\begin{equation}
    \ActionE [\FieldE ] = :\NActionE [\MFAFieldE , \FlucFieldE ]:
\label{NOSymb}
\end{equation}

The precise definition and properties of the normal ordered form $\NActionE $
of the action $\ActionE$  will be described presently.
In the second step, the fluctuation field integration is performed, neglecting
normal products of higher order than second  in the fluctuation  field.
The integral is again Gaussian, and the result reads
\begin{equation}
  \ActionZ [\FieldZ ] = \NActionE [\MFAFieldE , 0 ]
  -\half \tr \ln \FlucPropE [\MFAFieldE ] , \label{Normalaction}
\end{equation}
The trace is again over the space $\HCE$
 of functions $\FlucFieldE $ on
$\LatE $ with vanishing block average, $\AvOpE \FlucFieldE =0$.

The method of self-consistent normal ordering
is very old - see \cite{Ruehl}; sometimes it is called
the Feynman Bogoliubov method. Therefore we will refer to our method here 
as Feynman Bogoliubov approximation. The definition and properties of the
normal ordered action $\NActionE $ are as follows.

Let $d\mu_{\Gamma }$ be the normalized Gaussian measure (free field measure)
with covariance (propagator) $\Gamma $. 
 Given the 
fluctuation propagator $\FlucPropE $,
the normal ordered amplitude $\NActionE $ is defined as
\begin{eqnarray}
  \NActionE [\MFAFieldE , \FlucFieldE ] &=&
  \int d\mu_{\FlucPropE [\MFAFieldE] }(\xi )
 \Action [\MFAFieldE + \FlucFieldE + \xi ]
  \label{defNO}\\
  &=&
  \exp \left( \half \scalar{\frac{\delta}{\delta \xi}} 
                           {\FlucPropE [\MFAFieldE ] \frac{\delta}{\delta \xi}}
       \right)
  \Action [\MFAFieldE + \FlucFieldE + \xi ]\Einsch{\xi = 0}\label{pregap0}
\end{eqnarray}
The second formula is valid when $\ActionE$ is differentiable. The first 
formula is more general, it can be used to define a normal product expansion
also for discontinuous actions $\ActionE$ (cp. the section \ref{DS}).

We use the following notation for functionals with two arguments:
$$
    \frac{\delta }{ \delta \MFAFieldE (\xE)} A[\MFAFieldE,\FlucFieldE]
        = A_{,\xE}[\MFAFieldE,\FlucFieldE]\,,
    \qquad \frac{\delta }{ \delta \FlucFieldE (\xE)} A[\MFAFieldE,\FlucFieldE]
        = A_{;\xE}[\MFAFieldE,\FlucFieldE]\,.
$$
The notation is a little sloppy. Because $\AvOpE \FlucFieldE =0$, the
second expression only makes sense after smearing with a test function $f$ 
which obeys $\AvOpE f = 0$, viz.
$$ \int_z f(z) \frac{\delta }{ \delta \FlucFieldE (\xE)}
 A[\MFAFieldE,\FlucFieldE] = \frac {d}{d\tau }
 A[\MFAFieldE,\FlucFieldE + \tau f]\Einsch{\tau =0}.
$$
The background field $\MFAFieldE = \MFAFieldE [\FieldZ ] $ will be determined 
by the condition that the expansion of the  action  $\NActionE $ 
 in powers of the fluctuation field has no linear
term:
\begin{equation}
  \NActionE [\MFAFieldE , \FlucFieldE ]
  = \NAction [\MFAFieldE , 0 ] 
    + \half \scalar{\FlucFieldE} 
                   {\NActionppE [\MFAFieldE , 0 ]\FlucFieldE} 
    + ... \qquad \mbox{ for} \ \AvOpE \FlucFieldE = 0. \label{pregap1}
\end{equation}
That is the saddle point condition
\begin{equation}
    \int_{\xE}\FlucFieldE (\xE) \NActionE_{;\xE} [\MFAFieldE ] = 0
     \ \ \mbox{if} \ \AvOpE \FlucFieldE = 0 \label{saddleN}
\end{equation}
which is equivalent to
\begin{eqnarray}
    \NActionpE [\MFAFieldE] &=& \Adj {\AvOpE }\E\lambda , \label{MFAFieldEq1N} \\
    \AvOpE \MFAFieldE &=& \FieldZ \ . \label{MFAFieldEq2N}
\end{eqnarray}
$\E\lambda[\FieldZ] (\xZ ) $ are  Lagrange multipliers, $\xZ\in \LatZ $.

Now we specify the fluctuation propagator $\FlucPropE$ as pseudo inverse
of $\NActionppE$ so that
\begin{eqnarray}
    \FlucPropE [\MFAFieldE ]\NActionppE [\MFAFieldE ]
    \FlucPropE [\MFAFieldE ]&=&  \FlucPropE [\MFAFieldE ].\label{gap}\\
\AvOpE  \FlucPropE [\MFAFieldE ] =  \FlucPropE [\MFAFieldE ]\Adj{\AvOpE} &=&0.
\end{eqnarray}
As in  the saddle point approximation the fluctuation propagator satisfies
the constraints (\ref{GammaConstraint0}), is positive definite on
$\HCE$, and equal to the inverse of $\NActionppE$ on $\HCE$ 
But $\NActionppE $ depends itself on the propagator $\FlucPropE$. Therefore
(\ref{gap}) leads to a gap equation for $\FlucPropE$.

One sees that the background field $\MFAFieldE $ is the minimum of the
normal ordered form $\NActionE $ of the action $\ActionE $ on the surface with
prescribed block-spin $\FieldZ = \AvOpE \FieldE $, and
 $\FlucPropE $ is  the inverse of the Hessian $\NActionppE $ at
this minimum; it depends on $\MFAFieldE $. (The Hessian is a quadratic form
on the tangent space $\HCE $ to the surface.)

The determination of the fluctuation field propagator $\FlucPropE $ involves
solving a gap equation. The gap equation is obtained by differentiating
eq. (\ref{pregap0}) twice at $\FlucFieldE = 0$.
\begin{eqnarray}
  \NActionppE &=&
  \expdiff \ActionPPE [\MFAFieldE+\xi]\Einsch{\xi=0} \label{gapeqn}\\
  \FlucPropE [\MFAFieldE ] &=& \NActionppE [\MFAFieldE , 0 ]^{-1} \ \mbox{on }
  \HCE .
\end{eqnarray}
The gap equation is shown in graphical form in figure \ref{BildI}.
Figure \ref{BildAb}  explains the graphical notation.
Further derivatives can be applied  and we 
obtain a generalization of eq.(\ref{gapeqn}) as 
 shown in figure \ref{BildB}.

\BildAb{Graphical notation for normal ordered amplitudes}
\BildB{Definition of the normal ordered amplitudes. Further legs
generated by $\MFAField$-derivatives will be marked by primes}

\BildI{Gap equation for the fluctuation propagator $\FlucPropE$ in
  Feynman-Bogoliubov approximation}

This completes the discussion of the improved saddle point approximation.
Eq.(\ref{Normalaction}) is still a formidable functional recursion formula.
We need to make it practical by our second approximation of localization.
The localization approximation has to be made for the second derivative
$\NActionppE $ and not for $\ActionPPE$.
In addition to the formula for the second derivative $\ActionPPE$ 
 there is a formula which expresses
the first derivative $ \ActionPE$ of the ordinary action in terms of the
first derivative $\NActionpE $ of the previous normal ordered action; this
furnishes the potential up to an additive constant.

We have a similar formula for the derivative of the background field:
\begin{equation}
    \MFAFieldE_{,\xZ }(\xE ) =
        (1 - \FlucPropVarE \NActionpPE )\Adj{ \tilde \AvOpE}(\xE,\xZ).
\end{equation}
where  $\FlucPropVarE$ is defined as inverse of $\NActionpPE$ on $\HCE$.
In contrast to $\FlucPropE$ it is defined as inverse of the mixed 
derivative $\NActionpPE$ which involves derivatives with respect to
the fluctuation field and the background field.

{\em Remark:} It can be shown with some effort that the two propagators are 
actually equal, i.e.  $\FlucPropVar [\MFAFieldE ] = \FlucPropE [\MFAField ]$
when $\MFAFieldE $ is the background field. See appendix \ref{derRelSimple}.

\subsection{Gaussian integration}

We wish to evaluate the integral (\ref{defEffective}) for $\ActionZ $, using the
 quadratic approximation  (\ref{NOSymb},\ref{pregap1}) for $\ActionE $, viz.
$$
    \ActionE [\MFAFieldE + \FlucFieldE ]
    = \NActionE [\MFAFieldE , 0 ] 
      + \half :\scalar{\FlucFieldE} {\NActionppE[\MFAFieldE,0]\FlucFieldE}:
$$
We undo the normal ordering again. Dropping arguments $\xi=0$ we obtain 
\begin{equation} 
  \ActionE [\MFAFieldE + \FlucFieldE ]
  = \NActionE [\MFAFieldE ] 
    + \half \scalar{\FlucFieldE} {\NActionppE [\MFAFieldE]\FlucFieldE}
    - \half \tr \NActionppE [\MFAFieldE ]
 \FlucPropE [\MFAFieldE ] . \label{newActionTemp}
\end{equation}
The trace is over $\HCE$. By arguments similar to appendix \ref{GammaFormAp}
the fluctuation propagator has the explicit form 
\begin{equation}
    \FlucPropE = {\NActionppE}^{-1} -
      {\NActionppE}^{-1}\Adj{\AvOpE}(\AvOpE{\NActionppE}^{-1}\Adj{\AvOpE})^{-1}
           \AvOpE{\NActionppE}^{-1}\,.
\label{GammaExplicit}
\end{equation}
As in the saddle point approximation
there is an alternative formula for  the fluctuation propagator 
\begin{equation}
  \FlucPropE[\MFAFieldE [\FieldZ ] ] = \lim_{\kappa \rightarrow \infty }\left(
  \NActionppE  [\MFAFieldE [\FieldZ ]] + \kappa \Adj{\AvOpE}\AvOpE   
  \right)^{-1}\,.
\end{equation}
We argue that the last term in (\ref{newActionTemp}) is a constant
independent of $\MFAFieldE $. Multiplying $\NActionppE$ from the right of
(\ref{GammaExplicit}) gives
\begin{equation}
    \FlucPropE\NActionppE = 1 -
      {\NActionppE}^{-1}\Adj{\AvOpE}(\AvOpE{\NActionppE}^{-1}\Adj{\AvOpE})^{-1}
           \AvOpE \,.
\label{GammaMalT}
\end{equation}
We take the trace of (\ref{GammaMalT}) 
\begin{equation*}
   \begin{split}
        \tr(\FlucPropE\NActionppE) &=\ \tr 1 
            - \tr\Bigl({\NActionppE}^{-1}\Adj{\AvOpE}
            (\AvOpE{\NActionppE}^{-1}\Adj{\AvOpE})^{-1} \AvOpE\Bigr)  \\
        &=\ \tr\ \delta_{\xE}(0) - \tr\ \delta_{\xZ}(0) \\
        &=\ \sum\limits_{\xE} 1 - \sum\limits_{\xZ} 1 = \text{const.} 
        \qquad\qquad \text{q.e.d.} \\
   \end{split}
\end{equation*} 
Now we can again evaluate the functional integral (\ref{defEffective}) in Gaussian
approximation with the result
\begin{equation}
        \ActionZ [\FieldZ ] = \NActionE [\MFAFieldE[\FieldZ] , 0 ]
                -\half \tr \ln \FlucPropE [\MFAFieldE[\FieldZ] ] . 
        \label{Normalaction1}
\end{equation}

\subsection{Relations between derivatives of the normal ordered action}
\label{derRel}

We derive a relation between the 
$\MFAFieldE$- and $\xi $-derivative by using the following well known change
of covariance lemma for Gaussian measures \cite{GlimmJaffe}.  
If the covariance $\Gamma_s$ depends parametrically on a variable $s$,
then 
\begin{equation}
    \frac {d}{ds} \int d\mu_{\Gamma_s }(\Field ) A[\Field ] =
    \half \int d\mu_{\Gamma_s }(\Field )
    \scalar{\frac{\delta}{\delta \Field }}
           {\dot{\Gamma}_s \frac{\delta}{\delta\Field}A[\Field]}\,,
    \label{ChangeCov}
\end{equation}
where    $\dot{\Gamma}_s = d\Gamma_s /ds $. 

We denote $\MFAFieldE$-derivatives by commas ``,'' and $\xi $-derivatives 
by semicolons ``;'' as before. 
Applying the change of covariance lemma to the definition (\ref{defNO})
of the normal ordered amplitude we obtain the relation
\begin{equation}
    \NActionE_{,\xE }[\MFAFieldE ] = \NActionE_{;\xE }[\MFAFieldE ] +  \half
    \tr \FlucProp{}_{, \xE }[\MFAFieldE ] \NActionppE [\MFAFieldE ].
\label{;to,}
\end{equation}
After another $\xi$-derivative we get
\begin{equation}
    \NActionE_{,\xE\,;\yE }[\MFAFieldE ] = \NActionE_{;\xE\yE }[\MFAFieldE ]
    +  \half
    \tr \FlucProp{}_{, \xE }[\MFAFieldE ] \NActionppE{}_{;\yE } [\MFAFieldE ].
\label{mixedReplacement}
\end{equation}


\subsection{Recursion relation for the first and second derivatives of the 
action}

We differentiate the recursion relation (\ref{Normalaction1}) 
to obtain a formula 
for the first derivative of the action $\ActionZ$. The derivative of the 
$\tr \ln $ is computed in the same way as before. The result reads
\begin{equation} 
    \ActionZ_{,\xZ} = \int_{\xE} \MFAFieldE_{,\xZ }(\xE ) \left(
    \NActionE_{,\xE} [\MFAFieldE ] -
    \half \tr \FlucPropE_{,\xE} [\MFAFieldE ]
    \NActionppE\right) .
\end{equation}  
We re-express the $\MFAFieldE$-derivative in the first term using 
eq.~(\ref{;to,}). A cancelation occurs and we
obtain the 
final form of the recursion relation for the first derivative of the action.
\begin{equation} 
  \ActionZ_{,\xZ} = \int_{\xE} \MFAFieldE_{,\xZ }(\xE ) 
  \NActionE_{;\xE } [\MFAFieldE].
  \label{actionP}
\end{equation}

Next we turn to the second derivative. We differentiate eq.~(\ref{actionP})
once more. Since $ \NActionE_{;\xE }$ is the $\xi$-derivative 
of the normal
ordered action we can use eq.~(\ref{;to,}) again to evaluate the derivative
of $\NActionE_{;\xE}$ with the help of 
\begin{equation}
  \FlucPropE_{,\xE}[\MFAFieldE ] 
  = - \FlucPropE[\MFAFieldE ] \NActionppE{}_{\,, \xE }[\MFAFieldE ] 
  \FlucPropE[\MFAFieldE ]. 
\end{equation}
Differentiating the factor $\MFAFieldE_{,\xZ }(\xE ) $
produces a term
\begin{equation}
    \int_{\xE}\MFAFieldE_{,\xZ \yZ }(\xE )\NAction_{;\xE} \label{1PI}.
\end{equation}
Differentiating the constraint 
(\ref{MFAFieldEq2N}) twice shows that $\MFAFieldE_{,\xZ\yZ}\in \HCE$ because
\begin{equation}
  \int\limits_{\xE} \AvOpE(\xZ',\xE) \MFAFieldE_{,\xZ}(\xE) 
  = \delta(\xZ'-\xZ)\,,  \label{constr} 
\end{equation}
implies
\begin{equation}
  \int\limits_{\xE} \AvOpE(\xZ',\xE) \MFAFieldE_{,\xZ\yZ}(\xE) = 0\,.
  \label{MFAFieldPPinHCE}
\end{equation}
The saddle point condition (\ref{saddleN}) implies therefore that expression 
(\ref{1PI}) vanishes. As a result we obtain the recursion formula for 
the second derivative of the action
\begin{eqnarray} 
  \ActionZ_{,\xZ \yZ} &=&
  \int_{\xE,\yE} \MFAFieldE_{,\xZ }(\xE )\NActionE_{;\xE\,,\yE}
  \MFAFieldE_{,\yZ}(\yE) \\
&=& 
  \int_{\xE,\yE} \MFAFieldE_{,\xZ }(\xE ) 
  \left(
    \NActionE_{;\xE \yE}  - \half \, \tr
    (\FlucPropE \NActionppE{}_{\,;\xE } \FlucPropE \NActionppE{}_{\,,\yE})
  \right) \MFAFieldE_{,\yZ}(\yE)
  \label{actionPP}
\end{eqnarray}
We neglected to indicate arguments $\MFAFieldE $ of propagators and normal
ordered action. We see that there is no 1-particle reducible contribution.  

A subtle feature of this formula is the appearance of two different 
vertices in the last term; one involves a $\MFAFieldE$-derivative and
the other a $\xi$-derivative to create the external leg. This feature is
familiar from Schwinger Dyson equations, cp. section \ref{DS} 
below. Neglecting 
the difference would only neglect a two loop correction, though. 

The recursion relations in Feynman Bogoliubov approximation are shown in 
figures \ref{BildN} and \ref{BildH}. 
They involve propagators which are to be determined as 
solutions of the gap equation (\ref{gapeqn}), shown graphically in figure
\ref{BildI}.

\BildN{Recursion relation for $\ActionPZ $ in the Feynman Bogoliubov 
  approximation}

\BildH{Recursion relations for the $\ActionPPZ$ in the Feynman Bogoliubov
  approximation. The second formula is obtained from the first with the help
  of eq.~(\ref{mixedReplacement}).}

\subsection{Derivative of the background field}

The recursion relations involve the derivative $\MFAFieldE_{,\xZ}$ of the
 background field. Differentiating the saddle point condition 
(\ref{saddleN}) we obtain 
\begin{equation}
    \int_{\xE,\yE}\FlucFieldE (\yE)\MFAFieldE_{,\xZ }(\xE )
    \NActionE_{;\yE,\xE} [\MFAFieldE ] = 0  \label{saddleConditionDerivative}
\end{equation}
We claim that this equation and the constraint (\ref{constr}) are solved by
\begin{equation}
    \MFAFieldE_{,\xZ }(\xE ) = \Adj{ \tilde \AvOpE}(\xE , \xZ )
    - \int_{\yE, u} \FlucPropVarE(\xE , \yE)
    \NActionE_{;\yE\,,u} \Adj{ \tilde \AvOpE}
    (u , \xZ ), \label{backDerN}
\end{equation}
where $\Adj {\tilde \AvOpE} =\Adj{\AvOpE} (\AvOpE \Adj{\AvOpE })^{-1}$
as before and $\FlucPropVarE$ is defined as the inverse of $\NActionpPE$
on $\HCE$. The proof proceeds as in the saddle point section.

The remark at the end of the last subsection leads to a simplification
of (\ref{backDerN})
because $\FlucPropE $ can be substituted for $\FlucPropVarE$.
 See appendix \ref{derRelSimple}.

\subsection{Discussion of the localization approximation}

There are new aspects of the localization approximation which need to be 
discussed. 
The consideration of this section assumed
that the  starting amplitude $\ActionE$ is accurately known.
Because of this, it is appropriate to regard the recursion relations as 
recursion relations for the normal ordered amplitudes $\NActionE$,
and to make the 
localization approximation for these amplitudes,
{\em not} for $\ActionE$. The localization approximation for 
$\NActionE$ is justified by
appealing to the same arguments as before:
In computing $\ActionZ$ we only need $\NActionppE $ for smooth fields. 
 
Thus
we may regard one step in the recursion relations as composed of three parts
\begin{enumerate}
 \item Localization approximation on $\NActionE $ for smooth fields . 
 \item Computation of $\ActionZ $ from $\NActionE $ for smooth fields
 \item computation of 
 $\NActionZ $ from $\ActionZ $
\end{enumerate}
When proceeding in this way, the result for $\ActionZ$ must
be substituted into the (gap) equations which determine the normal ordering. 
In general there 
result graphs which contain propagators from two successive length scales 
$i$ and $i+1$. This complication does not arise when only 1-loop graphs are 
retained.   

There is one further technical complication. In contrast with
$\ActionPPE$, 
$\NActionppE$ is not necessarily exactly invariant
under arbitrary translations by 
lattice vectors in $ \LatE $ but only under block
 lattice translations.
Therefore it  it is not easy to
calculate its pseudoinverse (the fluctuation field  propagator) exactly.
One can proceed as follows, however. 
   
Because $\NActionppE$ and $\FlucPropE$ are determined by a gap equation 
one can start with some guess for $\NActionppE$, e.g. with 
the pseudoinverse of $\ActionPPE$ which was used in the saddle point
 approximation.
 Then one iterates the gap equation.

The gap equations are here considered for reasons of accuracy
and not because there
are infrared problems. Quite on the contrary, the propagators must have 
exponential decay. Therefore the iteration can be expected to converge very
 fast.



%% file: DS.tex
\section{Schwinger Dyson Equations}
\label{DS}

In principle one can improve on the approximations for the fluctuation 
field integral if the  action $\ActionE $ is known.
  In any case, more accurate methods are applicable
in the first step, since the starting action $\Action^0$ 
is known by assumption. 

If $\ActionE$ is only known 
approximately, there can be problems because the correction terms might
depend  on details of the action which are neglected when a localization
approximation is made. This is the crucial point to  discuss.  It will turn out
that inclusion of two loop corrections requires that also the localization 
approximation is improved to the next order. One needs to consider
third derivatives of the 
normal ordered action evaluated at nearly constant fields.

Here we discuss the evaluation of the fluctuation field integral by
solution of Schwinger Dyson equations. 

Let $ \FlucProp $ be a positive semidefinite operator on  $\H $. Then 
the normalized Gaussian measure $d\mu_{\FlucProp }$ on $\H $ is defined. Its
characteristic function is
\begin{equation}
  \int d\mu_{\FlucProp }(\phi )\exp i\scalar{q}{\phi} = \exp\left( -\half
  \scalar{q}{\FlucProp q} \right) .  \label{qGauss}
\end{equation}
If $A[\phi ]$ is a function of $\phi $ which is integrable with respect
to the Gaussian measure $ d\mu_{\FlucProp }$,
it can be written in normal ordered form, as we know.  
\begin{eqnarray}
A[\phi ] &=& :B [\phi ]:\\
B[\phi ] &=& \int d\mu_{\FlucProp }(\xi ) A[\xi + \phi ]
\end{eqnarray}
The normal ordering prescription depends on $\FlucProp $.

If $\FlucProp $ is strictly positive and bounded away from zero,
 then $B [\phi ]$ is an entire analytic
function of $\phi $ and can therefore be expanded in  a power series. This
yields the expansion of $A$ in normal products of $\phi $. It exists 
whether or not $A$ is  continuous or differentiable. This is important 
when one wishes to deal with singular cases such as a 
discrete Gaussian model, where the action depends on the integer part of 
a real lattice field. 

Here we are interested in cases where $\FlucProp $ has zero eigenvalues. The 
1-di\-men\-sion\-al Dirac measure $ \delta (x)dx $ is the prototype of such a
measure; it is not a pathological case. In this degenerate case $\FlucProp $
projects on a subspace $\H_C$ of $\H $ and $B[\phi + \xi ]$ is an entire analytic
function of $\xi \in \H_C$, assuming the
restriction $\FlucProp_C$ of $\FlucProp $ to
$\H_C$ is bounded away from zero. 

The normal ordering commutes with shifts of the field
and with derivatives with respect to $\Field $.

We need the following general integration by parts formula
\begin{equation}
  \int d\mu_{\FlucProp }(\phi ) A[\phi ] e^{-\VPot[\phi]} =
  B[\FlucProp \frac {\delta}{\delta \xi } ] 
  \int d\mu_{\FlucProp }(\phi )  e^{-\VPot[\phi + \xi ]}\Einsch{\xi = 0}. 
  \label{PDiff}
\end{equation} 
This generalizes the well known special case \cite{GlimmJaffe}
\begin{equation}
  \int d\mu_{\FlucProp }(\phi )\phi (z) e^{-\VPot[\phi ]} =
  \int_w \FlucProp (z,w) \frac {\delta}{\delta \xi (w)  } 
  \int d\mu_{\FlucProp }(\phi )  e^{-\VPot[\phi + \xi]}\Einsch{\xi = 0}
\end{equation} 
and can be readily proven with the help of eq.~(\ref{qGauss}), as follows. 

{\sc Proof of the general integration by parts formula:}
We use the Taylor formula in compact form
\begin{equation}
e^{a \frac {\delta }{\delta \phi } }F(\phi ) = F(\phi + a). 
\end{equation}
Since an arbitrary functional $B[\phi]$ admits a functional Fourier transform,
it suffices to prove the formula (\ref{PDiff}) for the special case
$$
    B[\phi] =  e^{i\scalar{q}{\phi}}
$$
We compute 
\begin{equation}
\begin{split}
    & \int d\mu_{\FlucProp}(\phi) 
        e^{i\scalar{q}{\FlucProp\frac\delta{\delta\phi}}} A[\phi] \\
    & = N_{\FlucProp} \int D\phi A[\phi] 
        e^{-i\scalar{q}{\FlucProp\frac\delta{\delta\phi}}} 
        e^{-\half\scalar{\phi}{\FlucProp^{-1}\phi}} \\
   & = N_{\FlucProp} \int D\phi A[\phi] 
        e^{-\half\scalar{(\phi - iq\FlucProp )}
            {\FlucProp^{-1}(\phi -i\FlucProp q )}} \\
   & \int d\mu_{\FlucProp}(\phi) A[\phi]
        e^{i\scalar{q}{\phi}+\half\scalar{q}{\FlucProp q}}\\
   & \int d\mu_{\FlucProp}(\phi) A[\phi]
        :e^{i\scalar{q}{\phi}}:  \qquad\qquad \text{{\sc q.e.d.}}     \\  
\end{split}
\end{equation}

%
The general integration  by parts formula has many uses.
 Suppose we define expectation values of an  interacting
theory 
\begin{eqnarray}
  \mean{A} &=& Z^{-1}\int d\mu_{\FlucProp }(\phi ) A[\phi ] e^{-\VPot[\phi]}, \\
  Z &=&\int d\mu_{\FlucProp }(\phi ) e^{-\VPot[\phi]}, 
\end{eqnarray}
and the generating function $\GAction[\xi]$ of amputated  Green 
functions of the interacting theory
\begin{equation}
  e^{\GAction[\xi ]} = \int d\mu_{\FlucProp }(\phi ) e^{-\VPot[\phi+\xi]}.
\end{equation}
We preferred to introduce an unnormalized generating function,
so that $\GAction[0]=\ln Z$. The normalized generating function is
$$
    \KAction[\xi]= \GAction[\xi]-\GAction[0].
$$ 
The integration by parts formula tells us that the expectation values equal
\begin{equation}
  \mean{A} = e^{-\GAction[0]}B[\FlucProp \frac {\delta}{\delta \xi } ] 
  e^{\GAction[\xi]}\Einsch{\xi= 0}. 
\end{equation}
The Schwinger Dyson equation for the generating function  $\GAction $ is another
application of the integration by parts formula. It is obtained by
differentiating the definition of $e^\GAction $ with respect to $\xi $ and 
converting the resulting downstairs factor $ \frac {\delta}{\delta \xi }\VPot $
into a differential operator. Suppose
$$
  \VPot[\phi] = :\WPot[\phi]:.
$$
Then 
\begin{eqnarray}
        -\frac {\delta}{\delta \xi (z) }\GAction[\xi] &=& e^{-\GAction[\xi]} 
    \WPot_{;z }[\xi + \FlucProp \frac {\delta}{\delta \zeta } ]
        e^{\GAction[\xi+\zeta ]}\Einsch{\zeta=0}\
    \label{SD} \\
    \WPot_{;z }&\equiv&\frac {\delta}{\delta \xi (z) }\WPot.\nn
\end{eqnarray} 
If $\FlucProp $ has zero modes, 
this is only valid for the directional derivative in $\H_C$ directions, i.e.
when multiplied from the left with $\FlucProp $. Additive constants in $\GAction$
drop out. 

Let
\begin{eqnarray}
  \VPotE [\MFAField , \FlucField ] &=& \ActionE [\MFAField + \FlucField ] -
  \half \scalar{\FlucField}{\FlucPropE_C[\MFAField ]^{-1} \FlucField},   
  \label{VDef}  \\
e^{\GActionE[\MFAField , \xi ]}&=&
  \int d\mu_{\FlucProp }(\FlucField ) 
        e^{-{\VPotE [\MFAField ,\FlucField + \xi ]}} 
        \ \ \  \mbox{for }\ \xi \in \HCE , 
        \label{GDef} \\
\FlucProp &=& \FlucPropE [\MFAFieldE [\FieldZ]].
\end{eqnarray}
We chose to work with an unnormalized generating function for amputated
 Greens functions again; the normalized generating function is
$$
    \KActionE [\MFAField, \xi ]= \GActionE[\MFAField , \xi ]-\GActionE[\MFAField , 0 ]
$$
$\FlucPropE_C[\MFAField ]^{-1}$ is the inverse of the restriction of  
$\FlucPropE [\MFAField ] $ to its range $\HCE$.  
Dropping an additive constant, the exact effective action is 
\begin{eqnarray}
    \ActionZ [\FieldZ ] &=&
     -\GActionE[\MFAField , 0] -
     \half tr \ln \FlucPropE [\MFAField ], \label{GEffAction}\\
    \MFAField &=& \MFAFieldE [\FieldZ ].
\end{eqnarray}
The second term compensates for the transition from the unnormalized Gaussian
measure in definition (\ref{defEffective}) to a normalized one. It involves a 
trace over $\HCE$, cp. section \ref{GaussSection}.

The Schwinger Dyson equation (\ref{SD}) carries over literally; 
$\VPot,\WPot,G$ all
depend parametrically on $\MFAField $. 
The single power of $\FlucField $ which arises from differentiating
the second term in eq. (\ref{VDef}) converts into a differential operator. 
As a result one obtains
\begin{eqnarray}
    \ActionE [\MFAField + \xi ] &=&
     :\NActionE [\MFAField , \xi]:\\
    \xi (z) &=& e^{-\GActionE [\MFAField , \xi ]}
    \int_{\yE }\FlucPropE (\xE ,\yE )\NActionE_{;\yE }[\MFAField ,
    \xi + \FlucPropE \frac {\delta}{\delta \zeta }] e^{\GActionE [\MFAField ,\xi+\zeta ]}
     \Einsch{\zeta=0},\label{SDGeneral}\\
    \NActionE_{;\yE } &=& \frac {\delta}{\delta \xi (\yE )}\NActionE .  \label{SD1}
\end{eqnarray}

The above Schwinger Dyson equations 
are true for any choice of the background field $\MFAField $ and for any 
choice of the propagator $\FlucPropE $. A great simplification results if
they are chosen judiciously. 

It is customary to call the amputated Green functions {\em vertices}. $n$-point
vertices are obtained as $n$-th $\xi$-derivatives of $\KActionE$ at $\xi=0$. They
depend parametrically on the background field. In this terminology
the first and second derivative $\KActionpE$ and $\KActionppE$
of $\KActionE$ with respect to $\xi $ are the one- and two point vertices.
Only the directional derivatives in directions in $\HCE$ are defined. We
require
\begin{eqnarray}
    \FlucField_1 \KActionpE[\MFAField,\xi = 0] &=& 0, \label{C1}\\
    \FlucField_1 \KActionppE[\MFAField ,\xi = 0]\FlucField_2 &=& 0.\label{C2}
\end{eqnarray}
at $\MFAField = \MFAFieldE [\FieldZ ]$ and for arbitrary 
$\FlucField_1, \FlucField_2 \in \HCE$. These conditions are shown
graphically in figure \ref{BildE}.
In the figures, the vertices obtained by differentiating $\KActionE$ 
are cross hatched.
The Schwinger Dyson equations for the n-point functions 
are obtained  by differentiating 
eq. (\ref{SDGeneral}) with respect to 
$\xi$ at $\xi=0$. The resulting equations for one, two and three point
functions are shown in graphical form in figure \ref{BildF}.

\BildE{Condition on one- and two-point vertices in the Schwinger Dyson
 approach}
\BildF{Schwinger Dyson equations for the one-,
 two- and three-point vertices.
The first two equations serve to determine the background field and the 
fluctuation field propagator.}
\BildG{Schwinger Dyson Gap equation in 1-loop approximation}

 In principle the vertices can be computed from the
Schwinger Dyson equations 
 by iteration. As a consequence of conditions (\ref{C1},\ref{C2}) 
the equations for the one- and two point functions substitute for
the earlier equations for the
background field $\MFAField $ and  fluctuation propagator $\FlucPropE $. 
In zeroth approximation, the vertices with three and more legs are equal
to the corresponding normal ordered amplitudes (``bare vertices'').  
The iteration proceeds by inserting the previous approximation for the 
vertices on the right hand side of the 
equations for the three-point functions, and similarly in
higher ones. 

Because we have an infrared cutoff $\LatSpaceE^{-1}$, we can expect that
the iteration converges  fast. 

The effective action can be computed from eq. (\ref{GEffAction}). We 
determine its $\FieldZ$-derivative by differentiating the definition 
(\ref{GDef}) of $\GActionE[\MFAField, 0]$.
 Using the change of covariance lemma (\ref{ChangeCov})
again, we obtain
\begin{eqnarray}
\frac{\delta}{\delta \MFAField (\xE )} \GActionE[\MFAField ,0]
&=& e^{-\GActionE[\MFAField , 0]}\int d\mu_{\FlucProp }(\zeta )\nn \\
& &
  \left( \half \scalar{\frac{\delta}{\delta \zeta}} {\FlucProp_{,\xE }
  \frac{\delta}{\delta \zeta}} -  \VPotE_{,\xE}[\MFAField , \zeta ]\right) 
     e^{-\VPotE[\MFAField , \zeta ]} \label{DG}
\end{eqnarray}
where $\FlucProp = \FlucPropE[\MFAField ]$. 
We restrict attention to 
$\FieldZ$-derivatives.
From the definition of $\VPotE$ we get
\begin{eqnarray}
  \frac{\delta}{\delta \FieldZ (\xZ )} \VPotE[\MFAFieldE[\FieldZ ],\zeta ]
  &=& \int \MFAField_{,\xZ} (\xE ) \left( \ActionE_{,\xE }+\half
      \scalar{\zeta} {\FlucProp_C^{-1}\FlucProp_{,\xE }\FlucProp_C^{-1} \zeta} 
  \right) \\
  &=& \int \MFAField_{,\xZ} (\xE ) \left(: \NActionE_{;\xE }:+\half
     :\scalar{\zeta} {\FlucProp_C^{-1}\FlucProp_{,\xE }\FlucProp_C^{-1} \zeta}:
   + \half \tr \FlucProp_{,\xE }\FlucProp_C^{-1} \right)    \nn
\end{eqnarray}
This is inserted into eq. (\ref{DG}). We convert  again the downstairs factors 
of $\zeta $ into differential operators, using the integration by parts
formula. One term cancels against the second term in eq. (\ref{GEffAction}) 
and we obtain the final result
\begin{eqnarray}
    \frac {\delta}{\delta \FieldZ (\xZ )}
        \ActionZ [\FieldZ] &=&
        \int_{\xE} \MFAField_{,\xZ }(\xE) \QActionpE[\MFAField,0 ](z)
        \label{SDActionFirstDerivative} \\
    \QActionpE[\MFAField, \xi ](z)&\equiv &
        \NActionE_{;\xE }[\MFAField , \xi +
        \FlucProp \frac {\delta}{\delta \FlucField }]
        e^{\KActionE[\MFAField ,\xi + \FlucField ]}
        \Einsch{\FlucField=0}
 \label{SDAction1}
\end{eqnarray}
The result is shown in graphical notation in figure \ref{BildJ}.
The derivative of the effective action can be computed when the
$\KActionE$-vertices have been determined from the Schwinger Dyson equation.

We consider next the equation for the background field.
Comparing the above definition of $\QActionpE$ with  the Schwinger Dyson equation
(eq. (\ref{SDGeneral}) at $\xi = 0$) for the 1-point function we see that
the equation for the background field   takes the form
\begin{equation}
    \zeta \QActionpE[\MFAField ] =0 \ \ \mbox{for all } \ \zeta \in \HCE
\end{equation}
The Schwinger Dyson equation for the 2-point function
implies 
\begin{eqnarray}
    \FlucPropE \QActionppE \FlucPropE = \FlucPropE.
\end{eqnarray}
One deduces (in the by now familiar way) the formula
for the derivative of the background field, 
\begin{equation}
    \MFAFieldE_{,\xZ }(\xE ) = \Adj{ \tilde \AvOpE}(\xE , \xZ )
    - \int_{\yE, u} \FlucPropVarE(\xE , \yE)
    \QActionpE{}_{\,,u}(\yE) \Adj{ \tilde \AvOpE}(u , \xZ ).
\label{backDerQ}
\end{equation}
$\FlucPropVarE$ is defined as inverse of the mixed second derivative 
$\QActionpPE$ on $\HCE$. Also we have
\begin{equation}
  \FlucPropE_{,\xE } = -\FlucPropE \QActionppE{}_{\,,\xE} \FlucPropE.
  \label{Q}
\end{equation}

One may finally 
differentiate  (\ref{SDActionFirstDerivative})
once more to get an equation for the second derivative of
the effective action. The term involving the second derivative of the
background field vanishes again because the constraints imply that 
$\MFAFieldE_{,\xZ \yZ }\in  \HCE$ (see (\ref{MFAFieldPPinHCE}). As a result
\begin{multline}
   \frac{\delta}{\delta \FieldZ (\xZ )\delta \FieldZ (\yZ )} 
        \ActionZ [\FieldZ ] = \\
   \int_{\xE\yE} \MFAField_{,\xZ }(\xE)\nn \biggl(
        \NActionE_{;\xE\,,\yE } [\MFAField,
        \FlucProp \frac{\delta}{\delta \FlucField } ] +
        \int_u 
        \NActionE_{;\xE u } [\MFAField ,
        \FlucProp \frac{\delta}{\delta \FlucField } ]
        \left(\FlucProp_{,\yE} \frac{\delta}{\delta \FlucField }
                \right)(u)  \\
   + \NActionE_{;\xE } [\MFAField ,
        \FlucProp \frac{\delta}{\delta \FlucField } ]
        \KActionE_{,\yE}[\MFAField , \FlucField ] 
        \biggr)
        e^{\KActionE[\MFAField , \FlucField ]}\Einsch{\FlucField = 0}
        \ \MFAField_{,\yZ }(\yE) 
\label{SDAction2}
\end{multline}
One can insert (\ref{mixedReplacement})
and eq. (\ref{Q}) for the first term in order to eliminate a 
reference to the first $\xi-$derivative of $\NActionE $. 

The equations for the first and second derivative of the effective action 
are shown in Figures \ref{BildJ} and \ref{BildK}.
Again, there are no 1-particle reducible contributions. 
\BildJ{Recursion relation for 
  $\ActionPE$ in the Schwinger Dyson approach.}
\BildK{Recursion relation for 
  $\ActionPPE$ in the Schwinger Dyson approach. The prime denotes a 
  derivation with respect to the background field.}

From these equations one can see what are the leading corrections to the 
approximations which were considered in the previous section. To
make the comparison we take the zeroth loop order of the Schwinger Dyson 
equations. After inserting b) into c) in figure \ref{BildF} in zeroth order 
we get the simplified vertices shown in figure \ref{BildO}. Using these in
the recursion relation for the derivatives of the effective action and 
keeping only 1-loop graphs leads to
the familiar result of the previous section (figures \ref{BildN} and 
\ref{BildH}).
\BildO{Schwinger Dyson equations of the first three vertices in zeroth order}


\subsection{Two loop corrections}

When we make appropriate 
 localization approximations, the Schwinger Dyson equations should
become integro-differential equations for matrix functions of a single
 variable. We see, however, that the equations involve normal ordered 
amplitudes with more than two high frequency legs.

The Schwinger Dyson equations for the vertices and the formulae for  the
derivatives of the effective action require a separate discussion. 

We see that the two loop terms in the recursion relation for the second
derivative of the effective action involves  normal ordered amplitudes with 
three high-frequency (``hard'') legs.
 They are not known 
if we only know the second derivative of the normal ordered action 
evaluated at nearly constant field.

 Therefore we need 
 consider {\em third} derivatives of normal ordered actions 
evaluated at nearly constant fields. This introduces quantities 
$ \WPotlocE\left( \xE ,\yE , u |\eta \right) $ . They have exponential 
decay with the tree distance of $\xE , \yE , u$ with decay length not
bigger than one lattice spacing.

We will also need recursion relations for these 3-point quantities. They can be
obtained from eq.~(\ref{SDAction2}) by differentiating once more. We will not
write the result explicitly, the graphs involved are similar to those in the 
Schwinger Dyson equation (figure \ref{BildF}) for the 3-point vertices,
except that the external legs are soft ones. It appears 
that this equation involves a normal ordered amplitude with {\em four} 
hard legs (in the last graph), and so we seem to be in trouble again.
 But this is actually not really so. Because the external legs are soft, 
there are actually only {\em two} independent hard relative momenta 
in this graph as in the remaining graphs.
 The arguments of two hard legs which join to different 
soft external legs may therefore be identified, and the 
knowledge of up to third derivatives of the normal ordered 
action suffices to determine this vertex accurately enough.
 
The strategy for the Schwinger Dyson equations is different. There are 
vertices with still more hard legs. We regard the determination of the 
fluctuation propagator and of the (cross hatched) vertices as one task, 
whose input is the effective action $\ActionE$. We propose to make localization
approximations for the normal ordered actions $\NActionE $, but not for 
$\ActionE$. Therefore $\ActionE$ may be regarded as accurately known
for arbitrary fields from the previous application of the recursion formula.
Therefore one may take derivatives of any order to create hard legs.  
One needs to insert the resulting formulae into the Schwinger Dyson 
equations. As a result, one has to evaluate graphs involving propagators at
two different length scales. This complication arose already in the
Feynman Bogoliubov approximation in section \ref{FBapproximation}.

The practical evaluation of all the 
two loop graphs requires a serious effort in 
high performance computing, unless one is willing to approximate the 
rather complicated exact formulae for the bare propagators and vertices. 
We are not prepared to elaborate on this. 
In this paper we only discuss the actual evaluation of 1-loop graphs. 


%% file: rest.tex
\section{Relation to other approaches}

\subsection{Flow equation for one particle irreducible average actions}

In this paper we are chiefly interested in blocking to discrete ``flows'' 
from lattice
to lattice because we wish to investigate the accuracy of variants of the 
method by comparison with Monte Carlo simulations. But the method itself is
 not limited to this situation. We could consider actions $\Action^K$ which
depend on a  continuously variable
cutoff $K$ in place of the discrete length scales $\LatSpaceE$, and we may
leave $\kappa$ finite, or take the limit $\kappa \mapsto \infty$ at the end. 
The flow will depend on a substitute for the projector 
\begin{equation}
R_K= \kappa
\Adj{\AvOpE}\AvOpE .
\end{equation}
Differentiating the recursion relation 
 in saddle point approximation, eq.~(\ref{action})
 with respect to $t= \ln K$, we get
from the formula (\ref{GammaWithKappa}) for the fluctuation propagator
\begin{equation}
\frac {\partial}{\partial t}\Action^K
= \frac 12 tr (\Action^{K\prime \prime} + R_K)^{-1}
  \frac {\partial}{\partial t}R_K . \label{Wet}
\end{equation}
This is Wetterich's flow equation for average actions \cite{Wetterich}.
 The initial condition is also the same. We conclude that
 Wetterich's effective actions can be interpreted
as Wilson effective actions, considered as functions of the background
 field. 

This is a remarkable 
result because Wetterich's average actions are not by definition equal to
a Wilson effective action. They are one-particle irreducible while 
a computation of a Wilsonian effective action in perturbation theory contains
one particle reducible pieces. A one particle reducible piece 
 occurs also in Polchinski's version of a 
flow equation \cite{Polchinski}. 

However, there are no one-particle reducible pieces 
in the recursion relations when the Wilson effective action
is considered as a function of the background field.
For  given values of the block-spin, the background field depends not only on
the block-spin but also on the action.
When the background field is determined accurately enough (e.g. in Feynman 
Bogoliubov approximation) no one particle reducible graph appears.

\subsection{Perfect actions} 
Let us also clarify the relation of our approach with  the work of 
Hasenfratz and Niedermayer \cite{Hasenfratz} on perfect actions. They also use 
a background field $\MFAField$. Let $\sigma : \LatE \mapsto \LatZ $ be 
the map which scales the coordinates of every lattice point by a factor 
$s=\LatSpaceZ /\LatSpaceE $. This induces a map of fields 
$\sigma^{\ast }: \Z{\H}\mapsto \E{\H }$ defined by 
$\sigma^{\ast }\FieldZ (\xE ) = s^l \FieldZ( \sigma \xE )$, where $l$ is
determined by the dimension of the field. 
When seeking a perfect action in tree approximation, the $tr \ln \FlucPropE$
in the recursion relation is neglected, and the fixed point condition which 
defines a perfect action in tree approximation reads
\begin{equation}
 \ActionE (\MFAFieldE [\FieldZ ] )
 =
 \ActionE (\sigma^{\ast}\FieldZ ) . 
\end{equation}

\subsection{Legendre transforms of higher order}

It was pointed out  to us by Yu. Pismak  \cite{Pismak}
that our method invites the
application of the formalism of Legendre transform of second order
\cite{Haymaker},
and of higher order when higher order corrections are considered. 
This yields expansions in skeleton graphs similar to those familiar
from renormalization theory \cite{BD,MT}
with two-point functions and vertices which are 
determined as solutions of the Schwinger Dyson equations.

\section{Theories with fermions}

In principle the considerations of this paper apply also to theories  with 
fermions, if one knows how to block them. The $\delta$-function in
the defining recursion relation (\ref{defRG}) for the actions 
 has only a symbolic meaning in this case, but the split of the integration over Fermi fields
into integration over high and low frequencies is nevertheless possible in 
the same way as for Bose fields. But because of the different formula for 
Grassmannian Gaussian integrals, the second term in the recursion relation
(\ref{action}) changes sign. In the formulae for normal ordered actions and
for derivatives of actions, this is taken into account by the familiar rule
\begin{center}
{\em a factor (-1) for every closed loop }
\end{center} 
It is known how to block from fermions with the right number of flavors 
in the continuum to 
Kogut Susskind lattice fermions, and on from there.
How to do this in a way which preserves locality was first discovered by
G. Mai and is reviewed in \cite{KalkEtAl}; Bietenholz and Wiese found a 
similar scheme in their studies of perfect actions \cite{BW}. 

In fermionic theories the improvement of the saddle point method 
by self-consistent normal ordering can be important, see the next section. 

\section{How locality can fail}
\label{locSection} 

Things can go badly wrong when locality of the effective action 
fails due to a bad choice of block-spin. 
It is therefore essential to monitor whether the method can be expected
to give reliable results by monitoring the range
\begin{equation}   average 
  \frac{\int_{\yE } | (\xE -\yE )\WPot(\xE , \yE |\MFAField )|}
       {\int_{\yE } | \WPot(\xE , \yE |\MFAField )|}   
\end{equation}
of the interaction 
for each value of the constant field which enters into the computation of the
 desired quantity.

Failures of locality can occur for several reasons which require different  
remedies. 

One sees from the recursion relation for the derivatives of the action that 
good locality properties can only be expected if the fluctuation 
propagator $\FlucPropE (\xE ,\yE |\MFAField )$ decays with distance
$ (\xE -\yE) $ with decay length no more than one block lattice spacing .

For a $\phi^4$-action 
\begin{equation}
\FlucProp = \lim_{\kappa \mapsto \infty } \left(  
 -\Delta + m^2 + \half \lambda \phi (\xE)^2 +  
 \kappa \Adj{\AvOpE}\AvOpE \right)^{-1} .
\end{equation}
Following Balaban \cite{balaban}, one can estimate the decay properties by 
the lowest eigenvalue of the operator in brackets with $\Delta $ replaced
by the Laplacian  $\Delta^N$  
with Neumann boundary conditions on the block boundary;
basically this eigenvalue should be strictly positive and of order at least 
 one in units where $\LatSpaceZ = 1 $.
 We are interested chiefly in (nearly) constant
$\phi $; in this case one can also find the decay by Fourier analysis
cp. ref.  \cite{GawKup}.  
The  $ \kappa \Adj{\AvOpE}\AvOpE $-term effectively eliminates
the zero mode of $\Delta^N$ (constants) and the next eigenvalue of 
$\Delta^N$ has the desired magnitude. Therefore things go wrong when 
$ m^2 + \half \lambda \phi (\xE)^2 $ becomes too strongly negative. This can 
happen when $\phi \approx 0 $ and $m^2 $ is too strongly negative - 
i.e. near the top of a  Mexican hat which is too high, cp. 
figure \ref{mexican}.

\MexicanHat{Potential and its derivative. The middle region is unstable
  and unphysical. It is separated from the stable region by a metastable
  region. Our method cannot compute the constraint effective potential in
the unstable region because we assumed that translation invariance is
not spontaneously broken}

When $m^2$ is negative enough, something still more terrible happens. 
Numerical work revealed that the minimal action for constant block-spin zero
 is reached for background fields which have periodic domain walls of
 alternating slope. In other words, the 
auxiliary theory of section \ref{introSection} shows 
spontaneous breaking of translation symmetry by block lattice vectors. 

At intermediate RG-steps one could try to remedy this by choosing 
block-spins of fixed length - i.e. 
integrating out the modulus of the field - to obtain nonlinear 
$\sigma$-model type effective actions. 

But anyway we cannot calculate the dependence of the constraint 
effective potential on the magnetization in the unstable region 
(figure \ref{mexican}) because the assumption of no spontaneous 
breaking  of translational symmetry is physically wrong there. 

Nonlocalities can appear in another more interesting way which is relevant for 
studies of a dynamical Higgs mechanism, in particular {\em superconductivity.} 
The remedy in this case is the introduction of a composite Bose field  as 
a block-spin. In superconductivity it represents Cooper pairs. This 
mechanism was studied in some detail by Grabowski \cite{Grabowski}.

 Nonlocalities of
3-point functions will also 
induce nonlocalities of $\ActionPPE$. The formal solution 
of the Schwinger Dyson equation for the 3-point function 
 involves  (among others) the famous
 chain-of-bubble diagrams. The fluctuation propagators in these diagrams
have exponential decay. Nevertheless the sum of these diagrams can fail 
to have the desired locality properties due to a pole in momentum space 
below the UV-cutoff $\LatSpaceZ^{-1}$ of the new action. This occurs in 
fermionic theories when the coupling gets strong enough. In the 
2-dimensional Gross Neveu model and in models of superconductivity,
 one is always driven into this domain by the renormalization group flow. 
When this happens, one is forced to introduce a composite block-spin 
\cite{Grabowski}. 

\section{Motivation for the improvement of the saddle point method}

In a $\phi^4$-theory the simple saddle point method works quite well. 
But this is due to the simple form of the kinetic term in this theory.

Consider instead an action for an $n$-component field $\Field $  
in two dimensions of the 
form
\begin{equation} 
\Action [\Field ] = \int \left(\frac{2n}f 
(1  + \Field^2)^{-2}(\nabla \Field )^2 + n \delta (0) \ln (1  + \Field^2) 
\right)
\end{equation}
where $f$ is a coupling constant. In the continuum, $\delta (0) $ is 
quadratically divergent, on the lattice $\delta (0)=\LatSpace^{-2}$. 

This action comes from the $O(n+1)$-symmetric nonlinear $\sigma$-model 
in stereographic coordinates; the term proportional $\delta (0) $ comes 
from the measure on the sphere. 
 But let us forget where the action came from. 

Suppose  we wish to block from the continuum to the lattice. For simplicity,
consider first what happens if we apply a simple saddle point treatment to 
the whole theory. 
Expanding the action up to terms quadratic in the field, we get 
\begin{equation}
  \Action [\Field ] \approx \int \left( 
  \frac{2n}f (\nabla \Field )^2 + n\delta (0)\Field^2
  \right) + ...
\end{equation}
Evidently we met with disaster. There is a quadratically divergent coefficient 
in a quadratic action. 
The same problem appears if we do the high frequency integral to
compute an effective lattice action from the continuum action.
   
If we normal order first before expanding in the field, the situation 
is different. Let us imagine a finite volume and normal order with respect to  
the massless free propagator $v_{Cb} $
 without its zero mode. Normal ordering the
leading term  $f(\nabla \Field )^2\Field^2 $ produces a quadratic term
$- f \Delta v_{Cb}(0) \Field^2 = f\delta (0) \Field^2 $ 
which cancels against the quadratically divergent term from the measure. 

The disease of the simple saddle point method comes from its lack of
covariance under field reparameterization.  This can lead to catastrophic 
violations of Ward identities in theories with symmetries. 
\footnote{It is known 
that Ward identities guarantee a correct cancellation of divergences in the 
nonlinear $\sigma$-model \cite{sigmaModel}.}
Since we shall want to apply our method to such theories in the future, it 
was essential to go beyond a simple saddle point approximation. 

Another advantage of the normal ordered version of our method (section 
\ref{FBapproximation}) consists in the fact that it enlarges the class of
theories which can be dealt with to theories in which the field assumes a
{\em discrete} set of values, such as the discrete Gaussian model, Ising
and Potts models etc. This is true because any action can be expanded
in normal ordered products of the fields.


%% file: discreteModels.tex
\section{Models whose fields assume discrete values}
Here we wish to explain how normal ordering can help to deal with models whose
fields assume a discrete set of values. Only the main idea will be presented;
detailed numerical studies remain to be done.

Consider as an example the discrete Gaussian Model in two or three dimensions.
It lives on a lattice $\LatNull$. The field assumes integer 
multiples of $2\pi $ as its values.
 The starting action is 
\begin{equation} 
\ActionNull (n) = \frac 1 {2\beta} \int_{\xE} [ \nabla_{\mu }n (\xE )]^2 , 
\ \ \  n(\xE ) \in 2\pi {\bf Z} 
\label{discreteGauss}
\end{equation}
In two dimensions, this model is related by a duality transformation to the
plane rotator model (with Villain action \cite{Villain}) which has
 a Kosterlitz Thouless phase transition.
 In three dimensions it is the dual
transform of a U(1)-lattice gauge theory. The renormalization group flow of
this model is well understood and lead to a rigorous proof that
the three dimensional U(1)-lattice gauge theory shows linear confinement for 
arbitrary values of the coupling constant \cite{Confine}.
 The overall factor in the action  is written as
$1/\beta$ because the duality transformation
 interchanges high and low temperatures.

The basic idea is to regard the discrete field as a function of a continuous
 field $\Field $. One regards the action as a function of this continuous 
field. One normal orders it in a self-consistent way. This furnishes at the 
same time a split into a free action which is quadratic in $\Field $ 
and an interaction. In favorable cases, a good approximation to the 
self-consistent split can be guessed a priori. 

The normal ordered action is an entire function of the 
field and can be expanded in powers of the field components which one wishes
 to integrate out. 

Let $N(\xi ) $ be the periodic function of $\xi \in {\bf R}$ with period 
$2\pi \beta^{-1/2}$ 
which is defined by
$$
N(\xi ) = \xi \ \ \mbox{for } -\pi \beta^{-1/2}<\xi < \pi \beta^{-1/2}.
$$
We note that $N(\xi )$ is small for all values of $\xi $ if $\beta $ is large. 
Now we may write
$$ n(\xE ) = \beta^{1/2}\left[ \Field (\xE) - N(\Field (\xE)) \right]$$
We may substitute an integration over $\Field (\xE )$ for
the sum over $n(\xE)$. The action takes the form
\begin{eqnarray}
\ActionNull (\Field )&=&
  \frac 1 2  \int_{\xE} [ \nabla_{\mu }\Field (\xE )]^2 + 
\VPotNull_1(\Field )+\VPotNull_2(\Field ) ,\\
\VPotNull_1 (\Field ) &=& \int_{\xE}
  N(\Field (\xE))\Delta \Field (\xE) \\
\VPotNull_2 (\Field ) &=& \frac 1 2 \int_{\xE}
 [\nabla_{\mu }N(\Field (\xE)]^2 .
\end{eqnarray} 
Given the propagator $\FlucProp $ with respect to which we want to normal order, the normal ordered form of the action is determined by Gaussian integration
(\ref{defNO}). In the case at hand, the result can be computed by Fourier
transformation.  (There are also more direct methods which exploit the fact
 that the Lagrangian depends only on the field at two lattice points
 \cite{MackPordt}). 

The periodic function $N(\varphi )$ of $\varphi \in {\bf R} $ admits a Fourier expansion 
\begin{eqnarray}
  N(\varphi )&=&\sum_{m \in {\bf Z}\setminus\{0\}} N_m e^{i m\beta^{1/2} 
    \varphi } , \\
  N_m &=& \frac i{m}\beta^{-1/2} (-1)^m . 
\end{eqnarray}
 
We wish to normal order in the fluctuation field
$\FlucField $ which is to be integrated 
out in one RG-step. We decompose as usual 
$ \Field = \MFAField + \FlucField  $,
$\MFAField = \MFAField^0 [\FieldZ ]$.
%
Now we can write the three terms in the action in normal ordered form.
The quadratic term is trivial and remains quadratic. We use
 partial integration  and the formula  (\ref{qGauss}) for the 
characteristic function of a Gaussian measure to compute the normal ordered
form of $V^0_1$.
\begin{equation}
\begin{split}
& \int d\mu_{\FlucProp }(\xi ) \VPotNull_1 (\Field + \xi ) =\\
&  \int d\mu_{\FlucProp }(\xi ) \int_{\xE} \left(
 \Delta \Field (\xE)
N((\Field + \xi)(\xE))
+ \int_{\yE} \Delta \FlucProp (\yE , \xE ) \frac {\delta}{\delta \xi (\xE)} 
  N(( \Field + \xi )(\xE))\right)\\
& = \sum_{m>0}2 i N_m 
\int_{\xE}  e^{-\beta m^2\FlucProp (\xE , \xE)/2}\\
&
 \left(
\Delta \Field (\xE )
 \sin (m \beta^{1/2}\Field (\xE )) +  \beta^{1/2}m \int_w
\Delta \FlucProp (\yE , \xE ) \cos (m \beta^{1/2}\Field (\xE ))
\right) \\
& \Field = \MFAField + \FlucField  .\\
\end{split}
\label{V1}
\end{equation}
The second term $V^0_2$ can be treated in a similar fashion. First the 
lattice derivatives must be written  as  differences of two terms. 
After that, the Fourier expansions are inserted and the Gaussian integration
can be performed as before. We will not give the somewhat complicated result
in full,  but only an approximation to 
the resulting effective action which is valid when $\beta $ 
is large. Although it is not quantitatively correct when $\beta $ is not large,
it will suffice to discuss qualitative features for orientation. 
The $V_2^0$-term is small compared to the $V_1^0$-term in this limit. 

The crucial aspect of the model is the  breaking of the symmetry under 
field translations 
$$ \Field (z)\mapsto \Field(z) + 2\pi\beta^{-1/2} a. $$
The free action is symmetric under translations with $a \in {\bf R}$, but
the interaction breaks the symmetry down to translations with $a \in {\bf Z}$.

When $\beta $ is large, the expressions
 $\exp \left(- \beta m^2 \Gamma(\xE,\xE)/2 \right)$ are exponentially 
small for $m\not=0$, and the leading  symmetry breaking terms are those with 
$m=\pm 1$. We neglect to write the others, and also the small 
symmetry breaking corrections to the generalized kinetic term .

The effective action  $\Action^1$ 
after one step is given by eq. (\ref{Normalaction}). It depends
 on the normal ordered form of $\ActionNull$
at $\FlucField = 0$, on the fluctuation propagator and on 
the background field as we know. It takes the following form in $d$ dimensions
\begin{eqnarray}
\Action^1 (\FieldZ ) &=& \int_{\xE } 
\left( \frac 12 [ \nabla \MFAField (\xE ) ]^2  + \lambda (\xE ) 
 \cos ( \beta^{1/2} \MFAField (\xE ))  - \frac 12\ln \Gamma (\xE ,\xE)
+ ... \right) \\
{\lambda }(\xE ) &=&
2e^{ - \beta  \Gamma(\xE, \xE)/2} \int_w\Delta\FlucProp(\yE,\xE).\\
\MFAField &=& \MFAField^0[\FieldZ] \\
\Gamma &=& \Gamma^0 [\MFAField ].  \label{effGauss}
\end{eqnarray}
For a translation invariant propagator $\Gamma $, $\lambda (\xE )$ would be
 constant. The propagator depends on the dimension $d$.  

 Below
we will briefly discuss some qualitative implications on the basis of this
approximate expression,  
 neglecting  a possible $\MFAField $-dependence of $\Gamma $. 
\footnote{The standard free $\MFAField$-independent fluctuation propagator
of Kupiainen and Gawedzki \cite{GawKup} 
gives a good a priori guess for the self-consistent split in this
particular model.}
 Quantitative calculations could start from the exact
normal ordered expression for $\ActionNull$. 
The exact 1-loop effective action after one step is still given by 
eq. (\ref{Normalaction}), which can be differentiated. A localization approximation 
is made after the first RG-step, and the calculation proceeds in the usual way
from there on.

The fluctuation propagator depends on an infrared cutoff 
$M=\LatSpace_1^{-1} $. For orientation, we may imagine that the cutoff is
lowered by as much as we please in one step; we discuss what happens if 
$M\mapsto 0$.

 In $d=3$ dimensions, the propagator $\Gamma $ has a limit 
when $M\mapsto 0$. Therefore, also ${\lambda }$ has a limit. However, to 
judge the RG-flow, we need to rescale to the unit lattice. Therefore we need
to consider the dimensionless quantity $a_1^{d}{\lambda }$. It  goes to 
infinity. Therefore the cosine shaped potential wells become infinitely 
high in the infrared limit, and there will always be spontaneous breaking
of the ${\bf Z}$-symmetry with a nonvanishing surface tension. 
There is also a curvature at the minimum which produces a
mass. In the dual picture, the surface tension becomes the string tension
of the $U(1)$-lattice gauge theory.  

In $d=2$ dimensions, $\Gamma $ diverges logarithmically when $M\mapsto 0$. 
The asymptotic behavior of $a_1^{d}{\lambda }$ as $M\mapsto 0 $ depends then 
on $\beta $. There will be a critical value $\beta_c$.
Its precise value depends on details which were neglected in (\ref{effGauss}). 
For $\beta < \beta_c$, $a_1^2 {\lambda }$ tends to infinity, and for $\beta
>\beta_c $ it tends to zero. This corresponds to the two phases
of the plane rotator  which are
separated by the Kosterless-Thouless phase transition. The
 ${\bf Z}$-symmetry is either spontaneously broken or enhanced to ${\bf R}$. 
This phenomenon of symmetry enhancement was discovered long ago by Fr\"ohlich 
and Spencer \cite{FS}. 

It might also be of interest to consider the dual transform of the 
4-dimensional $U(1)$-lattice gauge theory which is a
 ${\bf Z}$-gauge theory. In principle it could be treated with our method. 
Discussions on how to deal with gauge theories are found in
\cite{balabanGauge,balabanJaffe,KMP}.


%% file: appendix.tex
\appendix
\section{Formulas for saddle point approximation}

\subsection{Form of $\FlucProp$}
\label{GammaFormAp}

We are looking for the propagator which corresponds to the quadratic term
$\half\scalar{\FlucFieldE}{\ActionPPE[\MFAFieldE[\FieldZ]]\FlucFieldE}$ in 
(\ref{MFAintegral}). The fluctuation fields $\FlucFieldE$ obey the 
constraint $\AvOpE\FlucFieldE=0$.
Let\footnote{$\ActionPPE$ has in general no zero modes. If there are some
             zero modes (e.g. for $\FieldE=0$) then they can be eliminated
             with the help of projection operators before building the inverse
             ${\ActionPPE}^{-1}$. This is well known (see \cite{GawKup}).}
\begin{equation}
    \FlucPropE = {\ActionPPE}^{-1} -
      {\ActionPPE}^{-1}\Adj{\AvOpE}(\AvOpE{\ActionPPE}^{-1}\Adj{\AvOpE})^{-1}
           \AvOpE{\ActionPPE}^{-1}\,.
    \label{GammaForm}
\end{equation}
It obeys the constraint $\AvOpE\FlucPropE=0$. 
Therefore $\FlucPropE:\E\H\rightarrow\HCE$. We now want to show that
\begin{equation*}
  (\FlucPropE)^{-1} = \ActionPPE
\end{equation*}
This means that 
$$
   \FlucPropE\ActionPPE\Einsch{\HCE} = \idE_{\HCE} 
   = \ActionPPE\FlucPropE\Einsch{\HCE}\,.
$$
The first equation is equivalent to 
$$
   \FlucPropE\ActionPPE\FlucFieldE = \FlucFieldE 
   \qquad \forall\ \FlucFieldE\in\HCE\,.
$$
Applying $\ActionPPE\FlucFieldE$ to (\ref{GammaForm}) from the right 
confirms this. \\
The second equation is true because it is the adjoint of the first.


\subsection{Proof of the relation for the background field}
\label{proofOfLemma}

Differentiating the defining equation  (\ref{MFAFieldEq1}) for the
background field by the chain rule, we obtain
\begin{equation}
    \ActionPPE [\MFAFieldE] \MFAFieldPE = \Adj{\AvOpE} \E\lambda{}^{\prime}
      \label{varfluc}
\end{equation}
for some $\E\lambda{}^{\prime} \in \Z\H$. Multiplying (\ref{varfluc})
with ${\ActionPPE}^{-1}$
\begin{equation}
  \MFAFieldPE = {\ActionPPE}^{-1}\Adj{\AvOpE}
     \E\lambda{}^\prime  \label{psiprime}
\end{equation}
and $\AvOpE$ gives
\begin{equation*}
  \AvOpE\MFAFieldPE = \AvOpE{\ActionPPE}^{-1}\Adj{\AvOpE}
     \E\lambda{}^\prime
\end{equation*}
By virtue of $\AvOpE\MFAFieldPE=\idZ$
it follows
\begin{equation*}
  \idZ = (\AvOpE{\ActionPPE}^{-1}\Adj{\AvOpE}) \E\lambda{}^\prime
\end{equation*}
Therefore
\begin{equation}
  (\AvOpE{\ActionPPE}^{-1}\Adj{\AvOpE})^{-1} = \E\lambda{}^\prime
  \label{lambdaprime}
\end{equation}
We have now one possible form of $\MFAFieldPE$:
\begin{equation}
    \MFAFieldPE = {\ActionPPE}^{-1}\Adj{\AvOpE}
                  (\AvOpE{\ActionPPE}^{-1}\Adj{\AvOpE})^{-1}
       \label{psiprimeVariant}
\end{equation}
Multiplying $\FlucPropE$ in (\ref{GammaForm})
with ${\ActionPPE}$ from right and using (\ref{lambdaprime}) gives
\begin{align*}
  \FlucPropE\ActionPPE &= \idE -
  {\ActionPPE}^{-1}\Adj{\AvOpE}(\AvOpE{\ActionPPE}^{-1}\Adj{\AvOpE})^{-1}
           \AvOpE \\
&= \idE -
  {\ActionPPE}^{-1}\Adj{\AvOpE}\E\lambda{}^\prime \AvOpE
\end{align*}
Inserting (\ref{psiprime}) leads to
\begin{equation}
    \FlucPropE \ActionPPE = 1 - \MFAFieldPE \AvOpE.
    \label{lemmaEquation}
\end{equation}
By multiplying $\Adj{\AvOpE}$ from the right we get the result
\begin{equation}
    \MFAFieldPE = (1-\FlucPropE \ActionPPE) \Adj{\widetilde{\AvOpE}}.
\end{equation}

\subsection{Derivative of $\FlucPropE$}
\label{FlucPropDer}

Multiplying (\ref{lemmaEquation}) with $\FlucPropE$ from the right and using
$\AvOpE\FlucPropE=0$ gives
$$
    \FlucPropE\ActionPPE\FlucPropE = \FlucPropE.
$$
Differentiation leads to
$$
    \FlucPropPE = \FlucPropPE\ActionPPE\FlucPropE
            + \FlucPropE\ActionPPPE\FlucPropE
            + \FlucPropE\ActionPPE\FlucPropPE\,.
$$
$\ActionPPE\FlucPropE$ and $\FlucPropE\ActionPPE$ can be replaced now by
(\ref{lemmaEquation}). Using $\AvOpE\FlucPropPE=0$ leads to the result
$$
    \FlucPropPE  = - \FlucPropE\ActionPPPE\FlucPropE
$$
or expressed with $\WPotE$
\begin{equation}
   \FlucPropE_{,\xE} = \FlucPropE\WPotE_{,\xE}\FlucPropE\,.
   \label{GammaDerivative}
\end{equation}

\subsection{Determination of the background field}

An equation for $\MFAFieldE_{,\xZ }[\FieldZ ] (\xE)$ was already derived in
\ref{proofOfLemma}. It involves quantities which are obtained  together with the
propagator. Expressed in terms of $\WPotE$ it looks
\begin{equation*}
    \MFAFieldPE = (1+\FlucPropE \WPotE ) \Adj{\widetilde{\AvOpE}}.
\end{equation*}
To determine the recursion relation of $\WPotE$ we need also the second derivative.
Using the formula for the  derivative of the fluctuation propagator
(\ref{GammaDerivative}) we find
\begin{equation}
  \MFAFieldE_{,\xZ\yZ}[\FieldZ](\xE) =
  \int\limits_{\yE,\xE_1,\xE_2\in\LatE}
      \FlucPropE(\xE,\xE_1)\, \WPotE_{,\yE}(\xE_1,\xE_2)\,
      \MFAFieldE_{,\yZ}[\FieldZ](\yE)\, \MFAFieldE_{,\xZ}[\FieldZ](\xE_2).
  \label{secondDerMFA}
\end{equation}

\subsection{Recursion relation for $\WPotE$}
\label{recursionRelation}

To get $\WPotZ$ out of (\ref{actionPrel}) we need the second derivative of
$\ActionZ$.

 Consider first
\begin{equation*}
\begin{split}
\dZx {\xZ }
 \tr \ln \FlucPropE [\MFAFieldE [\FieldZ ]] = &
  \,\tr \,\FlucPropE [\MFAFieldE [\FieldZ ]] \,
    \dZx {\xZ }  \WPotE [\MFAFieldE [\FieldZ ]]  \\
 =& \int_{\xE} \tr \Bigl(\FlucPropE [\MFAFieldE [\FieldZ ]] \,
    \WPotE_{,\xE} [\MFAFieldE [\FieldZ ]]\Bigr)
    \MFAFieldE_{,\xZ }[\FieldZ](\xE ) \\
\end{split}
\end{equation*}
where we have used (\ref{GammaDerivative})
and $\WPotE_{,\xE}[\MFAFieldE]$ or $\MFAFieldE_{,\xZ }[\FieldZ]$
denote the functional derivatives
$$
    \frac{\delta}{\delta\MFAFieldE(\xE)}\WPotE[\MFAFieldE]\,,
    \qquad\qquad\dZx {\xZ} \MFAFieldE[\FieldZ]\,.
$$ 
We differentiate once more to obtain the second derivative and
omit the functional argument $\MFAFieldE[\FieldZ]$:
\begin{equation}
\begin{split}
 & \dZxy {\xZ} {\yZ} \tr \ln \FlucPropE [\MFAFieldE [\FieldZ] ] \\[8pt]
  = & \, \int\limits_{\xE,\yE\in\LatE} \!\!\!\! \tr \Bigl(
     \FlucPropE \,\WPotE_{,\xE}  
     \FlucPropE \,\WPotE_{,\yE} 
   + \FlucPropE \WPotE_{,\xE\yE} \Bigr) 
     \MFAFieldE_{,\xZ} [\FieldZ] (\xE)
     \MFAFieldE_{,\yZ} [\FieldZ] (\yE) 
   \\[5pt]
&  + \int\limits_{\xE\in\LatE} \tr \Bigl( \FlucPropE \WPotE_{,\xE}  \Bigr)
           \MFAFieldE_{,\xZ\yZ} [\FieldZ] (\xE) 
\\[3pt]
= & \int\limits_{\xE,\yE,\xE_i\in\LatE } 
         \FlucPropE (\xE_1,\xE_2)
         \WPotE_{,\xE} (\xE_2,\xE_3) 
         \FlucPropE (\xE_3,\xE_4)
         \WPotE_{,\yE} (\xE_4,\xE_1)   
    \\[-8pt] 
& \qquad\qquad\qquad
         \MFAFieldE_{,\xZ} [\FieldZ] (\xE)  \MFAFieldE_{,\yZ} [\FieldZ] (\yE)
    \\[8pt]
& + \int\limits_{\xE,\yE,\xE_1,\xE_2\in\LatE } 
    \FlucPropE (\xE_1,\xE_2)
        \WPotE_{,\xE\yE}(\xE_2,\xE_1)
        \MFAFieldE_{,\xZ }[\FieldZ ](\xE)
        \MFAFieldE_{,\yZ }[\FieldZ ](\yE) \\
& + \int\limits_{\xE,\xE_1,\xE_2\in\LatE} 
        \FlucPropE (\xE_1,\xE_2)
        \WPotE_{,\xE} (\xE_2,\xE_1)
        \MFAFieldE_{,\xZ\yZ}[\FieldZ](\xE). \\
\end{split}
 \label{WrecLast}
\end{equation}

Now we consider $\ActionE[\MFAFieldE[\FieldZ]]$.
By the chain rule
\begin{equation}
\dZxy {\xZ}{\yZ} \ActionE [\MFAFieldE [\FieldZ ]]    =
 \,\Adj {\MFAFieldE_{,\xZ } [\FieldZ ]} \,
 \ActionPPE [\MFAFieldE ] \,\MFAFieldE_{,\yZ } [\FieldZ ]
  + \ActionPE [\MFAFieldE ] \,\MFAFieldE_{,\xZ \yZ } [\FieldZ ].
\label{Wsaddle}
\end{equation}  
By the saddle point condition,
 $\ActionPE [\MFAFieldE] =  \Adj {\AvOpE}\E\lambda $.
Therefore the last term looks in coordinates
$$
    \int\limits_{\xE\in\LatE} \dZxy {\xZ}{\yZ} \MFAFieldE [\FieldZ](\xE)
    \int\limits_{\xZ_1\in\LatZ}\Adj{\AvOpE}(\xE,\xZ_1)\E\lambda(\xZ_1)\,.
$$
Inserting (\ref{secondDerMFA}) shows that this term vanishes because of
$\FlucPropE\Adj{\AvOpE}=0$.
Therefore
\begin{equation*}
    \WPotZ[\FieldZ](\xZ,\yZ) = \,\Adj {\MFAFieldE_{,\xZ } [\FieldZ ]}\,
       \WPotE[\MFAFieldE[\FieldZ]] \,\MFAFieldE_{,\yZ } [\FieldZ ]
      + \half\dZxy {\xZ} {\yZ} \tr \ln \FlucPropE [\MFAFieldE [\FieldZ]]\,.
\end{equation*}
Using (\ref{WrecLast}) and (\ref{secondDerMFA}) leads to the result
(\ref{recursionW}).


\section{Simplification of mixed derivatives}
\label{derRelSimple}

In the case of Feynman Bogoliubov approximation we defined
the propagator $\FlucPropE $ for arbitrary fields $\FieldE$ as
inverse of $\ActionPPE$ on $\HCE$ and derived some formulas in a general
manner. Afterwards we used special fields namely the background field
which depends on the block-spin. Now we regard it only
as a function of the background field. 
It is convenient to extend this definition to arbitrary
fields by specifying that it is independent of the 
high frequency components of the field,
\begin{equation}
  \FlucPropE [\MFAFieldE + \FlucFieldE ] =  \FlucPropE [\MFAFieldE]
  \label{sumDepP}
\end{equation} 
for $\FlucFieldE \in \HCE$.

It is important that the formula for the derivative of the propagator
  \begin{eqnarray} 
\FlucPropE_{,\xE}[\MFAFieldE ] \equiv 
\frac {\delta}{\delta \MFAFieldE (\xE )}\FlucPropE [\MFAFieldE ]
&=& - \FlucPropE \NActionppE{}_{\,, \xE } \FlucPropE \label{FlucPropDerN} \\
 \NActionppE{}_{\,, \xE }
&=& \frac {\delta}{\delta \MFAFieldE (\xE )}\NActionppE  \label{GammaDerN}
\end{eqnarray}
is then only valid in the restricted sense that it can be used 
for evaluating $ \FieldZ$-derivatives of the propagator. 

\sticky{
The equation is derived from eq.~(\ref{gap}) 
in the same way as in appendix \ref{FlucPropDer}.
}

With this specification of the propagator, the definition (\ref{pregap0})
of the normal ordered amplitude implies that 
\begin{equation}
\NAction [\MFAFieldE + \FlucFieldE , \xi - \FlucFieldE ] =
\NAction [\MFAFieldE  , \xi ] \label{sumDep}
\end{equation}   
for $\FlucFieldE \in \HCE$. In other words,
 $\NAction [\MFAFieldE  , \xi ]$ depends actually only on 
$\MFAFieldE + \xi $ so long as $\xi \in \HCE $.

We find it convenient, however, to extend the definition (\ref{defNO}) 
of $\NAction [\MFAFieldE  , \xi ]$ to arbitrary $\xi $. 
Eq.~(\ref{sumDep}) remains valid, of course, but 
$\NAction [\MFAFieldE,\xi]$ is no longer only a function of the
sum of its arguments.

When evaluating $\FieldZ $ derivatives, eq. (\ref{FlucPropDerN}) can be used 
to express the derivative $\FlucPropE_{,\xE }$ of the propagator. 
On the other hand, $\MFAFieldE $-derivatives in $\HCE$ directions 
are the same as $\xi$-derivatives; the last term does not contribute in this 
case because of eq. (\ref{sumDepP}).

Also the equation (\ref{backDerN}) for the derivative of the background 
field simplifies, i.e. $\FlucPropVarE=\FlucPropE$:
\begin{equation}
\MFAFieldE_{,\xZ }(\xE ) = \Adj{ \tilde \AvOpE}(\xE , \xZ )
 - \int_{\yE, u} \FlucPropE(\xE , \yE)  
 \NActionE_{;\yE\,,u} \Adj{ \tilde \AvOpE}
(u , \xZ ), \label{backDerNsimple}
\end{equation}
It fulfills also (\ref{saddleConditionDerivative}) and the constraint
(\ref{constr}). One needs the identity
\begin{equation} 
\int_{\xE ,\yE}\FlucPropE(*,\xE )\PsiderivativeE{\xE}
\NActionE_{;\yE } \FlucPropE (\yE , *) = \FlucPropE (*,*).
\end{equation}
$\MFAFieldE $- and $\xi$-derivatives of $\NActionE$ agree in $\HCE$-directions, 
and the range of $\FlucPropE$ is $\HCE$. Therefore we may convert
the $\MFAFieldE$-derivative into a $\xi$-derivative. The assertion follows  
now from the gap equation (\ref{gap}).

\section{Momentum space representation}
\label{GammaLattice}

Let $\LatE$ be a $\Dim$-dimenional position space lattice with 
lattice spacing $\LatSpaceE$ and length $\LatLengthE=\LatExtE\LatSpaceE$.
The corresponding dual lattice $\DualLatE$ has a lattice spacing 
$\DualLatSpaceE=2\pi/\LatLengthE$ and length $\DualLatLengthE=2\pi/\LatSpaceE$.
The block factor between two successive lattices shall be denoted by 
$\BlockFactor=\LatSpaceZ/\LatSpaceE$. 
It is convenient to introduce an additonal momentum space lattice
$\DualLatEZ$ with spacing $\DualLatSpaceEZ=\DualLatLengthE$ and length
$\DualLatLengthEZ=s\DualLatLengthE$.
Then any $\qE\in\DualLatE$ can be decomposed into $\qE=\qZ+\lZ$ with
$\qZ \in \DualLatZ$ and $\lZ \in \DualLatEZ$.
The definitions
\begin{alignat}{3}
  \int \limits_{\qE} &:= \frac{\DualLatSpaceE^D}{(2\pi)^D} 
\sum\limits_{\qE \in \DualLatE}
&\quad,\quad
 \int \limits_{\qZ} &:= \frac{\DualLatSpaceZ^D}{(2\pi)^D} 
\sum\limits_{\qZ \in \DualLatZ}
& \quad,\quad
\int \limits_{\lZ} & :=  \sum \limits_{\lZ \in \DualLatEZ} 
\end{alignat}
imply the split
\begin{equation}
   \int \limits_{\qE} =  \int \limits_{\qZ}  \int \limits_{\lZ}.
\end{equation}

We now want to derive the momentum space representation (\ref{flucmorep}) of
the fluctuation propagator.
First note that any linear operator $L:\E\H\rightarrow\E\H$ with 
$L(\xE,\xE')=L(\xE-\xE')$ can
be written as
\begin{equation}
  L(\xE-\xE') = \int \limits_{\qE} L(\qE) e^{i\qE(\xE-\xE')}.
\label{fourierrep}
\end{equation}
Examples are $\WPotlocE(\xE,\xE'|\FieldlocE)$ in the saddle point approximation
and the laplacian $\E\Laplace(\xE,\xE')$. In the latter case one gets
\begin{alignat}{2}
  \E\Laplace(\qE) & = - \hat{\qE}^2  &
  \qquad,\qquad \hat{\qE}_{\mu} 
  &= \frac{2}{\LatSpaceE}\sin\left(\frac{\qE_{\mu}\LatSpaceE}{2}\right).
\end{alignat}

The representation (\ref{fourierrep}) also holds for the 
averaging operator $\AvOpE(\xZ,\xE)$ since $\xZ \in \Z\H \subset \E\H$ 
with
\begin{equation}
\AvOpE(\qE) =
  \prod \limits_{\mu=1}^{D}
\left(\frac{\sin(p_\mu s \LatSpaceE/2)}{s\sin(p_\mu \LatSpaceE/2)} 
e^{+i p_\mu (s-1) \LatSpaceE/2}
\right)
\end{equation}

The fluctuation propagator $\FlucPropE$ on the other hand
is only invariant under translations
 $\FlucPropE(\xE,\xE')= \FlucPropE(\xE-\xZ,\xE'-\xZ)$ 
with   $\xZ \in \LatZ$ and therefore has the more complicated Fourier 
representation
\begin{equation}
          \FlucPropE(\xE,\xE') =\int\limits_{\lM,\q,\lM'} 
                \FlucPropE(\lM,\q,\lM') e^{i(\q+\lM)\xE-i(\q+\lM')\xE'}
\end{equation}
with
\begin{eqnarray}
   \FlucPropE(\lM,\qZ,\lM')&=&
   \E{v}(\qZ+\lZ)\delta_{\lM,\lM'} \nonumber\\
     & &   - \E{v}(\qZ+\lZ)\Adj\AvOpE(\qZ+\lZ)  \E{u}^{-1}(\qZ)
             \AvOpE(\qZ+\lZ') \E{v}(\qZ+\lZ') \\
  \E{u}(\qZ) &=& \int \limits_{\lZ} \AvOpE(\qZ+\lZ)\E{v}(\qZ+\lZ)\Adj\AvOpE(\qZ+\lZ)
\\
\E{v}(\qE) & :=& \WPotlocE(\FieldlocE)^{-1}(\qE)
\end{eqnarray}



%% file: paper.bbl
\begin{thebibliography}{1}

\bibitem{balaban}
    T. Balaban, Regularity and Decay of Lattice Green's Functions, 
    Commun. Math. Phys. {\bf 89} (1983), p. 571

\bibitem{balabanGauge}  
    T. Balaban,  Commun. Math. Phys. {\bf 109} (1983) 571

\bibitem{balabanJaffe} 
    T. Balaban, A. Jaffe, Constructive Gauge Theory,
    in: Erice School Math. Phys (1985) 207

\bibitem{BW} 
    W. Bietenholz and U. Wiese, Nucl.Phys.Proc.Suppl. 34, 516-521 (1994)

\bibitem{BD}
    J. D. Bjorken and S. D. Drell, vol 2, Relativistic Quantum Fields,
    Mc Graw-Hill Book Inc., New York, 1965

\bibitem{sigmaModel}
    E. Br\'{e}zin, J. Zinn-Justin and J. C. Le Guillou, Phys. Rev {\bf B14},
    4976 (1976) 

\bibitem {FS} 
    J. Fr\"ohlich, T. Spencer,
    Phase diagrams and critical properties of (classical) Coulomb systems.
    Erice lectures 1980\\
    J. Stat.Phys {\bf 24} (1981) 617 

\bibitem{GawKup}
    K. Gawedzki and A. Kupiainen, A Rigorous Block Spin Approach to Massless 
    Lattice Theories, Commun. Math. Phys. {\bf 77}, 31-64 (1980)

\bibitem{GlimmJaffe}
    J. Glimm and A. Jaffe, Quantum Physics: A Functional Integral Point of 
    View, Springer Verlag, New York 1987,  second edition


\bibitem{Confine}
    M. G\"opfert, G. Mack, Proof of confinement of static quarks in 
    3-dimensional $U(1)$-lattice gauge theory for all values of the 
    coupling constant, Commun. Math. Phys. {\bf 82} (1982) 545

\bibitem{Grabowski}
    M. Grabowski, Low Energy Effective Actions with Composite 
    Fields, PhD thesis Hamburg, DESY 94-146

\bibitem{Hasenfratz}
    P. Hasenfratz, F. Niedermayer et al. , Nucl. Phys. B414 (1994), 
    785-814,
    Nucl. Phys. B454 (1995), 615-637,
    Nucl. Phys. B454 (1995), 638-644,
    Nucl. Phys. B454 (1995), 587-614

\bibitem{Haymaker} Richard W. Haymaker, Variational Methods for Composite 
    Operators, Riv.Nuovo Cim.14, no.8: 1-89, 1991

\bibitem{KMP}
    U. Kerres, G. Mack, G. Palma, Nucl. Phys. (Proc. Suppl.) B42 (1995), 584-586, and: 
    Perfect 3-dimensional lattice actions
    for 4-dimensional quantum field theories at finite temperatures,
    DESY 94-226, submitted to Nucl. Phys. B. 

\bibitem{MackCargese}
    G. Mack, Multigrid methods in Quantum Field Theory, {\em in}
    G. 't Hooft et al (eds), Nonperturbative Quantum Field Theory
    (Cargese Proceedings 1987) Plenum Press New York 1988
   

\bibitem{MackPordt}
    G. Mack, A. Pordt, Convergent weak coupling expansions for lattice field
    theory that look like perturbation series, Rev. Math. Phys {\bf 1}, 47-87, 
    (1989),  esp. Appendix A

\bibitem{KalkEtAl}
     G. Mai, Ein Blockspin f\"ur $2^d/2$ Fermionen, Diplomarbeit 
     Universit\"at Hamburg,  1989 .\\
     G. Mack, T. Kalkreuter, G. Palma, M. Speh, Effective Field Theories, {\em in}
     H. Gausterer, C.B. Lang (eds.) Computational Methods in Field Theory , 
     Lecture Notes in Physics {\bf 409}, Springer Heidelberg 1992 

\bibitem{MT} 
     G. Mack. I.Todorov,  Conformal invariant Green functions without 
     ultraviolet divergences, Phys. Rev. D8 (1971) 1764

\bibitem{Pismak}
    Yu. Pismak, private communication

\bibitem{Polchinski}
    J. Polchinski, Renormalization and Effective Lagrangians, 
    Nucl. Phys. B231 (1984) 269-295

\bibitem{Pordt}
    A. Pordt and C. Wieczerkowski, Nonassociative Algebras and Nonperturbative 
    Field Theory for Hierarchical Models, 74 pages, report MS-TPI-94-04   

\bibitem{Ruehl}
    W. R\"uhl, Mean Fields and Self Consistent Normal Ordering of Lattice
    Spin and Gauge Field Theories, Z. Phys. C (1986), 32, p. 265--278

\bibitem{Villain}  
    J. Villain, J. Phys (Paris) {\bf 36} (1975) 631

\bibitem{Wetterich}
    C. Wetterich, Exact Evolution Equation for the Effective Potential,
    Physics Letters B301 (1993) 80-94

\end{thebibliography}
